\documentclass[reprint,amsmath,amssymb,aps]{article}
\usepackage{authblk}
\usepackage{graphicx}
\usepackage{amsmath,amssymb}
\usepackage{mathrsfs}
\usepackage{booktabs} 
\usepackage{subfigure}  
\usepackage{bbm}
\usepackage{slashed}
\usepackage{dcolumn}
\usepackage{multirow}
\usepackage{bm}
\usepackage[title]{appendix}
\usepackage{makecell}
\allowdisplaybreaks[4]
\usepackage{url}
\usepackage[colorlinks=true,citecolor=blue,urlcolor=blue,linkcolor=blue]{hyperref}
\usepackage[numbers,sort&compress]{natbib}
\usepackage[left=0.8in,right=0.8in,top=1in,bottom=1in]{geometry}
\begin{document}
	
	\title{Form factors of $\Lambda_b^0 \to \Lambda_c(2595)^+$ within light-cone QCD sum rules}
	\author{Hui-Hui Duan$^1$\footnote{duanhuihui@htu.edu.cn}}
\author{Yong-Lu Liu$^2$\footnote{yongluliu@nudt.edu.cn}}
\author{Qin Chang$^1$\footnote{changqin@htu.edu.cn}}
\author{Ming-Qiu Huang$^2$\footnote{mqhuang@nudt.edu.cn}}
\affil{$^1$Institute of Particle and Nuclear Physics, Henan Normal University, Xinxiang 453007, Henan, People's Republic of China}
\affil{$^2$Department of Physics, National University of Defense Technology, Changsha 410073, Hunan, People's Republic of China}
\renewcommand*{\Affilfont}{\small\it}
                 
    \date{}
\maketitle

    \begin{abstract}
     In this work, we calculated the form factors of the weak decay process $\Lambda_b^0 \to \Lambda_c(2595)^+$, where the final charm baryon represents an excited state with spin-parity $\frac{1}{2}^-$. Utilizing the light-cone QCD sum rules approach, we incorporated the contributions of the lowest two charm baryon states: the ground state $\Lambda_c$ with $J^P=\frac{1}{2}^+$ and the excited state $\Lambda_c(2595)^+$ with $J^P=\frac{1}{2}^-$ in the hadronic representation of the $\Lambda_b \to \Lambda_c(2595)^+$ transition correlation function. This approach allows us to extract the form factors of the $\Lambda_b^0 \to \Lambda_c(2595)^+$ from $\Lambda_b^0 \to \Lambda_c^+$ transition. During the light-cone QCD sum rules procedure, we employed the light-cone distribution amplitudes (LCDAs) of the $\Lambda_b$ baryon. Furthermore, by combining these form factors with the helicity amplitudes of the bottom baryon transition matrix elements, we calculated the differential decay widths for the processes $\Lambda_b^0 \to \Lambda_c(2595)^+\ell^-\bar{\nu}_\ell$ and provided the optimal choice of the interpolating current for $\Lambda_c$ in this process. Additionally, within the lifetime of $\Lambda_b^0$, we obtained the absolute branching fractions for the semileptonic decays $\Lambda_b^0 \to \Lambda_c(2595)^+ \ell^- \bar{\nu}_\ell$. With the branching fractions of $\Lambda_b^0 \to \Lambda_c(2595)^+ \ell^- \bar{\nu}_\ell$ calculated in this work, we also determined the parameter $\mathcal{R}(\Lambda_c(2595)^+)$ which tests the lepton flavor universality. This parameter is defined as the ratio of branching fractions $\mathcal{B}r(\Lambda_b^0 \to \Lambda_c(2595)^+\tau^-\bar{\nu}_\tau)$ and $\mathcal{B}r(\Lambda_b^0 \to \Lambda_c(2595)^+\mu^-\bar{\nu}_\mu)$. Our results provide a valuable theoretical test for these decay channels and offer insights into the LCDAs of bottom baryons, paving the way for further in-depth investigations.  \end{abstract}

\section{Introduction} \label{sec:I}  
	
    Form factors play a crucial role in the understanding of hadron weak decay processes. They are essential in investigating the decay widths of weak decay processes, providing insights into the underlying dynamics inside the hadrons. Within the Standard Model, the matrix element $|V_{cb}|$ of Cabibbo-Kobayashi-Maskawa (CKM) remains an imprecisely determined quantity. Traditionally, the values of $|V_{cb}|$ have been primarily derived from both exclusive and inclusive semileptonic decays of $B$ mesons to charm. However, to obtain a complementary and more robust determination, it is imperative to incorporate data from semileptonic decays of bottom baryons to charm. Currently, the only experimental data on semileptonic decays of bottom baryons to charm focus on the decays of the $\Lambda_b^0$ baryon to the ground state $\Lambda_c$ and its excited states~\cite{ParticleDataGroup:2022pth,ParticleDataGroup:2024cfk}. Various models have investigated the decays of $\Lambda_b^0$ to the ground state $\Lambda_c^+$ with spin-parity quantum number $J^P=\frac{1}{2}^+$. Moreover, excited states of $\Lambda_c$, such as $\Lambda_c(2595)^+$ (which we will refer to as $\Lambda_c^*$ in the following context for brevity) and $\Lambda_c(2625)^+$, with spin-parity $J^P=\frac{1}{2}^-$ and $J^P=\frac{3}{2}^-$, also contribute to the semileptonic decay of $\Lambda_b^0$. The research of weak decay processes of bottom baryons with $\Lambda_c$ excited states as final states will further enrich our understanding of these decay processes, and offer valuable insights into the dynamics of bottom baryon decays.  

	In our previous research, we calculated the weak decay form factors of the $\Lambda_b^0$ to $\Lambda_c^+$ transition, as well as the branching fractions of its semileptonic decay modes, utilizing the light-cone QCD sum rules approach and employing the LCDAs of the $\Lambda_b$ baryon \cite{Duan:2022uzm}. The application of light-cone QCD sum rules in assessing hadron weak decay form factors comprises two aspects: the calculation of the hadron weak decay correlation function using LCDAs of the final state hadron, and the employment of LCDAs of the initial state hadron. Notably, by incorporating the LCDAs of the initial $\Lambda_b$ baryon, we can concurrently determine the weak decay form factors for final states with different parity quantum numbers.  
	
	The form factors associated with the transition $\Lambda_b^0 \to \Lambda_c^*$ involve the spin-parity quantum number of baryon transition from $\frac{1}{2}^+$ to $\frac{1}{2}^-$. Within the framework of light-cone QCD sum rules, these form factors can be derived from the analysis of the bottom baryon weak decay process involving the $\frac{1}{2}^+$ to $\frac{1}{2}^+$ transition. In our prior work on the calculation of $\Lambda_b^0 \to \Lambda_c^+$ form factors, we developed a methodology to isolate the contributions of the negative parity final state within the light-cone QCD sum rules framework. This methodology can be effectively applied in the current investigation to explore the phenomenology of the $\Lambda_b^0 \to \Lambda_c^*$ transition.
	
	Experimentally, the CDF Collaboration reported their measurements of the ratio of branching fractions $\mathcal{B}(\Lambda_b^0 \to \Lambda_c^+ \mu^- \bar{\nu}_\mu)/\mathcal{B}(\Lambda_b^0 \to \Lambda_c^+ \pi^-)$ in 2009. In the same report, they also presented the branching fractions of semileptonic decays $\Lambda_b^0 \to \Lambda_c(2595)^+\mu^-\bar{\nu}_\mu$ and $\Lambda_b^0 \to \Lambda_c(2625)^+\mu^-\bar{\nu}_\mu$ relative to the branching fraction of the $\Lambda_b^0\to\Lambda_c^+\mu^-\bar{\nu}_\mu$ decay \cite{CDF:2008hqh}. Notably, the spin-parity quantum numbers of the $\Lambda_c(2595)^+$ and $\Lambda_c(2625)^+$ baryons are $J^P=\frac{1}{2}^-$ and $J^P=\frac{3}{2}^-$ respectively \cite{ParticleDataGroup:2022pth, ParticleDataGroup:2024cfk}. These are the only two negative parity weak decay final states that have been experimentally determined in bottom baryon decays to date. In this work, we focus specifically on the $\Lambda_b^0 \to \Lambda_c(2595)^+$ weak decay process.  
	
	Theoretically, the form factors of $\Lambda_b^0 \to \Lambda_c(2595)^+$ and the semileptonic decay widths of $\Lambda_b^0 \to \Lambda_c^{*+}\ell^-\bar{\nu}_\ell$ have been investigated using various theoretical models and methods, including the light-front quark model (LFQM) \cite{Li:2021qod}, the constituent quark model \cite{Pervin:2005ve}, the covariant confined quark model (CCQM)\cite{Gutsche:2018nks}, lattice QCD (LQCD) \cite{Meinel:2021rbm}, heavy quark symmetry \cite{Nieves:2019kdh}, and the Bakamjian-Thomas approach with quark model \cite{Becirevic:2020nmb}. These theoretical frameworks have provided their predictions within their respective validity ranges.
	
	In this paper, we employ the light-cone QCD sum rule method to calculate the form factors of $\Lambda_b^0 \to \Lambda_c^*$ transition. By considering all Lorentz structures within the sum rules and isolating the $\Lambda_c$ and $\Lambda_c^*$ states in the correlation function, including all the contributions of excited and continuum states of $\Lambda_c$ into the spectral integral of dispersion relation, we are able to dissociate the $\Lambda_b^0 \to \Lambda_c^*$ decay channel from the $\Lambda_b^0 \to \Lambda_c^+$ channel within the framework of the sum rules. In the main body of the text, we utilize the LCDAs of $\Lambda_b$ discussed in Ref.~\cite{Ali:2012pn, Ball:2008fw, Bell:2013tfa}. The LCDAs of $\Lambda_b$ were initially explored in Ref.~\cite{Ball:2008fw} and subsequently extrapolated to ground-state bottom baryons in Ref.~\cite{Ali:2012pn}. The application of $\Lambda_b$ baryon LCDAs in heavy hadron weak decays has been extensively researched, yielding promising results~\cite{Wang:2009yma, Lu:2009cm, Wang:2009hra, Feldmann:2011xf, Wang:2015ndk, Shi:2019fph, Han:2022srw, Zhang:2022iun, Huang:2022lfr, Aliev:2022maw, Rui:2022sdc, Miao:2022bga, Rui:2022jff, Aliev:2023tpk, Feldmann:2023plv, Shi:2024uqs}, yet compared to the study and application of B meson LCDAs, this field still requires further development.  
	
	In the subsequent sections of this article, we present the theoretical framework for the form factors of $\Lambda_b^0 \to \Lambda_c^*$ within the light-cone QCD sum rules in Section~\ref{sec:II}. The numerical analysis of these form factors, as well as the semileptonic decay width and branching fractions of $\Lambda_b^0 \to \Lambda_c^*\ell^-\bar{\nu}_\ell$, is carried out in the helicity amplitude formalism and detailed in Section~\ref{sec:III}; the ratio $\mathcal{R}(\Lambda_c^*)$ of lepton flavor universialty is also given in this section. Finally, we conclude this work and discuss our results in the last section.  
	
   \section{Weak decay form factors of $\Lambda_b^0 \to \Lambda_c(2595)^+$ } \label{sec:II}
		
	In the context of the $\Lambda_b^0 \to \Lambda_c^+$ ($\frac12^+ \to \frac12^+$) transition, form factors of weak decay matrix element are generally defined as  	
	\begin{align}  
		\langle \Lambda_c |  \bar{c}\gamma_\mu(1-\gamma_5)b | \Lambda_b \rangle =& \bar{u}(p,s_1) \left[ \gamma_\mu f_1(q^2) + i\frac{f_2(q^2)}{M_{\Lambda_b}}\sigma_{\mu\nu}q^\nu  + \frac{f_3(q^2)}{M_{\Lambda_b}}q_\mu \right] u_{\Lambda_b}(p+q,s_2) \notag \\  
		-& \bar{u}(p,s_1) \left[ \gamma_\mu g_1(q^2) + i\frac{g_2(q^2)}{M_{\Lambda_b}}\sigma_{\mu\nu}q^\nu + \frac{g_3(q^2)}{M_{\Lambda_b}}q_\mu \right] \gamma_5 u_{\Lambda_b}(p+q,s_2).  
	\end{align} 
	Here, the spin-parity quantum number of the final state charm baryon $\Lambda_c$ is $J^P = \frac{1}{2}^+$. 
	
	For the $\Lambda_b^0 \to \Lambda_c^*$ ($\frac12^+ \to \frac12^-$) transition, where $\Lambda_c^*$ possesses $J^P = \frac{1}{2}^-$, the form factors can be analogously defined as  	
	\begin{align}  
		\langle \Lambda_c^* |  \bar{c}\gamma_\mu(1-\gamma_5)b | \Lambda_b \rangle  =& \bar{u}(p,s_1) \left[ \gamma_\mu f_1^*(q^2) + i\frac{f_2^*(q^2)}{M_{\Lambda_b}}\sigma_{\mu\nu}q^\nu +  \frac{f_3^*(q^2)}{M_{\Lambda_b}}q_\mu \right] \gamma_5 u_{\Lambda_b}(p+q,s_2) \notag \\  
		  &- \bar{u}(p,s_1) \left[ \gamma_\mu g_1^*(q^2) + i\frac{g_2^*(q^2)}{M_{\Lambda_b}}\sigma_{\mu\nu}q^\nu 
		  +  \frac{g_3^*(q^2)}{M_{\Lambda_b}}q_\mu \right] u_{\Lambda_b}(p+q,s_2).  \label{FFs2}
	\end{align}
	This definition represents a generalized form for the decay of a $J^P = \frac{1}{2}^+$ baryon to a $J^P = \frac{1}{2}^-$ state. It is different from the expression in Eq.~(7) of Ref.~\cite{Duan:2022uzm} through the substitution $f_i^*(q^2) \leftrightarrow -g_i^*(q^2)$.

    In the following context, we will employ the light-cone QCD sum rules method to calculate the form factors of the $\Lambda_b^0 \to \Lambda_c^*$ transition. This method provides a valuable framework for calculating weak decay form factors of hadrons in regions of small momentum transfer squared $q^2$. The starting point for determining the $\Lambda_b^0 \to \Lambda_c^*$ form factors using light-cone QCD sum rules involves evaluating the $\Lambda_b^0 \to \Lambda_c^+$ weak decay correlation function,
    \begin{equation}  
    	T_\mu(p, q) = i\int d^4x e^{ip\cdot x}\langle 0|\mathcal{T}\{j_{\Lambda_c}(x), j_\mu(0)\}|\Lambda_b(p+q)\rangle. \label{cf}
    \end{equation}  
    Here, $p$ represents the momentum of the final state charm baryon, and $q$ stands for the momentum transfer in the weak decay. $j_\mu(x)=\bar{c}(x)\gamma_\mu(1-\gamma_5)b(x)$ is the weak decay $V-A$ current of $b$ quark decay to $c$ quark. The interpolating current $j_{\Lambda_c}(x)$ pertains to the $\Lambda_c$ baryon with spin-parity $J^P=\frac{1}{2}^+$. It has three types: 
    
     \emph{\rm{(1) Scalar}}
    	\begin{gather}  
    		j_{\Lambda_c}^S(x) = \epsilon_{ijk}[u^{iT}(x)Cd^j(x)]\gamma_5c^k(x), 
    	\end{gather} 
    	
    \emph{\rm{(2) Pseudoscalar}}
    	\begin{gather}  
    		j_{\Lambda_c}^P(x) = \epsilon_{ijk}[u^{iT}(x)C\gamma_5d^j(x)]c^k(x), 
    	\end{gather}
    	
    \emph{\rm{(3) Axial-vector}}
    	\begin{gather}
    		j_{\Lambda_c}^A(x) = \epsilon_{ijk}[u^{iT}(x)C\gamma_5\gamma_\nu d^j(x)]\gamma^\nu c^k(x).  
    	\end{gather} 
   In the subsequent calculations, the scalar interpolating current $j_{\Lambda_c}^S$ does not contribute to the sum rules, and only $j_{\Lambda_c}^P(x)$ and $j_{\Lambda_c}^A$ make contributions.
	
   As a standard calculation procedure in the light-cone QCD sum rules, the correlation function Eq. (\ref{cf}) should be computed on both the hadronic and quark-gluon levels, and we will obtain a hadronic representation and a quark-gluon representation. From different calculation schemes, they can be decomposed into twelve different Lorentz structures for both hadronic and quark-gluon representations. Matching the coefficients of these Lorentz structures in the hadronic and quark-gluon representations of the correlation function, we can isolate and distinguish the contributions of $\Lambda_c$ baryons with $J^P=\frac{1}{2}^+$ and $\Lambda_c^*$ with $J^P=\frac{1}{2}^-$ separately by solving linear equations. Actually, in our previous work, we have calculated the correlation function on the hadronic level, which will be given in the appendix for completeness~\cite{Duan:2022uzm}. On the quark-gluon level, we substitute the $\Lambda_c$ interpolating current and the weak decay $V-A$ current of $b \to c$ into the correlation function. After applying Wick's theorem and treating the charm quark mass as finite, we obtain the correlation function on the quark-gluon level. For the different types of $\Lambda_c$ baryon interpolating current, the correlation function on the quark-gluon level is different. 
	
	\emph{\rm{For the pseudoscalar current $j_{\Lambda_c}^P$, the correlation function $T_\mu(p,q)$ is given by: }} 
	\begin{align}  
		T_\mu(p,q) =& i\int d^4 x e^{ip\cdot x} \langle 0|\mathcal{T}\{j_{\Lambda_c}^P(x),j_\mu(0)\}|\Lambda_b(p+q)\rangle \notag \\  
		=& i\int d^4 x e^{ip\cdot x} (C\gamma_5)_{\alpha\beta} S_{\sigma\tau}(x) [\gamma_\mu(1-\gamma_5)]_{\tau\gamma}   \langle 0|\epsilon_{ijk}u_\alpha^{iT}(x)d_\beta^j(x)b_\gamma^k(0)|\Lambda_b(p+q)\rangle. \label{CF_P}  
	\end{align}  
	
	\emph{\rm{Similarly, for the axial-vector current $j_{\Lambda_c}^A$, the correlation function $T_\mu(p,q)$ can be expressed as:}}  	
	\begin{align}  
		T_\mu(p,q) =& i\int d^4 x e^{ip\cdot x} \langle 0|\mathcal{T}\{j_{\Lambda_c}^A(x),j_\mu(0)\}|\Lambda_b(p+q)\rangle \notag \\  
		=& i\int d^4 x e^{ip\cdot x} (C\gamma_5\gamma_\nu)_{\alpha\beta} S_{\sigma\tau}(x) [\gamma_\mu(1-\gamma_5)]_{\tau\gamma} \gamma^\nu_{\rho\sigma}  \langle 0|\epsilon_{ijk}u_\alpha^{iT}(x)d_\beta^j(x)b_\gamma^k(0)|\Lambda_b(p+q)\rangle. \label{CF_A}  
	\end{align}  
	Here, $S(x)$ represents the free charm quark propagator. The matrix element $\langle 0|\epsilon_{ijk}u_\alpha^{iT}(x)d_\beta^j(x)b_\gamma^k(0)|\Lambda_b(p+q)\rangle$ can be decomposed into four $\Lambda_b$ baryon light-cone distribution amplitudes (LCDAs), which have been studied using various models \cite{Ball:2008fw, Feldmann:2011xf, Ali:2012pn, Bell:2013tfa}. These LCDAs are defined through the $\Lambda_b$ current in heavy quark effective theory (HQET),  
	$$
	\bar{J}(x) = \epsilon_{abc}\left[\bar{d}^a(x)(A+B\slashed{v})\gamma_5\mathcal{C}^T\bar{u}^b(x)\right]\bar{h}_v^c(x), 
	$$
	which is a combination of two nonlocal heavy flavor baryon currents $\bar{J}_1(x)=\epsilon_{abc}[\bar{d}^a(x)\gamma_5\mathcal{C}^T\bar{u}^b(x)]\bar{h}_v^c(x)$ and $\bar{J}_2(x)=\epsilon_{abc}[\bar{d}^a(x)\slashed{v}\gamma_5\mathcal{C}^T\bar{u}^b(x)]\bar{h}_v^c(x)$, where $B=1-A$. LCDAs of $\Lambda_b$ are given by:  
	\begin{align}  
		\frac{1}{v_+}\langle 0 |\left[u(t_1)\mathcal{C}\gamma_5\slashed{n}d(t_2)\right]b_\gamma |\Lambda_b\rangle &= \psi^n(t_1,t_2)f_{\Lambda_b}^{(1)}u_\gamma, \notag \\  
		\frac{i}{2}\langle 0|\left[u(t_1)\mathcal{C}\gamma_5\sigma_{\bar{n}n}d(t_2)\right]b_\gamma|\Lambda_b\rangle &= \psi^{n\bar{n}}(t_1,t_2)f_{\Lambda_b}^{(2)}u_\gamma, \notag \\  
		\langle 0|\left[u(t_1)\mathcal{C}\gamma_5d(t_2)\right]b_\gamma|\Lambda_b\rangle&=\psi^{\mathbbm{1}}(t_1,t_2)f_{\Lambda_b}^{(2)}u_\gamma, \notag \\
    	v_+\langle 0|\left[u(t_1)\mathcal{C}\gamma_5\slashed{\bar{n}}d(t_2)\right]b_\gamma|\Lambda_b\rangle&=\psi^{\bar{n}}(t_1,t_2)f_{\Lambda_b}^{(1)}u_\gamma.	\label{LCDAs}	 
	\end{align}
	
	 In the above equations, $v_+=v\cdot n$, $v^\mu$ is the four-velocity of the $\Lambda_b$ baryon. $n^\mu=(1,0,0,-1)$ represents a light-cone vector, while $\bar{n}^\mu=(1,0,0,1)$ is another light-cone vector, satisfying the conditions $n^2=\bar{n}^2=0$ and $n_\mu \bar{n}^\mu=2$. If we consider the $\Lambda_b$ baryon in rest, $v^\mu=\frac12\left(n^\mu+\bar{n}^\mu \right)$, it will inherently possess $v_+=1$ as stated in \cite{Grozin:1996pq}. Combining the coordinate $x_\mu$ and velocity vector $v_\mu$ into the light-cone vectors $n_\mu$ and $\bar{n}_\mu$, we have $$n_\mu=\frac{x_\mu}{v\cdot x}$$ and $$\bar{n}_\mu=2 v_\mu-\frac{x_\mu}{v\cdot x}.$$ Additionally, $\mathcal{C}$ denotes the charge conjugation matrix and $\sigma_{\bar{n}n}=\sigma_{\mu\nu}\bar{n}^\mu n^\nu$. Utilizing these relations and the inherent properties of the charge conjugation matrix and gamma matrices, the aforementioned Eq. (\ref{LCDAs}) can be succinctly reformulated as:  
	\begin{align}  
		\epsilon_{ijk}\langle 0| & u^i_\alpha (t_1n)d^j_\beta(t_2n)b^k_{\gamma}| \Lambda_b\rangle \notag \\ =& \frac{1}{8}v_+f_{\Lambda_b}^{(1)}\Psi^n(t_1,t_2)(\slashed{\bar{n}}\gamma_5C^T)_{\beta\alpha}u_{\Lambda_b\gamma}(v) +\frac{1}{4}f_{\Lambda_b}^{(2)}\Psi^{\mathbbm{1}}(t_1,t_2)(\gamma_5C^T)_{\beta\alpha}u_{\Lambda_b\gamma}(v) \notag \\  -&
		 \frac{1}{8}f_{\Lambda_b}^{(2)}\Psi^{n\bar{n}}(t_1,t_2)(i\sigma_{n\bar{n}}\gamma_5C^T)_{\beta\alpha}u_{\Lambda_b\gamma}(v) + \frac{1}{8}\frac{1}{v_+}f_{\Lambda_b}^{(1)}\Psi^{\bar{n}}(t_1,t_2)(\slashed{n}\gamma_5C^T)_{\beta\alpha}u_{\Lambda_b\gamma}(v).  
	\end{align}  
	
	In this context, $t_1$ and $t_2$ represent the positions of the light quarks within the diquark. Each LCDA $\{\psi^n, \psi^{n\bar{n}}, \psi^{\mathbbm{1}}, \psi^{\bar{n}}\}$ possesses a distinct twist (dimension -- spin), which we denote using the subscript numbers $\{\psi_2, \psi_3^\sigma, \psi_3^s, \psi_4\}$. As discussed in \cite{Duan:2022uzm}, we have 
	\begin{gather}  
		\Psi_i(t_1,t_2)=\int_{0}^{\infty}\omega d\omega\int_{0}^{1}due^{-i\omega(t_1u+t_2(1-u))}\tilde{\psi}_i(\omega,u).  
	\end{gather}  	
	Here, $\omega$ and $u$ represent the energy of the diquark and the fraction of each light quark, respectively. There are five different LCDAs models in the literature, and all of them are considered in this work. Specifically, the LCDAs models used in this work will be categorized into Type I, Type II, Type III, Type IV and Type V. For brivity, we will list the LCDAs models Type II to Type V in the appendix, and here we give an example of one of the $\Lambda_b$ LCDAs models discussed in \cite{Ali:2012pn} (Type I):  
	\begin{align}
		\tilde{\psi}_2(\omega,u)=&\omega^2u(1-u)\sum_{n=0}^{2}\frac{a_n}{\epsilon_n^4}\frac{C_n^{3/2}(2u-1)}{|C_n^{3/2}|^2}e^{-\omega/\epsilon_n}, \\
		\tilde{\psi}_3(\omega,u)=&\frac{\omega}{2}\sum_{n=0}^{2}\frac{a_n}{\epsilon_n^3}\frac{C_n^{1/2}(2u-1)}{|C_n^{1/2}|^2}e^{-\omega/\epsilon_n}, \\
		\tilde{\psi}_4(\omega,u)=&\sum_{n=0}^{2}\frac{a_n}{\epsilon_n^2}\frac{C_n^{1/2}(2u-1)}{|C_n^{1/2}|^2}e^{-\omega/\epsilon_n}.
	\end{align}
	The subscript numbers of the above equations indicate the twist of LCDAs. $C_n^{3/2}(2u-1)$ and $C_n^{1/2}(2u-1)$ are Gegenbauer polynomials. The squared magnitudes of these polynomials are given as $|C_0^{1/2}|^2 = |C_0^{3/2}|^2 = 1$, $|C_1^{1/2}|^2 = \frac{1}{3}$, $|C_1^{3/2}|^2 = 3$, $|C_2^{1/2}|^2 = \frac{1}{5}$, and $|C_2^{3/2}|^2 = 6$.  
	
	Table~\ref{tab1} summarizes the parameters that characterize the LCDAs of the $\Lambda_b$ baryon with Type I. Of particular interest is the parameter $A$, which varies from 0 to 1. In the case of twist-2 LCDAs, we observe that $A$ cannot be set to 1. This constraint is indicative of the specific choice of interpolating current for the $\Lambda_b$ baryon, and we will delve deeper into this aspect in the final section of our discussion.  
	
	\begin{table*}[h]
		\centering  
		\caption{Parameters in the light-cone distribution amplitudes of the $\Lambda_b$ baryon.}  
		\label{tab1} 
		\begin{tabular}{ccccccc} \hline
			Twist & $a_0$ & $a_1$ & $a_2$ & $\epsilon_0$ $[\rm{GeV}]$ & $\epsilon_1$ $[\rm{GeV}]$ & $\epsilon_2$ $[\rm{GeV}]$ \\  
			\midrule  
			2 & 1 & - & $\frac{6.4-6.4A}{1.44-A}$ & $\frac{2-1.4A}{6.7-A}$ & - & $\frac{0.32-0.32A}{0.83-A}$ \\  
			3s & 1 & - & $\frac{0.12A-0.04}{A+0.4}$ & $\frac{0.56A+0.21}{A+1.6}$ & - & $\frac{0.09-0.25A}{1.41-A}$ \\  
			3a & - & 1 & - & - & $\frac{0.08+0.35A}{A+0.2}$ & - \\  
			4 & 1 & - & $\frac{0.07A-0.12}{1.34-A}$ & $\frac{0.87-0.65A}{2-A}$ & - & $\frac{9.3-5.5A}{30-A}$  \\  \hline
		\end{tabular}
	\end{table*}  
	
   By substituting the LCDAs into the quark-gluon correlation function, we obtain a set of Lorentz structures that are analogous to those encountered at the hadronic level. Specifically, the correlation function $T_\mu(p, q)$ can be expressed as:  
   \begin{align}  
    	T_\mu(p, q)=& \Pi_{v_\mu}v_\mu + \Pi_{\gamma_\mu}\gamma_\mu + \Pi_{q_\mu}q_\mu + \Pi_{v_\mu\slashed{q}}v_\mu\slashed{q} + \Pi_{\gamma_\mu\slashed{q}}\gamma_\mu\slashed{q} \notag \\ 
    	            &+ \Pi_{q_\mu\slashed{q}}q_\mu\slashed{q} + \Pi_{v_\mu\gamma_5}v_\mu\gamma_5 + \Pi_{\gamma_\mu\gamma_5}\gamma_\mu\gamma_5  \notag \\ 
    	            &+ \Pi_{q_\mu\gamma_5}q_\mu\gamma_5 + \Pi_{v_\mu\slashed{q}\gamma_5}v_\mu\slashed{q}\gamma_5 + \Pi_{\gamma_\mu\slashed{q}\gamma_5}\gamma_\mu\slashed{q}\gamma_5 \notag \\ 
    	            &+ \Pi_{q_\mu\slashed{q}\gamma_5}q_\mu\slashed{q}\gamma_5.  
   \end{align}
   To extract the desired form factors for the transitions $\Lambda_b^0 \to \Lambda_c^+$ and $\Lambda_b^0 \to \Lambda_c^*$, we match the coefficients of each Lorentz structure, disregarding the integral over the continuum spectrum (which is suppressed after Borel transformation, and will be eliminated by using the quark-hadron duality). By solving the resulting linear equations, we arrive at the expressions for the form factors: 
	\begin{align}
  	f_1(q^2)=&\frac{e^{M_{\Lambda_c}^2/M_B^2}}{2f_{\Lambda_c}M_{\Lambda_b}(M_{\Lambda_c}+M_{\Lambda_c^*})}\Big\{(M_{\Lambda_b}+M_{\Lambda_c})\Pi_{v_\mu}  \notag \\
  	         &+2M_{\Lambda_b}\Pi_{\gamma_\mu}+(M_{\Lambda_b}+M_{\Lambda_c})(M_{\Lambda_b}-M_{\Lambda_c^*})\Pi_{v_\mu\slashed{q}} \notag \\
             &-2M_{\Lambda_b}(M_{\Lambda_c}-M_{\Lambda_c^*})\Pi_{\gamma_\mu\slashed{q}} \Big\}, \\ \label{FFf1}
	f_2(q^2)=&\frac{e^{M_{\Lambda_c}^2/M_B^2}}{2f_{\Lambda_c}(M_{\Lambda_c}+M_{\Lambda_c^*})} \Big\{\Pi_{v_\mu}+(M_{\Lambda_b}-M_{\Lambda_c^*})\Pi_{v_\mu\slashed{q}} \notag \\
	         &-2M_{\Lambda_b}\Pi_{\gamma_\mu\slashed{q}} \Big\}, \\
	f_3(q^2)=&\frac{e^{M_{\Lambda_c}^2/M_B^2}}{2f_{\Lambda_c}(M_{\Lambda_c}+M_{\Lambda_c^*})}\Big\{\Pi_{v_\mu}+2M_{\Lambda_b}\Pi_{q_\mu} \notag \\
	         &+(M_{\Lambda_b}-M_{\Lambda_c^*})\Pi_{v_\mu\slashed{q}}+2M_{\Lambda_b}\Pi_{\gamma_\mu\slashed{q}} \notag \\ 
	         &+2M_{\Lambda_b}(M_{\Lambda_b}-M_{\Lambda_c^*})\Pi_{q_\mu\slashed{q}}\Big\}, \\
	f_1^*(q^2)=&\frac{-e^{M_{\Lambda_c^*}^2/M_B^2}}{2f_{\Lambda_c^*}M_{\Lambda_b}(M_{\Lambda_c}+M_{\Lambda_c^*})}\Big\{(M_{\Lambda_b}-M_{\Lambda_c})\Pi_{v_\mu} \notag \\ 
	           &+2M_{\Lambda_b}\Pi_{\gamma_\mu}+(M_{\Lambda_b}+M_{\Lambda_c})(M_{\Lambda_b}-M_{\Lambda_c^*})\Pi_{v_\mu\slashed{q}} \notag \\
               &-2M_{\Lambda_b}(M_{\Lambda_c}-M_{\Lambda_c^*})\Pi_{\gamma_\mu\slashed{q}}\Big\}, \\
	f_2^*(q^2)=&\frac{e^{M_{\Lambda_c^*}^2/M_B^2}}{2f_{\Lambda_c^*}(M_{\Lambda_c}+M_{\Lambda_c^*})}\Big\{\Pi_{v_\mu}+(M_{\Lambda_b}-M_{\Lambda_c})\Pi_{v_\mu\slashed{q}} \notag \\
	           &-2M_{\Lambda_b}\Pi_{\gamma_\mu\slashed{q}}\Big\}, \\
	f_3^*(q^2)=&\frac{e^{M_{\Lambda_c^*}^2/M_B^2}}{2f_{\Lambda_c^*}(M_{\Lambda_c}+M_{\Lambda_c^*})}\Big\{\Pi_{v_\mu}+2M_{\Lambda_b}\Pi_{q_\mu} \notag \\ 
	           &+(M_{\Lambda_b}+M_{\Lambda_c})\Pi_{v_\mu\slashed{q}}+2M_{\Lambda_b}\Pi_{\gamma_\mu\slashed{q}} \notag \\
	           &+2M_{\Lambda_b}(M_{\Lambda_b}+M_{\Lambda_c})\Pi_{q_\mu\slashed{q}}\Big\}, \\
	g_1(q^2)=&\frac{e^{M_{\Lambda_c}^2/M_B^2}}{2f_{\Lambda_c}M_{\Lambda_b}(M_{\Lambda_c}+M_{\Lambda_c^*})}\Big\{(M_{\Lambda_b}-M_{\Lambda_c})\Pi_{v_\mu\gamma_5} \notag \\
	         &-2M_{\Lambda_b}\Pi_{\gamma_\mu\gamma_5}-(M_{\Lambda_b}-M_{\Lambda_c})(M_{\Lambda_b}+M_{\Lambda_c^*})\Pi_{v_\mu\slashed{q}\gamma_5} \notag \\
	         &+2M_{\Lambda_b}(M_{\Lambda_c}-M_{\Lambda_c^*})\Pi_{\gamma_\mu\slashed{q}\gamma_5}\Big\}, \\
	g_2(q^2)=&\frac{e^{M_{\Lambda_c}^2/M_B^2}}{2f_{\Lambda_c}(M_{\Lambda_c}+M_{\Lambda_c^*})}\Big\{-\Pi_{v_\mu\gamma_5}+(M_{\Lambda_b}+M_{\Lambda_c^*})\Pi_{v_\mu\slashed{q}\gamma_5} \notag \\
	         &+2M_{\Lambda_b}\Pi_{\gamma_\mu\slashed{q}\gamma_5}\Big\}, \\
	g_3(q^2)=&\frac{-e^{M_{\Lambda_c}^2/M_B^2}}{2f_{\Lambda_c}(M_{\Lambda_c}+M_{\Lambda_c^*})}\Big\{\Pi_{v_\mu\gamma_5}+2M_{\Lambda_b}\Pi_{q_\mu\gamma_5} \notag \\ 
	         &-(M_{\Lambda_b}+M_{\Lambda_c^*})\Pi_{v_\mu\slashed{q}\gamma_5}+2M_{\Lambda_b}\Pi_{\gamma_\mu\slashed{q}\gamma_5} \notag \\
	         &-2M_{\Lambda_b}(M_{\Lambda_b}+M_{\Lambda_c^*})\Pi_{q_\mu\slashed{q}\gamma_5}\Big\}, \\
	g_1^*(q^2)=&\frac{-e^{M_{\Lambda_c^*}^2/M_B^2}}{2f_{\Lambda_c^*}M_{\Lambda_b}(M_{\Lambda_c}+M_{\Lambda_c^*})}\Big\{(M_{\Lambda_b}+M_{\Lambda_c^*})\Pi_{v_\mu\gamma_5} \notag \\
	           &-2M_{\Lambda_b}\Pi_{\gamma_\mu\gamma_5}-(M_{\Lambda_b}+M_{\Lambda_c^*})(M_{\Lambda_b}-M_{\Lambda_c})\Pi_{v_\mu\slashed{q}\gamma_5} \notag \\
	           &+2M_{\Lambda_b}(M_{\Lambda_c}-M_{\Lambda_c^*})\Pi_{\gamma_\mu\slashed{q}\gamma_5}\Big\}, \\
	g_2^*(q^2)=&\frac{-e^{M_{\Lambda_c^*}^2/M_B^2}}{2f_{\Lambda_c^*}(M_{\Lambda_c}+M_{\Lambda_c^*})}\Big\{\Pi_{v_\mu\gamma_5}-(M_{\Lambda_b}-M_{\Lambda_c^*})\Pi_{v_\mu\slashed{q}\gamma_5} \notag \\
	           &-2M_{\Lambda_b}\Pi_{\gamma_\mu\slashed{q}\gamma_5}\Big\}, \\
	g_3^*(q^2)=&\frac{-e^{M_{\Lambda_c^*}^2/M_B^2}}{2f_{\Lambda_c^*}(M_{\Lambda_c}+M_{\Lambda_c^*})}\Big\{\Pi_{v_\mu\gamma_5}+2M_{\Lambda_b}\Pi_{q_\mu\gamma_5} \notag \\
	           &-(M_{\Lambda_b}-M_{\Lambda_c})\Pi_{v_\mu\slashed{q}\gamma_5}+2M_{\Lambda_b}\Pi_{\gamma_\mu\slashed{q}\gamma_5} \notag \\
	           &-2M_{\Lambda_b}(M_{\Lambda_b}-M_{\Lambda_c})\Pi_{q_\mu\slashed{q}\gamma_5}\Big\}. \label{FFg3s}
    \end{align}	

   Where the coefficient quantities $\Pi_{\Gamma_i}$ ( the Lorentz structures $\Gamma_i = \{v_\mu, \gamma_\mu, q_\mu, v_\mu\slashed{q}, \gamma_\mu\slashed{q}, q_\mu\slashed{q}\}$) and their counterparts with $\gamma_5$, denoted as $\Pi_{\Gamma_i\gamma_5}$, have the following generalized expressions, they depend on the different types of interpolating currents of the $\Lambda_c$ baryon. For the pseudoscalar interpolating current $j_{\Lambda_c}^P$, the $\Pi_{\Gamma_i}$ are given by
    \begin{equation}  
	\Pi_{\Gamma_i} = \int_{0}^{1}du \int_{0}^{\infty}d\sigma \frac{\rho_{_{\Gamma_i}}}{(p-\omega v)^2-m_c^2} f_{\Lambda_b}^{(1)}.  
    \end{equation}  
   Similarly, for the axial-vector interpolating current $j_{\Lambda_c}^A$, the expressions for $\Pi_{\Gamma_i}$ read
    \begin{equation}  
	\Pi_{\Gamma_i} = \int_{0}^{1}du \int_{0}^{\infty}d\sigma \left[\frac{\rho_{_{\Gamma_i}}^1}{(p-\omega v)^2-m_c^2} + \frac{\rho_{_{\Gamma_i}}^2}{[(p-\omega v)^2-m_c^2]^2} \right] f_{\Lambda_b}^{(2)}.  
    \end{equation}  
   The detailed forms of $\rho_{_{\Gamma_i}}$, $\rho_{_{\Gamma_i}}^1$, and $\rho_{_{\Gamma_i}}^2$ will be listed in the appendix, the relation of $\sigma$ and $\omega$ is $\omega=\sigma M_{\Lambda_b}$.
 
   To suppress contributions from higher-twist LCDAs, excited states, and continuum states of the $\Lambda_b$ baryon, a Borel transformation is applied to both the hadronic and quark-gluon levels of the correlation function. In this work, we utilize the following Borel transformations:  
   \begin{equation}
   	\mathcal{B}_{M_B^2}\left(\frac{1}{(M_{\Lambda_c^{(*)}}^2-p^{(\prime)2})^n}\right)=\frac{1}{(n-1)!}\frac{e^{-M_{\Lambda_c^{(*)}}^2/M_B^2}}{M_B^{2(n-1)}}
   \end{equation}
   on the hadrnoic level, and
   \begin{align}  
	\int_{0}^{\infty}d\sigma &\frac{\rho_{_{\Gamma_i}}(\sigma)}{(p-\omega v)^2-m_c^2} \rightarrow  -\int_{0}^{\sigma_0}d\sigma \frac{\rho_{_{_{\Gamma_i}}}(\sigma)e^{-s/M_B^2}}{1-\sigma}, \notag \\  
	\int_{0}^{\infty}d\sigma &\frac{\rho_{_{\Gamma_i}}(\sigma)}{[(p-\omega v)^2-m_c^2]^2} \rightarrow \notag \\ & \int_{0}^{\sigma_0}d\sigma \frac{1}{(1-\sigma)^2} \frac{\rho_{_{\Gamma_i}}(\sigma)}{M_B^2} e^{-s/M_B^2} + \frac{\rho_{_{\Gamma_i}}(\sigma_0)}{(1-\sigma_0)^2} \eta(\sigma_0) e^{-s_0/M_B^2},  \label{Borel2}
   \end{align}
   on the quark-gluon level. Where the parameters relations are
   \begin{equation}
   	s=\sigma M_{\Lambda_b}^2+\frac{m_c^2-\sigma q^2}{1-\sigma},
   \end{equation}
   
   \begin{align}
   	\sigma_0=&\left((s_0+M_{\Lambda_b}^2-q^2)\right.  \notag \\
   	&\left.-\sqrt{(s_0+M_{\Lambda_b}^2-q^2)^2-4M_{\Lambda_b}^2(s_0-m_c^2)}\right)/2M_{\Lambda_b}^2, \label{s0}
   \end{align}
   and the function $\eta(\sigma)$ is defined as  
   \begin{equation}  
	\eta(\sigma) = \frac{d\sigma}{ds} = \frac{(1-\sigma)^2}{(1-\sigma)^2M_{\Lambda_b} + m_c - q^2}.  
   \end{equation}  
    This procedure ensures a more controlled extraction of the desired form factors and physical observables from the correlation functions. The threshold parameter $s_0$ in Eq. (\ref{s0}) will be discussed  in detail in the numerical analysis section. 

   In our prior work \cite{Duan:2022uzm}, the form factors $f_i$ ($i=1, 2, 3$) and $g_i$ ($i=1, 2, 3$) have been thoroughly examined. Therefore, in the current work, we shall focus solely on the form factors $f_i^*$ ($i=1, 2, 3$) and $g_i^*$ ($i=1, 2, 3$). Upon incorporating the LCDAs of the $\Lambda_b$ baryon into these form factors, we discovered intriguing relations. Specifically, for both the pseudoscalar interpolating current $j_{\Lambda_c}^P$ and the axial-vector interpolating current $j_{\Lambda_c}^A$ of the $\Lambda_c$ baryon, we obatin that $f_1^* = g_1^*$ and $f_2^* = f_3^* = g_2^* = g_3^*$. 
	
	\section{Numerical analysis} \label{sec:III}
	
	\subsection{Form factors} 
    Before analyzing the numerical variations of the form factors, it is crucial to establish the fundamental input parameters within the framework of light-cone sum rules. The mass parameters of baryons and leptons used in this analysis are taken from the Particle Data Group (PDG)~\cite{ParticleDataGroup:2022pth, ParticleDataGroup:2024cfk}. Specifically, the charm quark mass is the pole mass, while the other masses are the world averages reported by the PDG group. These values are summarized in Table~\ref{tab2}.  

    \begin{table}[h] 
	\centering  
	\caption{Mass parameters of baryons and leptons~\cite{ParticleDataGroup:2022pth, ParticleDataGroup:2024cfk}.}  
	\label{tab2}
		\begin{tabular}{cc}  \hline
		Parameters           &         Values              \\  
		\midrule  
		$M_{\Lambda_b^0}$    &        $5.6196$ GeV         \\  
		$M_{\Lambda_c}$      &        $2.28646$ GeV        \\ 
		$M_{\Lambda_c^*}$    &        $2.59225$ GeV        \\  
		$m_c$                &        $(1.67\pm0.07)$ GeV  \\  
		$m_e$                &        $0.51$ MeV          \\  
		$m_\mu$              &       $105.658$ MeV        \\  
		$m_\tau$             &       $1.77686$ GeV        \\  \hline
	\end{tabular}  
    \end{table}  

    Additional key parameters are the decay constants $f_{\Lambda_c}$ of $\Lambda_c$ and $f_{\Lambda_c^*}$ of $\Lambda_c^*$, defined by $\langle 0|j_{\Lambda_c}|\Lambda_c(p)\rangle=f_{\Lambda_c}u_{\Lambda_c}(p)$ and $\langle 0|j_{\Lambda_c^*}|\Lambda_c^*(p)\rangle=f_{\Lambda_c^*}\gamma_5u_{\Lambda_c^*}(p)$. For these parameters, we adopt the values obtained through QCD sum rules \cite{Wang:2010fq}, namely $f_{\Lambda_c}=0.022$ GeV$^3$ and $f_{\Lambda_c^*}=0.035$ GeV$^3$. The decay constants $f_{\Lambda_b}^{(1)}$ and $f_{\Lambda_b}^{(2)}$ of $\Lambda_b$, defined by the matrix elements $\epsilon^{abc}\langle 0 |[u^{aT}(0)\mathcal{C}\gamma_5\slashed{v}d^b(0)]b_\gamma^c(0)|\Lambda_b(v)\rangle=f_{\Lambda_b}^{(1)}u_{\Lambda_b\gamma}(v)$ and $\epsilon^{abc}\langle 0 |[u^{aT}(0)\mathcal{C}\gamma_5d^b(0)]b_\gamma^c(0)|\Lambda_b(v)\rangle=f_{\Lambda_b}^{(2)}u_{\Lambda_b\gamma}(v)$, have been discussed in \cite{Ball:2008fw}, it has $f_{\Lambda_b}^{(1)}\simeq f_{\Lambda_b}^{(2)}\simeq 0.030$ GeV$^3$.  
	
	Utilizing the given parameters, we are able to determine the numerical values of the form factors $f_i^*$ (with $i=1,2,3$) and $g_i^*$ (with $i=1,2,3$) by using the expressions from Eq. (\ref{FFf1}) to Eq. (\ref{FFg3s}) within the physical region, $m_\ell^2 \le q^2 \le (M_{\Lambda_b^0}-M_{\Lambda_c^*})^2$. However, it is important to note that light-cone QCD sum rules are not applicable across the entire physical region. Therefore, in this work, we restrict our analysis to the matching region for light-cone QCD sum rules, specifically $-15~\rm{GeV^2} \leq$ $q^2$ $\leq 2.5~\rm{GeV}^2$, and the regions higher than $q^2=2.5~\rm{GeV^2}$ will break down in our sum rules.  
	
\begin{figure*}
		\begin{center}
			\includegraphics[width=0.45\textwidth]{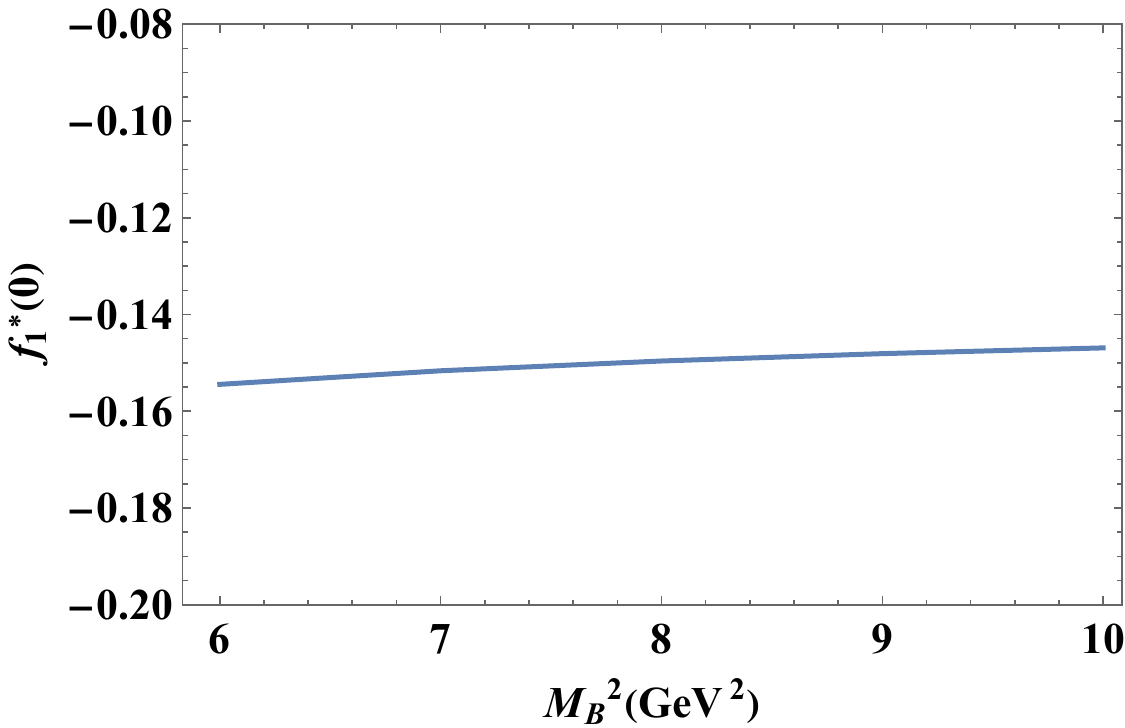}
			\qquad
			\includegraphics[width=0.45\textwidth]{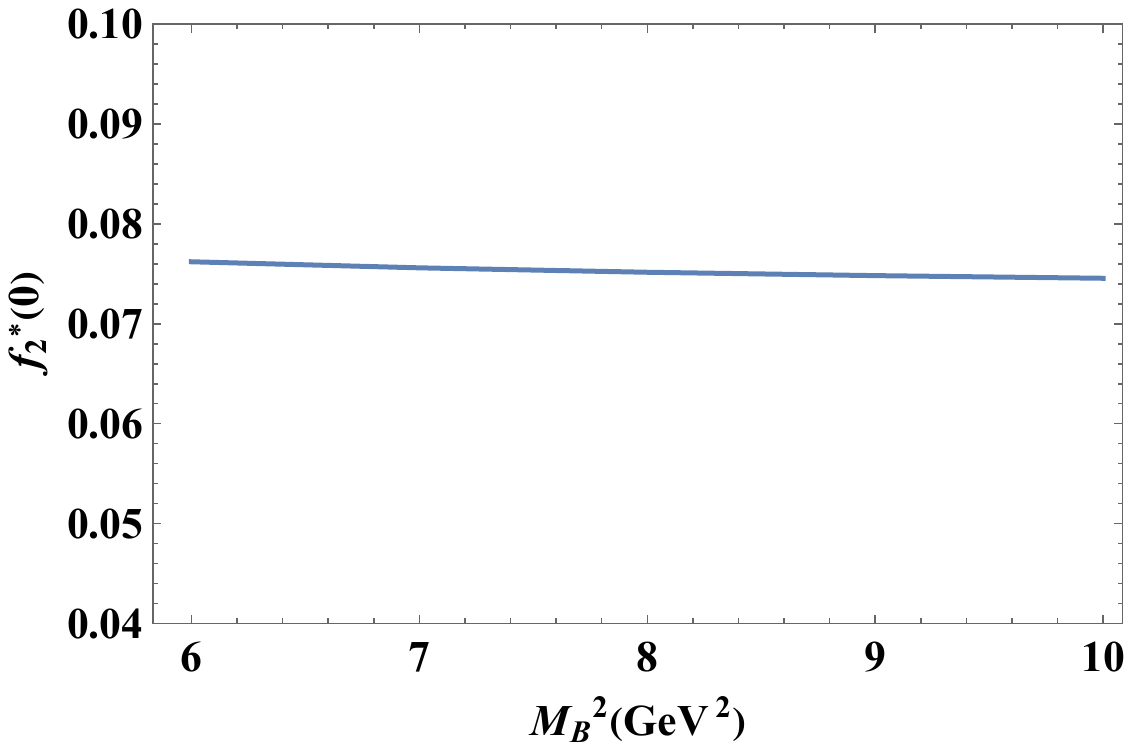}
		\end{center}
		\caption{Form factors $f_i^*(i=1,2)$ depend on the Borel Parameter $M_B^2$ within the free parton model $\Lambda_b$ LCDAs in $j^A$ interpolating current.}\label{figmb}
	\end{figure*}

	To suppress the unwanted impact of higher-twist LCDAs of $\Lambda_b$ and excited and continuum states of $\Lambda_c$, we have implemented Borel transformations at both the hadronic and theoretical levels. This introduces an additional Borel parameter, $M_B$, whose proper selection ensures the reliability and stability of our results. The chosen principle of $M_B$ is that it should make the lowest states contribute predominantly and suppress the excited and continuum states on both hadronic and QCD levels, so that it should neither be too small nor too large, while also ensuring minimal influence on the form factors. In consideration of these factors, we choose $M_B^2$ to lie within the range of $6 \text{ GeV}^2 \le M_B^2 \le 10 \text{ GeV}^2$, which is a suitbale adjustment range (as shown in Fig. \ref{figmb} for the Type III LCDAs of $\Lambda_b$ and the $j^A_{\Lambda_c}$ interpolating current for a sample). Additionally, we require the threshold $s_0$ of the heavy baryon, which its energy should exceed the $\Lambda_c^*$ baryon and lower than the excited state of $\Lambda_c$ with $J^P=\frac{1}{2}^+$ and $\Lambda_c^*$ with $J^P=\frac{1}{2}^-$. The contribution of states higher than $s_0$ on the quark-gluon level is cut off by $\sigma_0$ in the Borel transformation, and contributions from excited and continuum states on both hadronic and quark-gluon levels will be eliminated by the quark-hadron duality and Borel transformation Eq. (\ref{Borel2}). Therefore, we set a threshold higher than the energy of $\Lambda_c$ by $\Delta$ and set $s_0 = (M_{\Lambda_c} + \Delta)^2$ with $\Delta = (0.4 \pm 0.05) \text{ GeV}$, where $0.05$ serves as an uncertainty estimate in this work.  
	
	With these basic parameters in place, we aim to obtain the profiles of the $\Lambda_b \to \Lambda_c^{*}$ transition form factors across the entire physical region. To accomplish this, we extrapolate the form factors derived from light-cone sum rules to the entire physical region. For this purpose, we utilize $z$-series expansion fitting formulas, which are expressed as:  	
	\begin{equation}  
		f_i(q^2)/g_i(q^2)=\frac{f_i(0)/g_i(0)}{1-q^2/M_{B_c}^2} \Big\{b_1 \left[z(q^2)-z(0)\right] +b_2 \left[z(q^2)^2-z(0)^2\right]+1\Big\}.  \label{eqfiting}
	\end{equation}
    	Here, $b_1$ and $b_2$ are fitting parameters, which along with $f_i(0)/g_i(0)$, are listed in Table \ref{table3}. $M_{B_c} = 6.27447 \text{ GeV}$ represents the mass of the $B_c$ meson. The function $z(q^2, t_0)$ is defined by:  	
	\begin{equation}  
		z(q^2, t_0)=\frac{\sqrt{t_+-q^2}-\sqrt{t_+-t_0}}{\sqrt{t_+-q^2}+\sqrt{t_+-t_0}},  
	\end{equation}  
		where $t_+$ and $t_0$ are given by:  	
	\begin{align}  
		t_+ &= \left(M_{\Lambda_b} + M_{\Lambda_c^*}\right)^2, \notag  \\  
		t_0 &= \left(M_{\Lambda_b} + M_{\Lambda_c^*}\right)\cdot \left(\sqrt{M_{\Lambda_b}} - \sqrt{M_{\Lambda_c^*}}\right)^2.  
	\end{align}
	
	\begin{table*}[h] 
		\centering  
		\caption{Form factors $f_i^*$ ($i=0, 1)$ at zero momentum transfer square and the fitting parameters $b_1$ and $b_2$ corresponding to the $j_{\Lambda_c}^P$ and $j_{\Lambda_c}^A$ interpolating current.}  
		\label{table3} 
			\begin{tabular}{cccccccc}  \hline
				&&\multicolumn{3}{c}{$j^P_{\Lambda_c}$} &\multicolumn{3}{c}{$j^A_{\Lambda_c}$} \\ \midrule
				$\Lambda_b$ LCDAs & $f_i^{*}(q^2)$  &      $f_i^{*}(0)$           &           $b_1$            &             $b_2$      &$f_i^*(0)$   &   $b_1$  &    $b_2$  \\  
				\midrule  
				\multirow{2}*{Type I}&$f_1^*(q^2)$    & $0.087^{+0.045}_{-0.030}$   & $3.135^{+1.052}_{-1.827}$  & $-31.005^{+5.275}_{-1.394}$  &  $-0.250^{+0.090}_{-0.121}$ & $-2.070^{+2.287}_{-2.284}$ & $-18.173^{+13.416}_{-11.995}$ \\  
				&$f_2^*(q^2)$    & $-0.070^{+0.017}_{-0.013}$  & $-4.131^{+4.375}_{-4.592}$ & $-13.180^{+29.470}_{-19.423}$  & $0.199^{+0.098}_{-0.073}$ &$-15.683^{+2.208}_{-2.070}$& $70.549^{+16.979}_{-17.481}$ \\ 
				\midrule  
				\multirow{2}*{Type II}&$f_1^*(q^2)$    & $0.023^{+0.010}_{-0.008}$   & $-6.110^{+0.268}_{-0.504}$  & $5.244^{+2.826}_{-1.124}$  &  $-0.122^{+0.037}_{-0.046}$ & $-9.056^{+0.703}_{-0.863}$ & $27.885^{+5.771}_{-4.519}$ \\  
				&$f_2^*(q^2)$    & $-0.027^{+0.008}_{-0.010}$  & $-21.246^{+0.641}_{-0.728}$ & $118.441^{+6.101}_{-5.297}$  & $0.068^{+0.027}_{-0.022}$ &$-20.416^{+0.595}_{-0.189}$& $110.738^{+13.804}_{-4.832}$ \\
				\midrule
				\multirow{2}*{Type III}&$f_1^*(q^2)$    & $0.136^{+0.064}_{-0.047}$   & $-8.401^{+0.612}_{-0.707}$  & $17.115^{+5.484}_{-4.765}$  &  $-0.150^{+0.053}_{-0.070}$ & $-16.659^{+0.475}_{-0.441}$ & $83.488^{+3.694}_{-4.134}$ \\  
				&$f_2^*(q^2)$    & $-0.168^{+0.055}_{-0.068}$  & $-24.277^{+1.158}_{-0.995}$ & $143.305^{+9.632}_{-11.322}$  & $0.075^{+0.043}_{-0.032}$ &$-33.299^{+0.269}_{-0.238}$& $231.369^{+2.023}_{-2.507}$ \\
				\midrule
				\multirow{2}*{Type IV}&$f_1^*(q^2)$    & $0.024^{+0.012}_{-0.008}$   & $-8.513^{+0.619}_{-0.481}$  & $19.595^{+3.362}_{-4.418}$  &  $-0.107^{+0.036}_{-0.047}$ & $-14.680^{+0.503}_{-0.560}$ & $68.608^{+4.327}_{-3.907}$ \\  
				&$f_2^*(q^2)$    & $-0.030^{+0.010}_{-0.012}$  & $-24.590^{+0.657}_{-0.628}$ & $147.657^{+5.714}_{-6.045}$  & $0.049^{+0.026}_{-0.019}$ &$-28.373^{+0.186}_{-0.151}$& $183.921^{+1.035}_{-1.448}$ \\
				\midrule
				\multirow{2}*{Type V}&$f_1^*(q^2)$    & $0.057^{+0.022}_{-0.016}$   & $-0.521^{+0.339}_{-0.753}$  & $-24.239^{+2.607}_{-0.144}$  &  $-0.466^{+0.103}_{-0.116}$ & $4.634^{+1.186}_{-1.628}$ & $-50.227^{+5.668}_{-2.680}$ \\  
				&$f_2^*(q^2)$    & $-0.056^{+0.014}_{-0.022}$  & $-11.977^{+1.294}_{-1.584}$ & $40.732^{+12.077}_{-9.441}$  & $0.406^{+0.092}_{-0.085}$ &$-9.206^{+1.375}_{-1.750}$& $20.098^{+12.583}_{-9.251}$ \\
				 \hline
			\end{tabular} 
	\end{table*}  
	
	The $z$-series expansion provides an effective method for approximating form factors from the light-cone sum rules regions and extrapolating them to the entire physical domain.	Table \ref{table3} presents the values of the form factors at $q^2=0~\rm{GeV}^2$, along with their corresponding fitting parameters $b_1$ and $b_2$. These parameters are fitted within the $q^2$ range from $-15~\rm{GeV^2}$ to $2.5~\rm{GeV^2}$, with an interval of $0.5~\rm{GeV^2}$. The central values in Table \ref{table3} correspond the parameters $m_c=1.67~\rm{GeV}$, $M_B^2=8~\rm{GeV}^2$, and $s_0=(M_{\Lambda_c}+0.4)^2~\rm{GeV}^2$ (for type I LCDAs, $A=0.1$, and the corresponding $b_2$ of $f_1^*$ for $j^P_{\Lambda_c}$ is the lower error value). The upper error values of $f_1^*(0)$ and $b_2$ for $j^A_{\Lambda_c}$ interpolating current, and $f_2^*(0)$ and $b_2$ for $j^P_{\Lambda_c}$ interpolating current; as well as the lower error values of $f_2^*(0)$ and $b_1$ for $j^A_{\Lambda_c}$, and $f_1^*(0)$ and $b_1$ for $j^P_{\Lambda_c}$ correspond to the parameter set $m_c=1.74~\rm{GeV}$, $M_B^2=10~\rm{GeV}^2$, and $s_0=(M_{\Lambda_c}+0.35)^2~\rm{GeV}^2$ (for type I LCDAs, $A=0$ for $j^A_{\Lambda_c}$, $A=0.2$ for $j^P_{\Lambda_c}$, and the corresponding $b_2$ of $f_1^*$ for $j^P_{\Lambda_c}$ is the central value). The upper error values of $f_2^*(0)$ and $b_1$ for $j^A_{\Lambda_c}$, and $f_1^*(0)$ and $b_1$ for $j^P_{\Lambda_c}$; Lower error of $f_1^*(0)$ and $b_2$ for $j^A_{\Lambda_c}$, and $f_2^*(0)$ and $b_2$ for $j^P_{\Lambda_c}$ correspond to the parameter set $m_c=1.6~\rm{GeV}$, $M_B^2=6~\rm{GeV}^2$, and $s_0=(M_{\Lambda_c}+0.45)^2~\rm{GeV}^2$ (for type I LCDAs, $A=0.2$ for $j^A_{\Lambda_c}$, $A=0$ for $j^P_{\Lambda_c}$, and the corresponding $b_2$ of $f_1^*$ for $j^P_{\Lambda_c}$ is the upper error value).
   
   In order to compare our form factors with others (e.g., LQCD form factors), we should construct a relation to the form factors used in LQCD, such as~\cite{Meinel:2021rbm}:
   \begin{align}
   	\langle \Lambda_c^* | & \bar{c}\gamma_\mu(1-\gamma_5)b | \Lambda_b \rangle \notag \\
   	=& \bar{u}(p,s_1) \gamma_5\left\{ f_0^*(q^2)(M_{\Lambda_b}+M_{\Lambda_c^*})\frac{q^\mu}{q^2}\right. \notag \\&
   	 + f_+^*\frac{M_{\Lambda_b}-M_{\Lambda_c^*}}{s_-}[p^\mu+p'^\mu-(M_{\Lambda_b}^2-M_{\Lambda_c}^2)\frac{q^\mu}{q^2}] \notag \\ &+f_\perp^*[\gamma^\mu+\frac{2M_{\Lambda_c^*}}{s_-}p^\mu-\frac{2M_{\Lambda_b}}{s_-}p'^\mu] \notag \\ 
   	 &+ g_0^*\gamma_5(M_{\Lambda_b}-M_{\Lambda_c^*})\frac{q^\mu}{q^2} \notag \\&+g_+^*\gamma_5\frac{M_{\Lambda_b}+M_{\Lambda_c^*}}{s_+}[p^\mu+p'^\mu-(M_{\Lambda_b}^2-M_{\Lambda_c^*}^2)\frac{q^\mu}{q^2}] \notag \\  &\left.+g_\perp^*\gamma_5[\gamma^\mu-\frac{2M_{\Lambda_c^*}}{s_+}p^\mu-\frac{2M_{\Lambda_b}}{s_+}p'^\mu] \right\} u_{\Lambda_b}(p',s_2), \label{FFs3}
   \end{align}
   where $p'=p+q$ and $$s_\pm=(M_{\Lambda_b}\pm M_{\Lambda_c})^2-q^2.$$ The other dynamical variable used in both LQCD and HQET is
   \begin{equation}
   	{\it{w}}=\frac{M_{\Lambda_b}^2+M_{\Lambda_c^*}^2-q^2}{2M_{\Lambda_b}M_{\Lambda_c^*}}.
   \end{equation}
   The relations of form factors between Eq. (\ref{FFs2}) and Eq. (\ref{FFs3}) can be derived as:
   \begin{align}
   	f_0^*=&\frac{q^2}{M_{\Lambda_b}(M_{\Lambda_b}+M_{\Lambda_c^*})}f_3^*-f_1^*; \\
   	f_+^*=&-\frac{q^2}{M_{\Lambda_b}(M_{\Lambda_b}-M_{\Lambda_c^*})}f_2^*-f_1^*; \\
   	f_\perp^*=&\frac{M_{\Lambda_c^*}-M_{\Lambda_b}}{M_{\Lambda_b}}f_2^*-f_1^*; \\
   	g_0^*=&-\frac{q^2}{M_{\Lambda_b}(M_{\Lambda_b}-M_{\Lambda_c^*})}g_3^*-g_1^*; \\
   	g_+^*=&\frac{q^2}{M_{\Lambda_b}(M_{\Lambda_b}+M_{\Lambda_c^*})}g_2^*-g_1^*; \\
   	g_\perp^*=&\frac{M_{\Lambda_c^*}+M_{\Lambda_b}}{M_{\Lambda_b}}g_2^*-g_1^*.
   \end{align}
   
\begin{figure*}
		\centering  
		\includegraphics[width=0.3\textwidth]{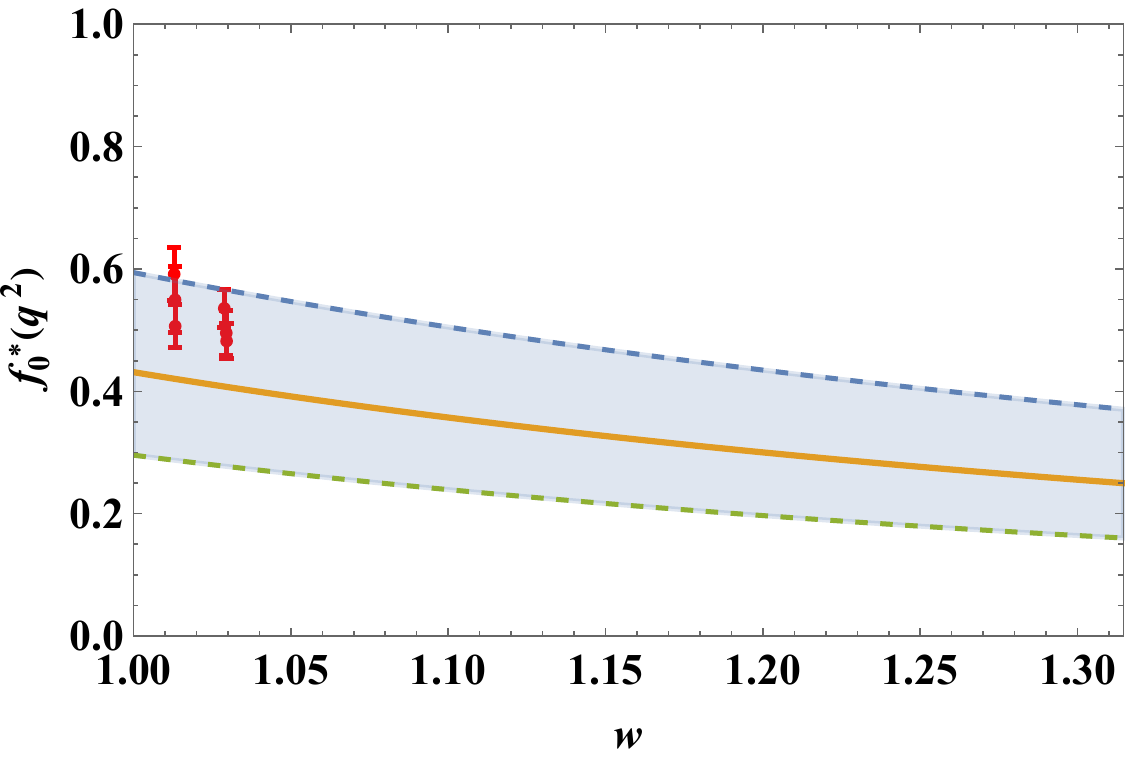} 
		\quad  
		\includegraphics[width=0.3\textwidth]{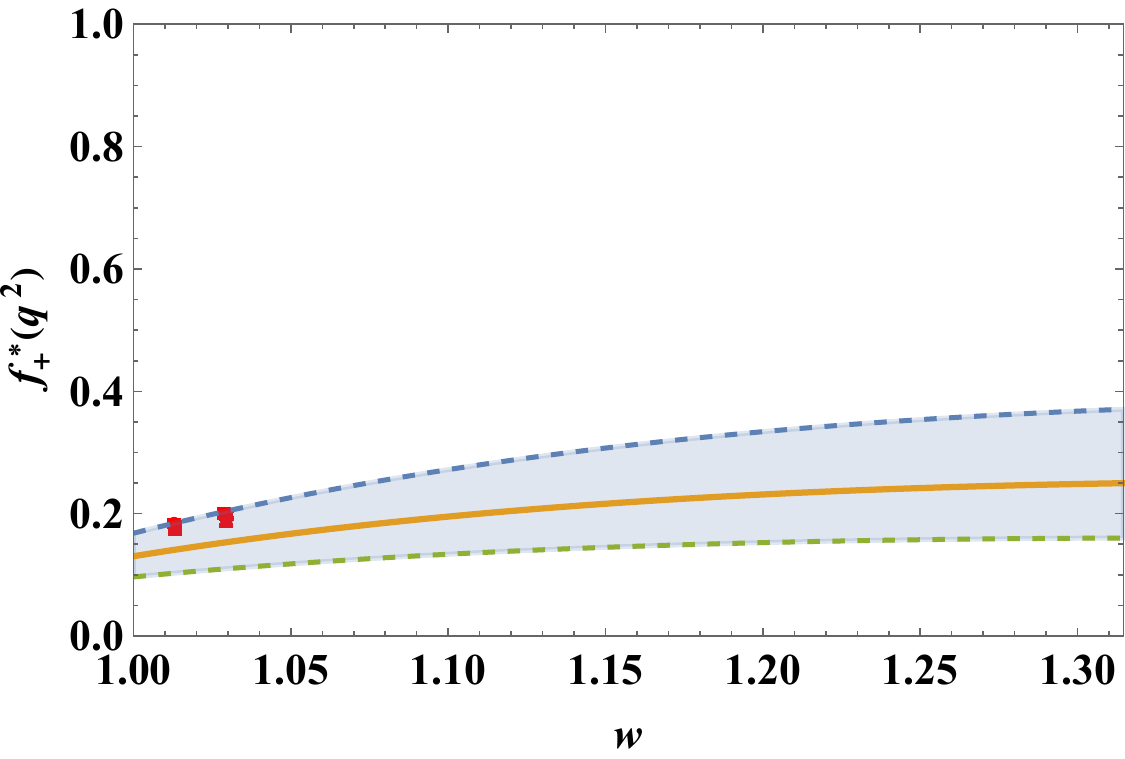} 
		\quad 
		\includegraphics[width=0.3\textwidth]{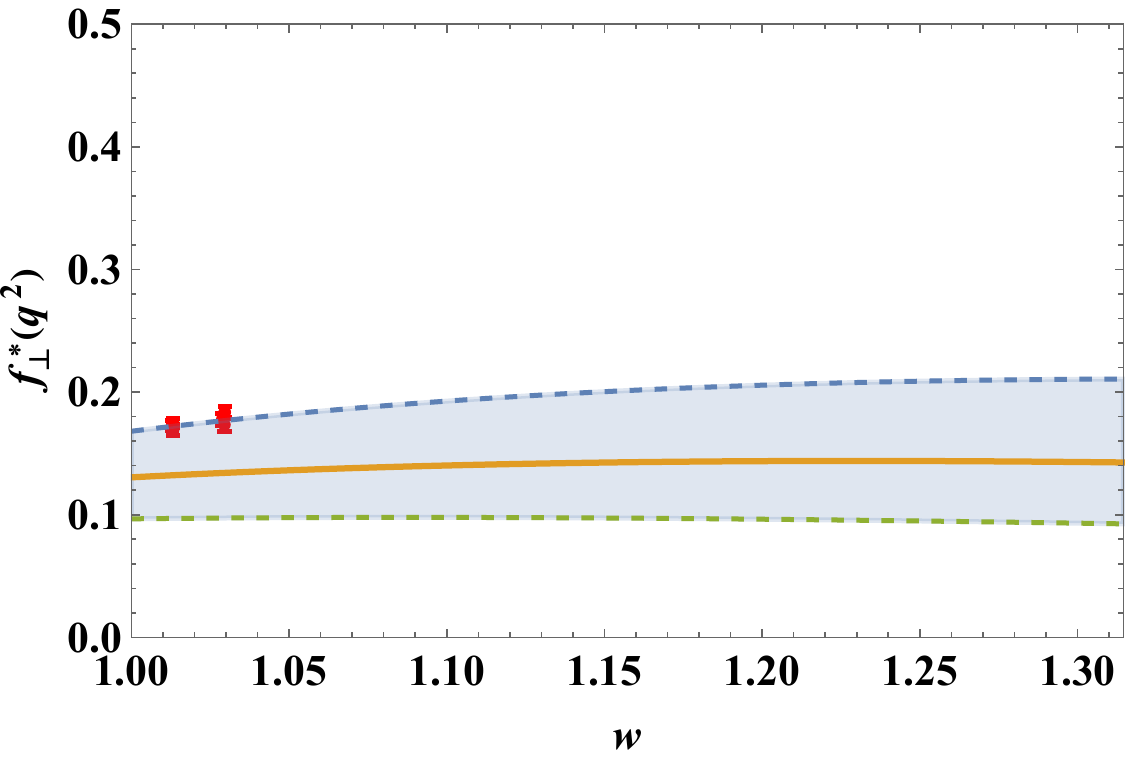} \\
		\includegraphics[width=0.3\textwidth]{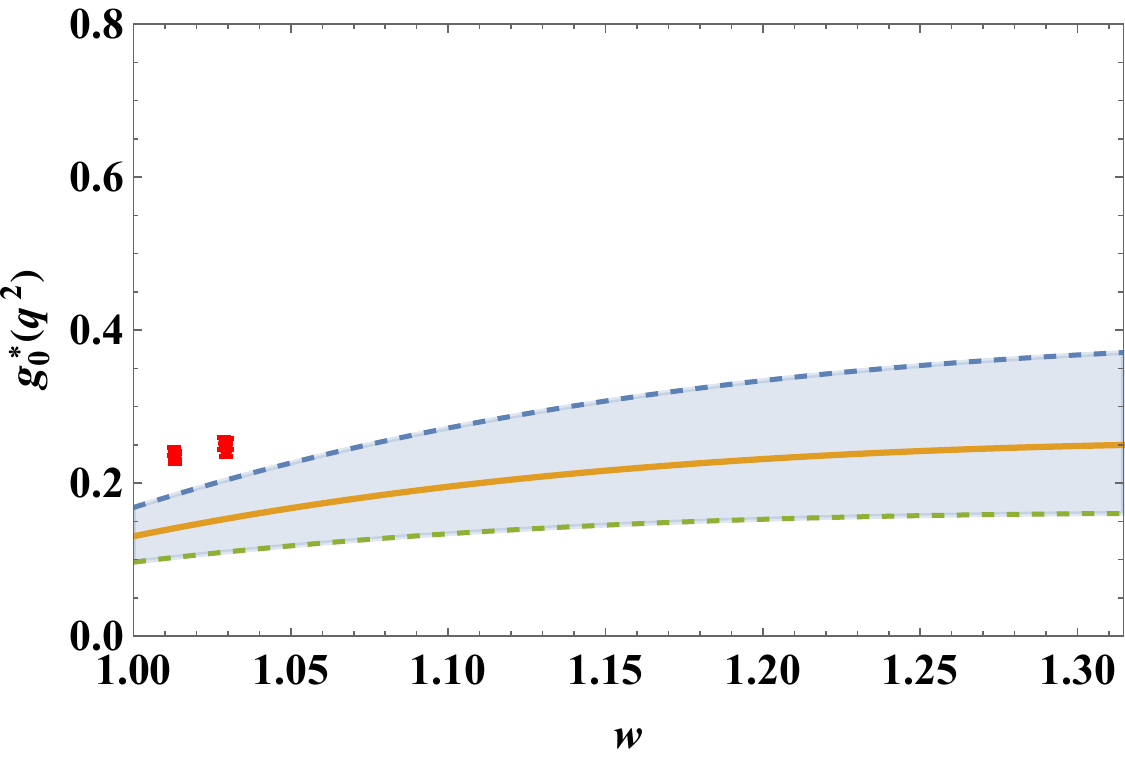} 
		\quad  
		\includegraphics[width=0.3\textwidth]{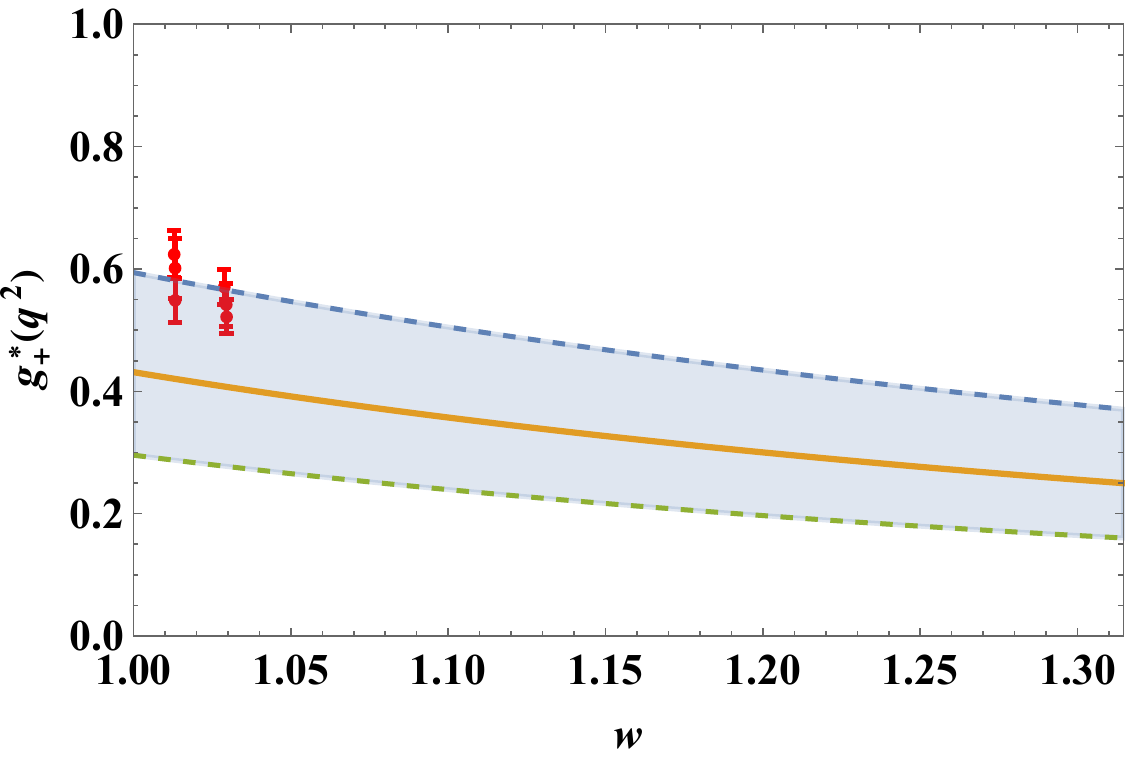} 
		\quad 
		\includegraphics[width=0.3\textwidth]{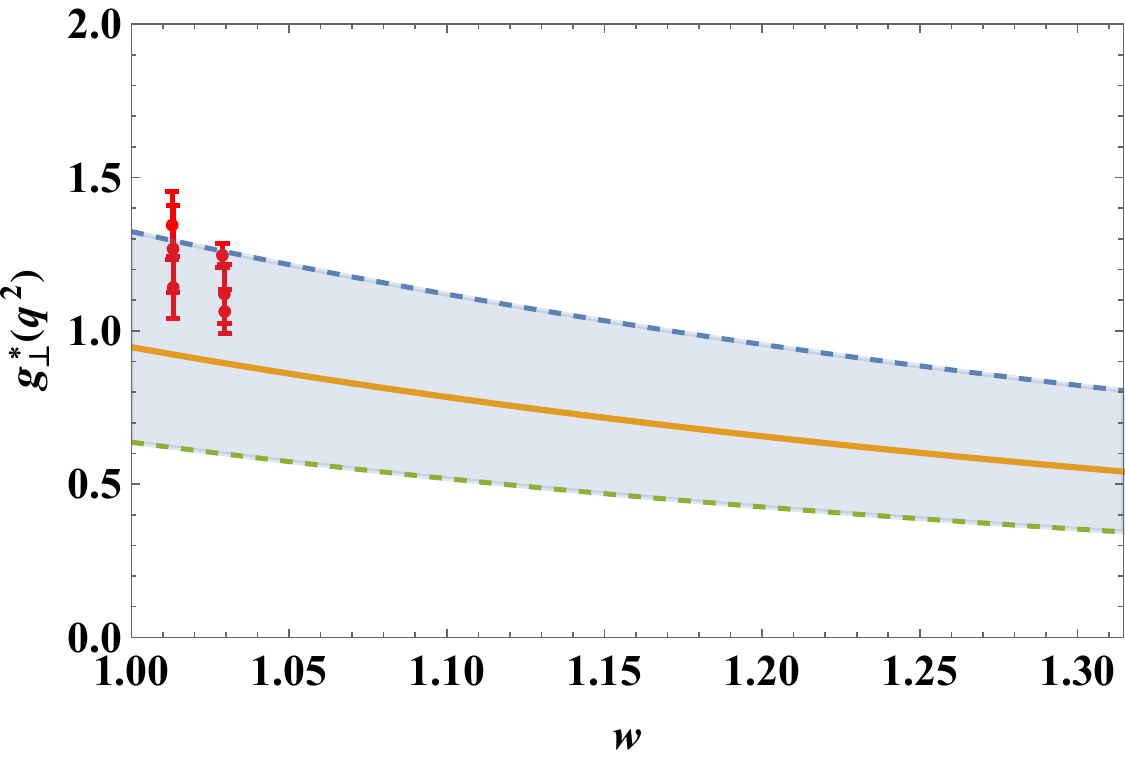}
		\caption{Form factors $f_i^*(i=1,2)$ on the entire physical region with the $j_{\Lambda_c}^A$ interpolating current. The central solid line represents the parameters $A=0.1$, $m_c=1.67~\text{GeV}$, and $s_0=(M_{\Lambda_c}+0.4)^2~\text{GeV}^2$ within the exponential LCDA models in reference \cite{Ali:2012pn} (Type I). The shaded regions contain the error bounds from the input parameters and the discrete points represent LQCD data from \cite{Meinel:2021rbm}.}\label{fig1.2}  
   \end{figure*}

\begin{figure*}
   	\centering  
   	\includegraphics[width=0.3\textwidth]{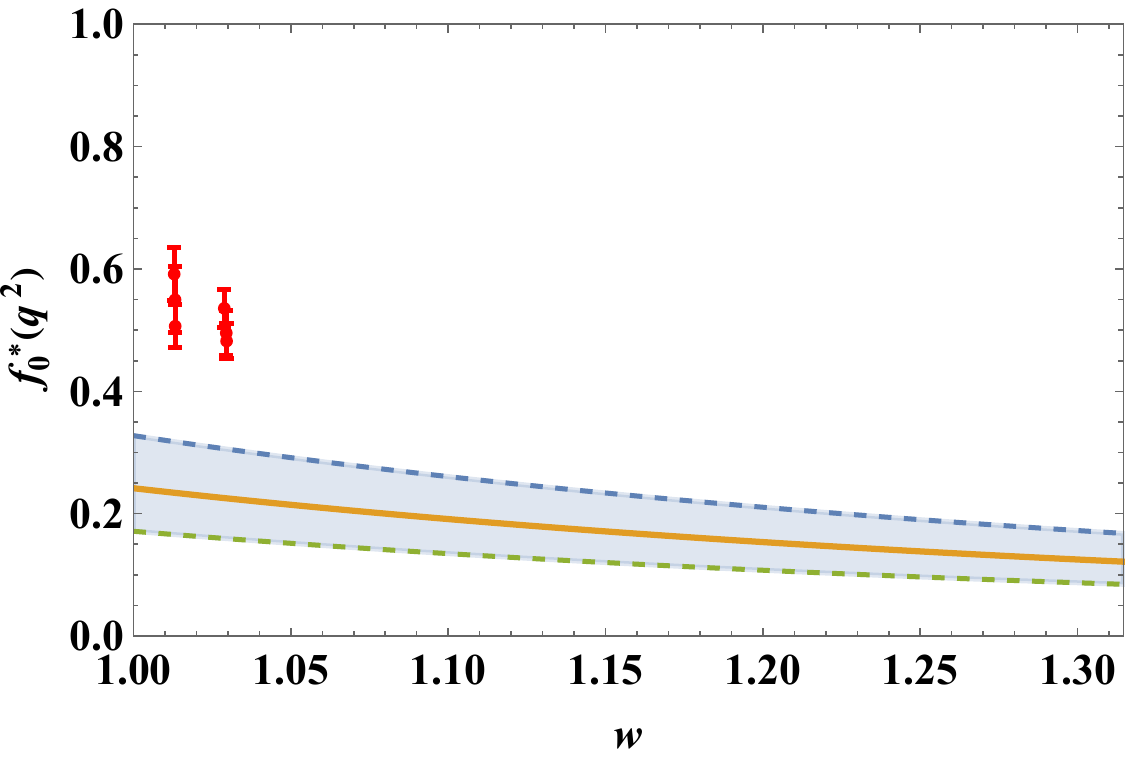} 
   	\quad  
   	\includegraphics[width=0.3\textwidth]{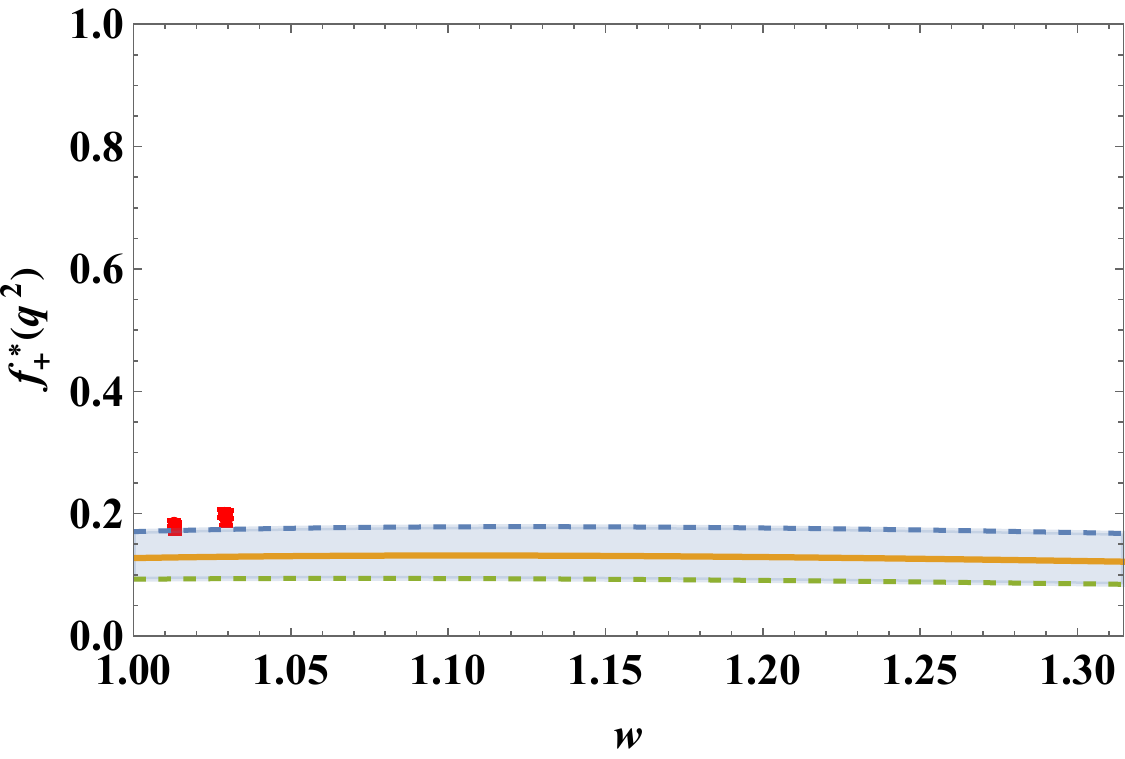} 
   	\quad 
   	\includegraphics[width=0.3\textwidth]{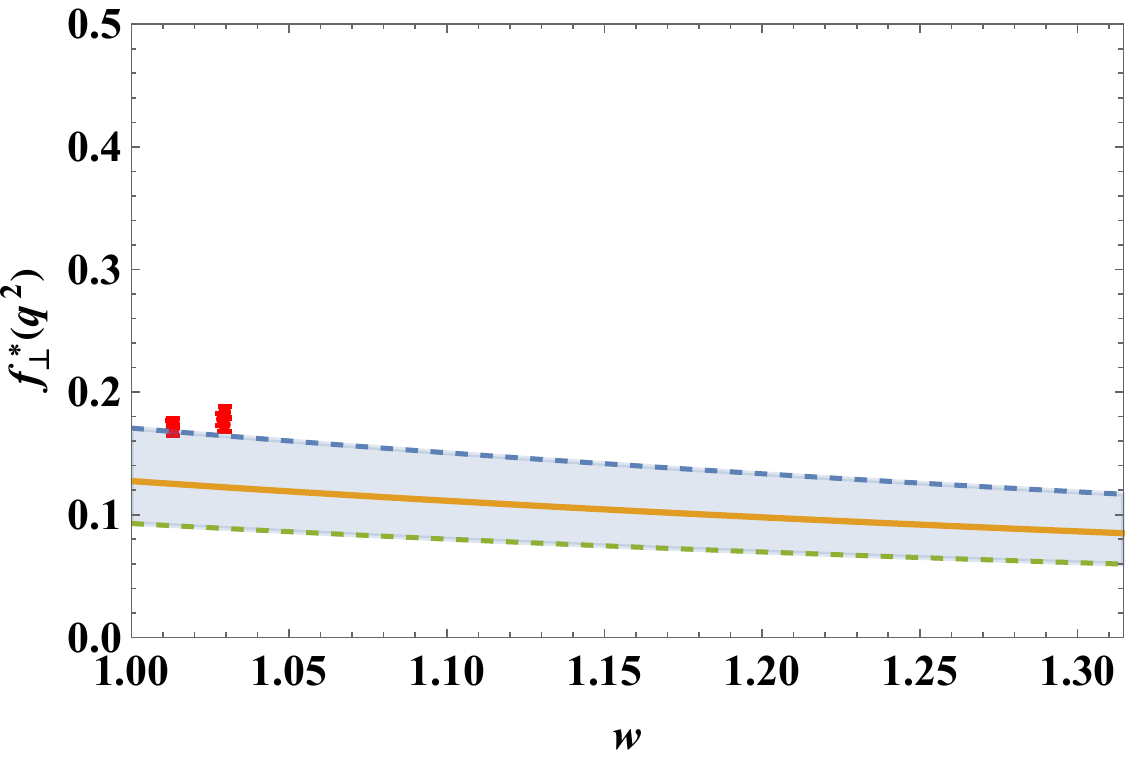} \\
   	\includegraphics[width=0.3\textwidth]{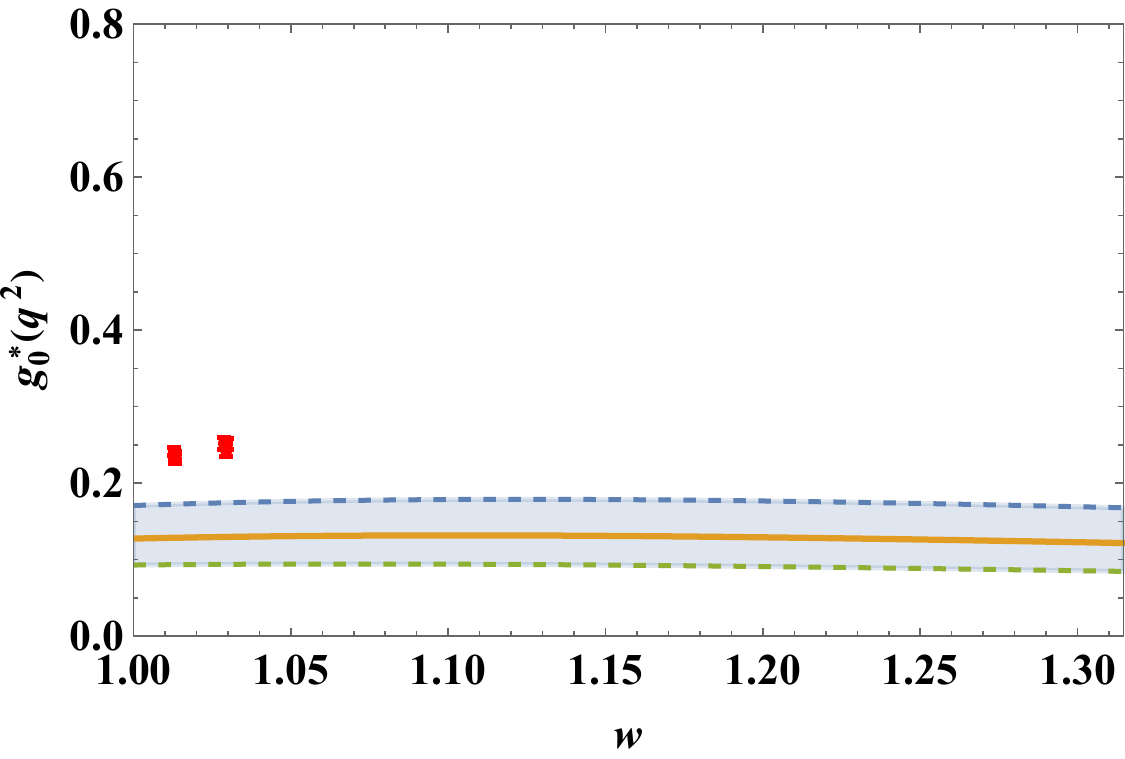} 
   	\quad  
   	\includegraphics[width=0.3\textwidth]{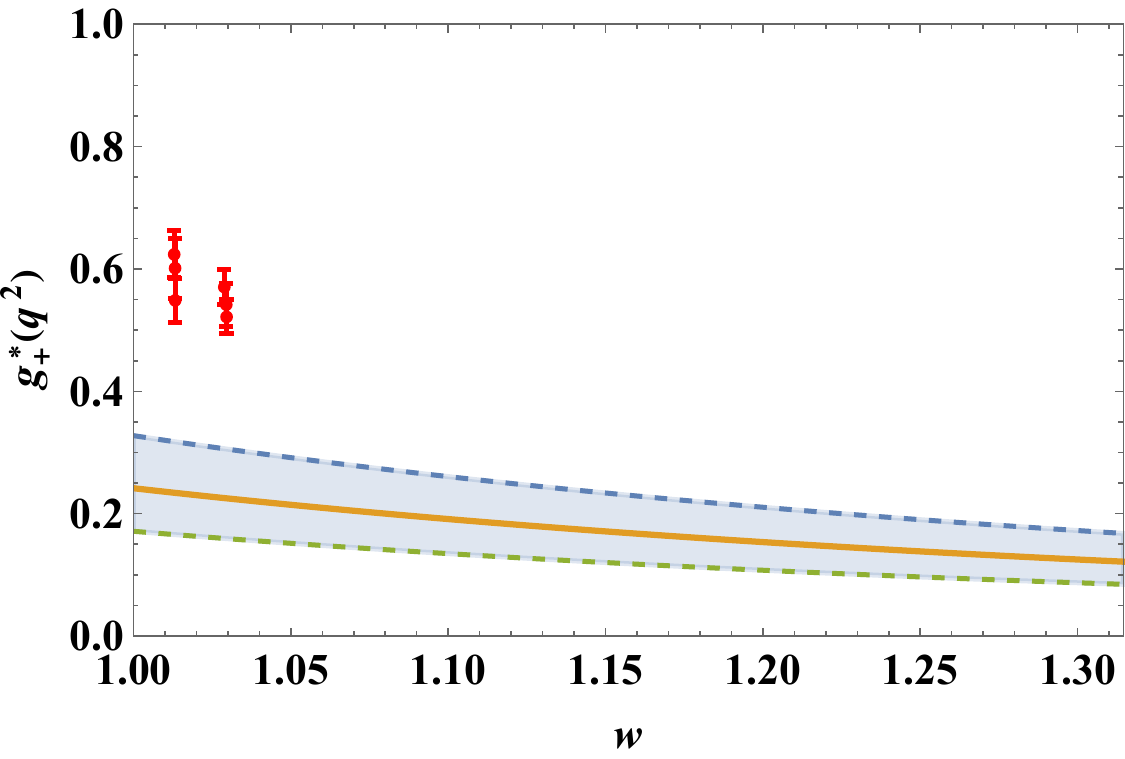} 
   	\quad 
   	\includegraphics[width=0.3\textwidth]{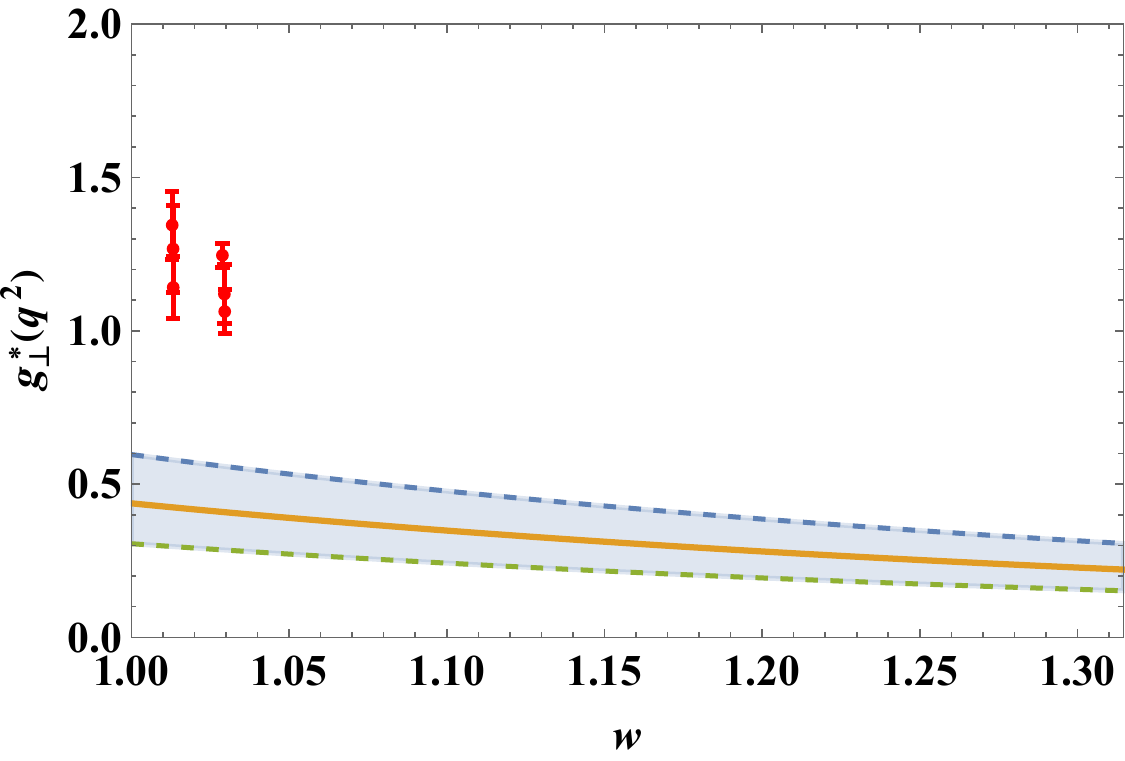}
   	\caption{Similar to Fig. \ref{fig1.2}, but for the exponential LCDA models described in reference \cite{Bell:2013tfa} (Type II).}\label{fig1.3}  
   \end{figure*}

\begin{figure*}
   	\centering  
   	\includegraphics[width=0.3\textwidth]{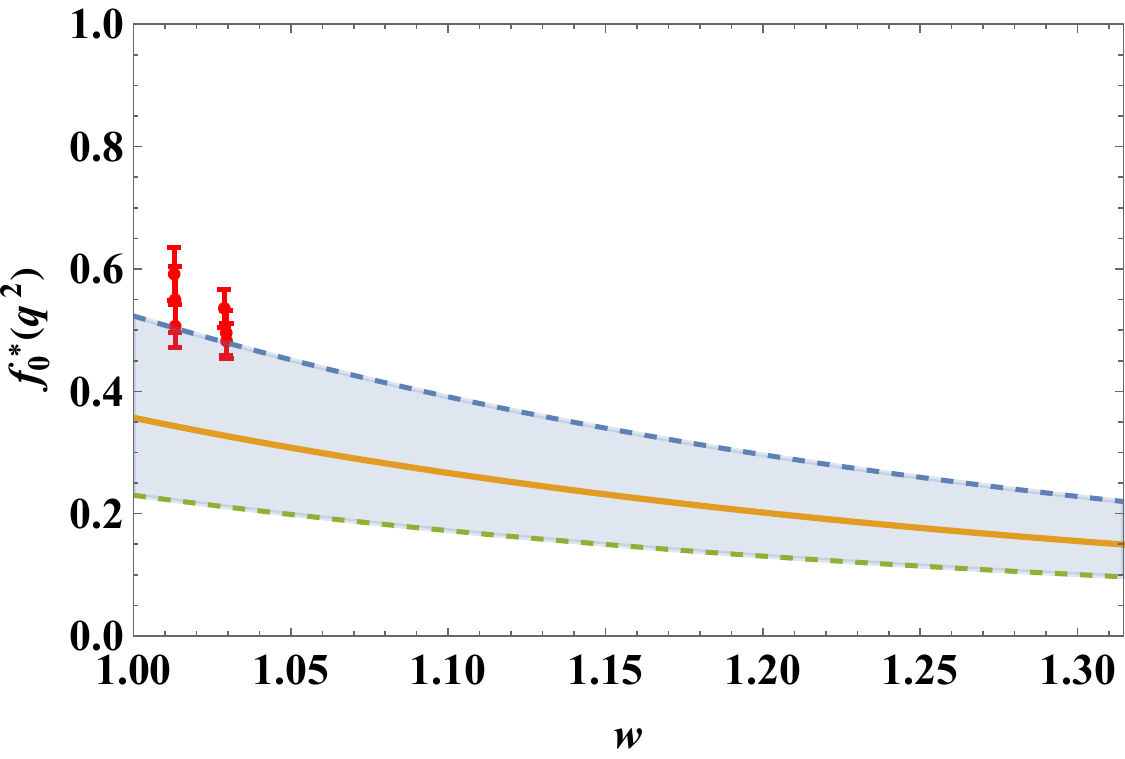} 
   	\quad  
   	\includegraphics[width=0.3\textwidth]{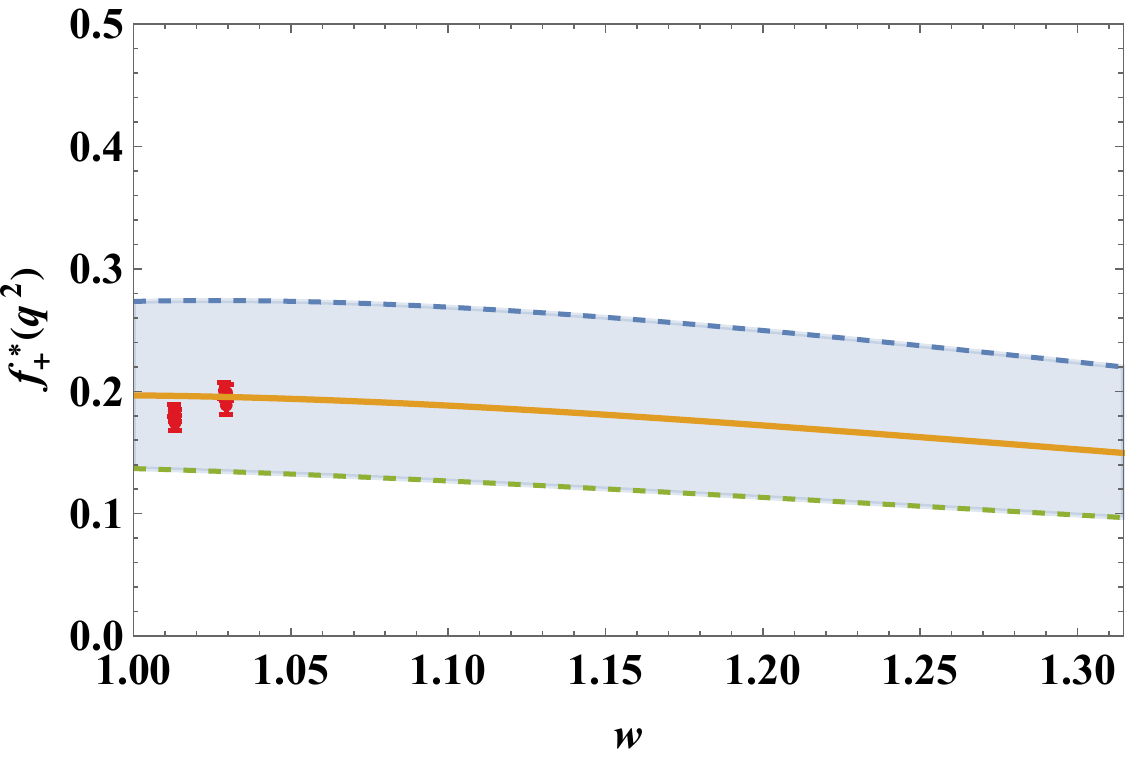} 
   	\quad 
   	\includegraphics[width=0.3\textwidth]{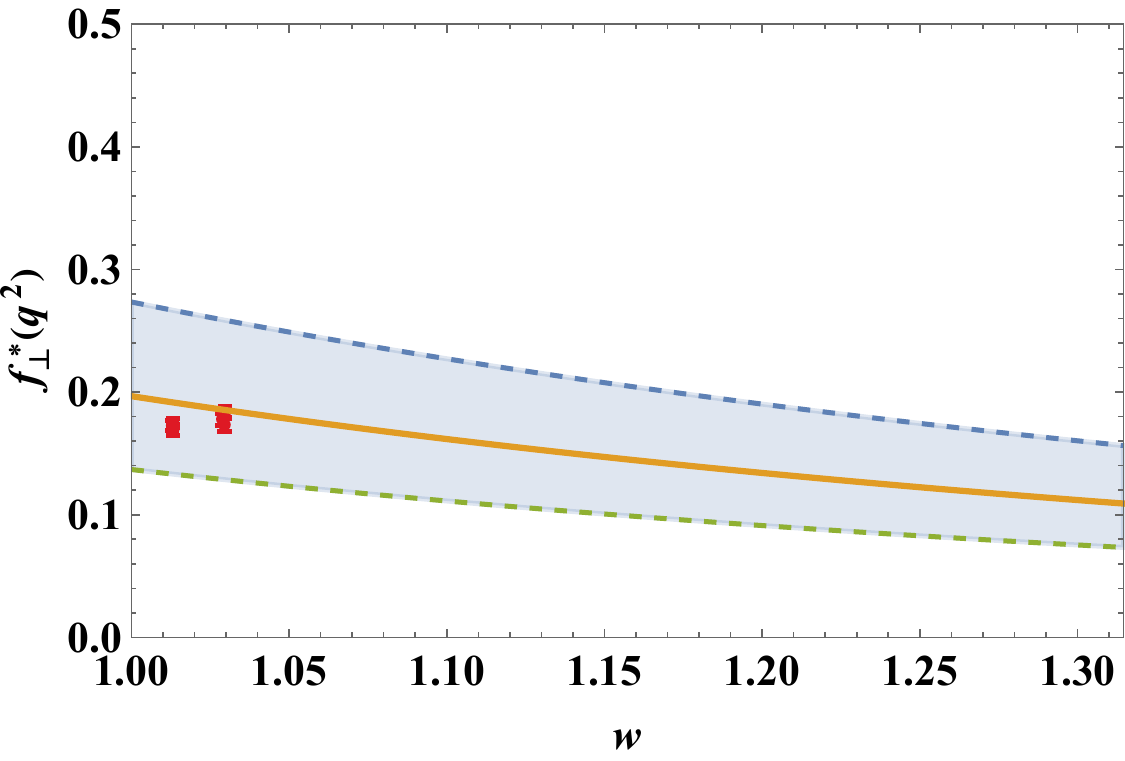} \\
   	\includegraphics[width=0.3\textwidth]{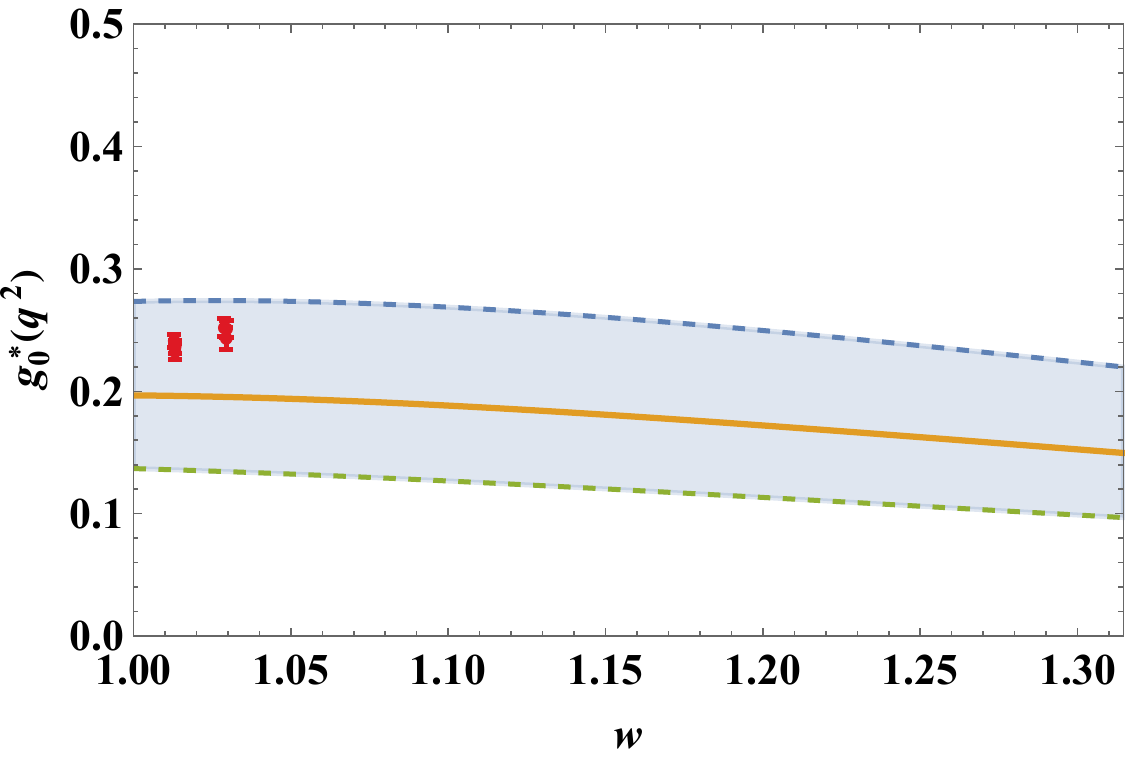} 
   	\quad  
   	\includegraphics[width=0.3\textwidth]{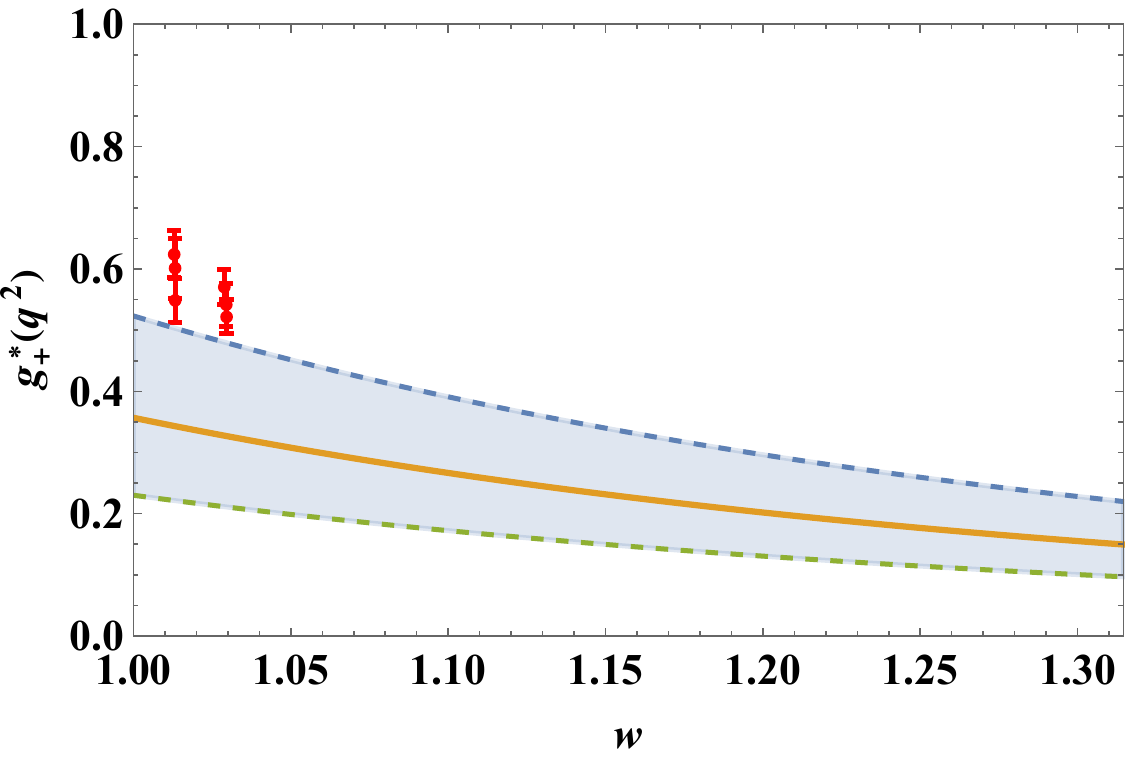} 
   	\quad 
   	\includegraphics[width=0.3\textwidth]{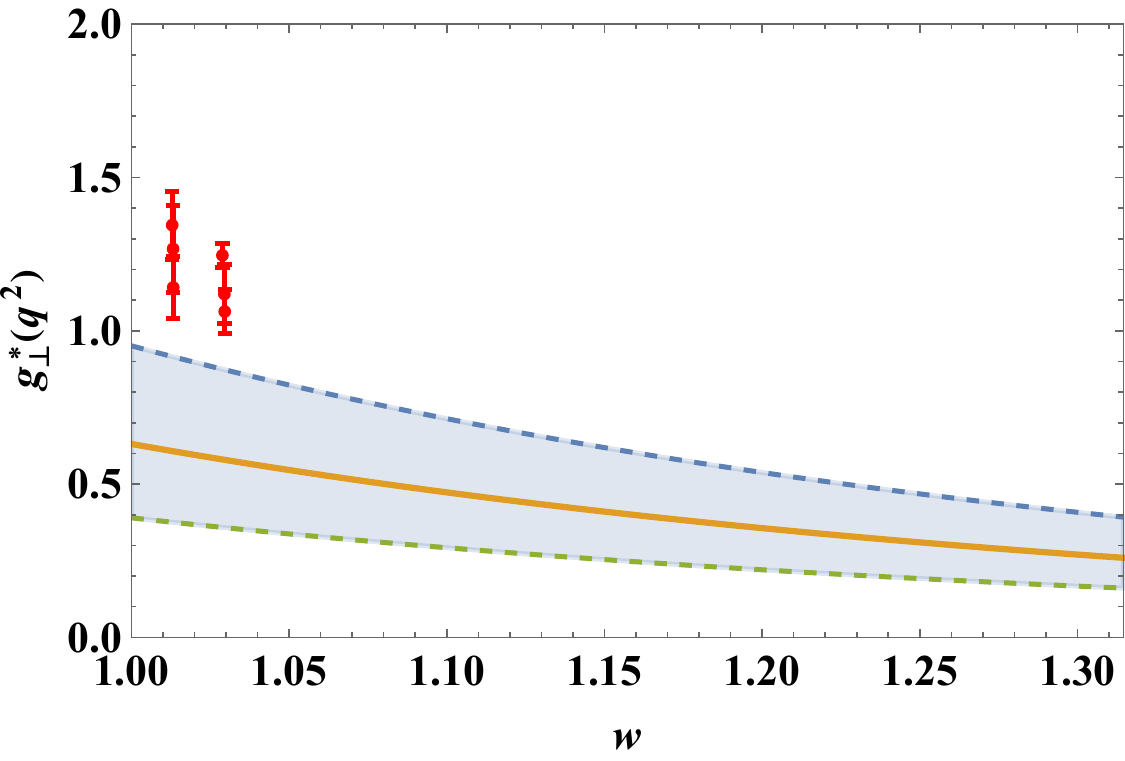}
   	\caption{Similar to Fig. \ref{fig1.2}, but for the free parton LCDA models described in reference \cite{Bell:2013tfa} (Type III).}\label{fig1.4}  
   \end{figure*}

\begin{figure*}
   	\centering  
   	\includegraphics[width=0.3\textwidth]{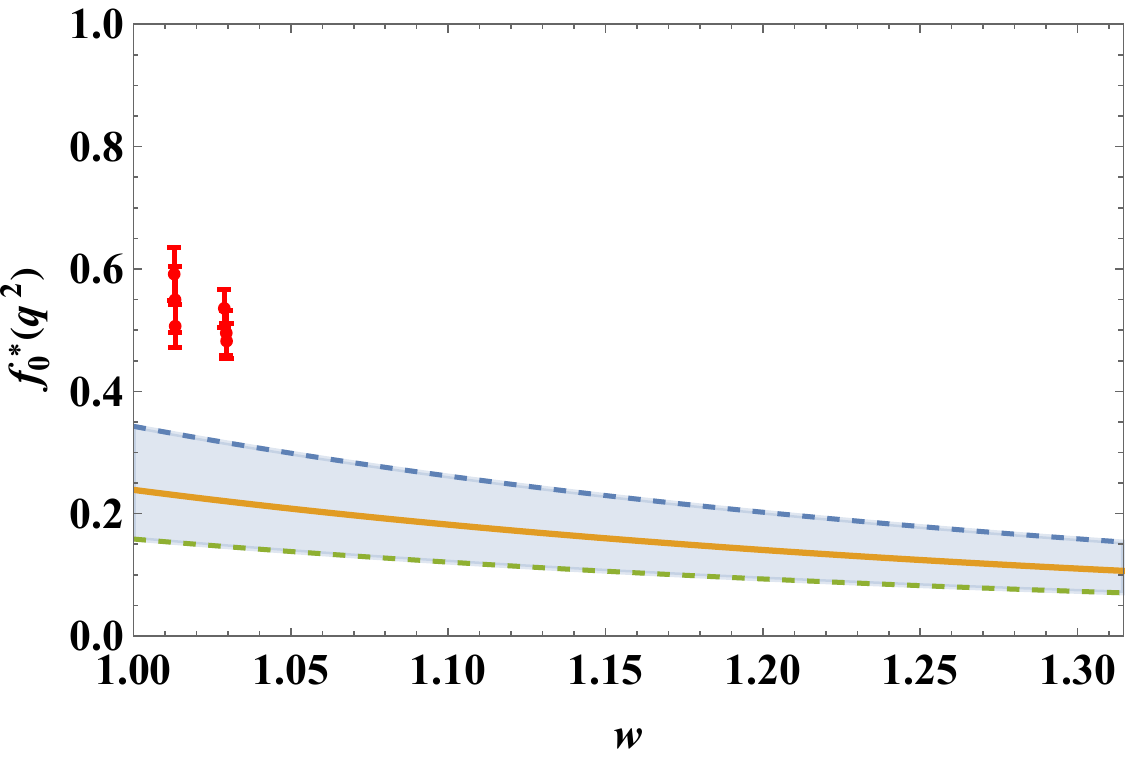} 
   	\quad  
   	\includegraphics[width=0.3\textwidth]{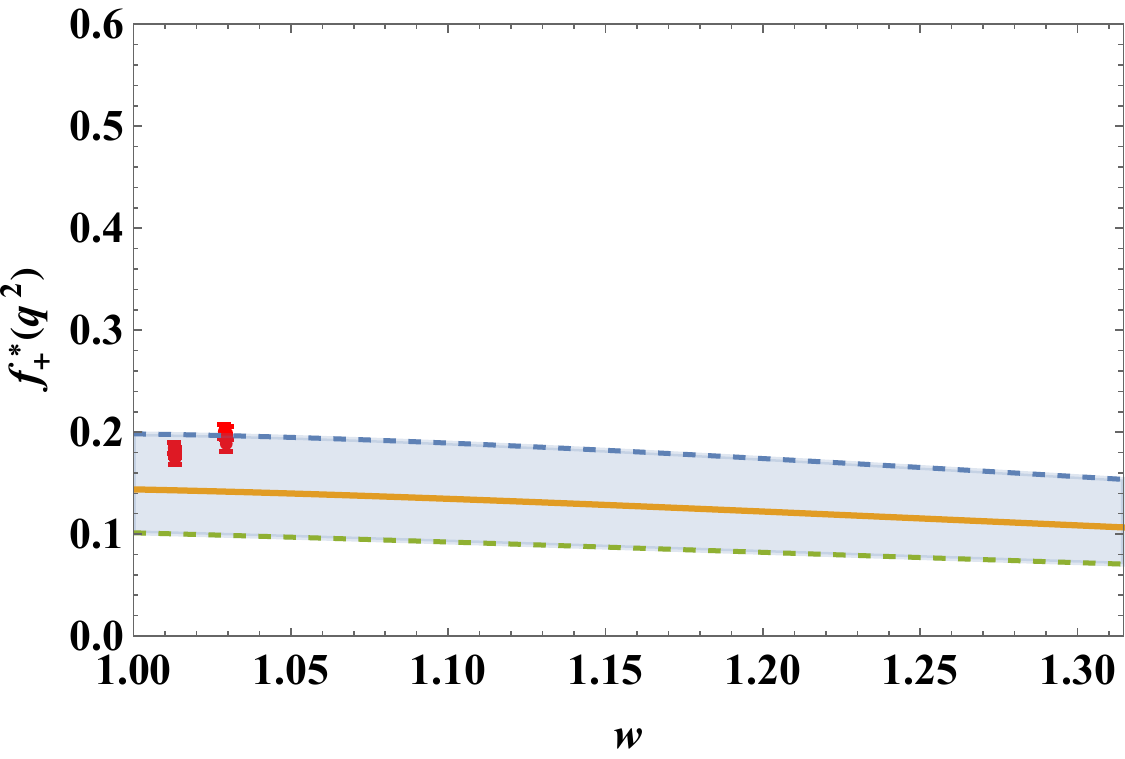} 
   	\quad 
   	\includegraphics[width=0.3\textwidth]{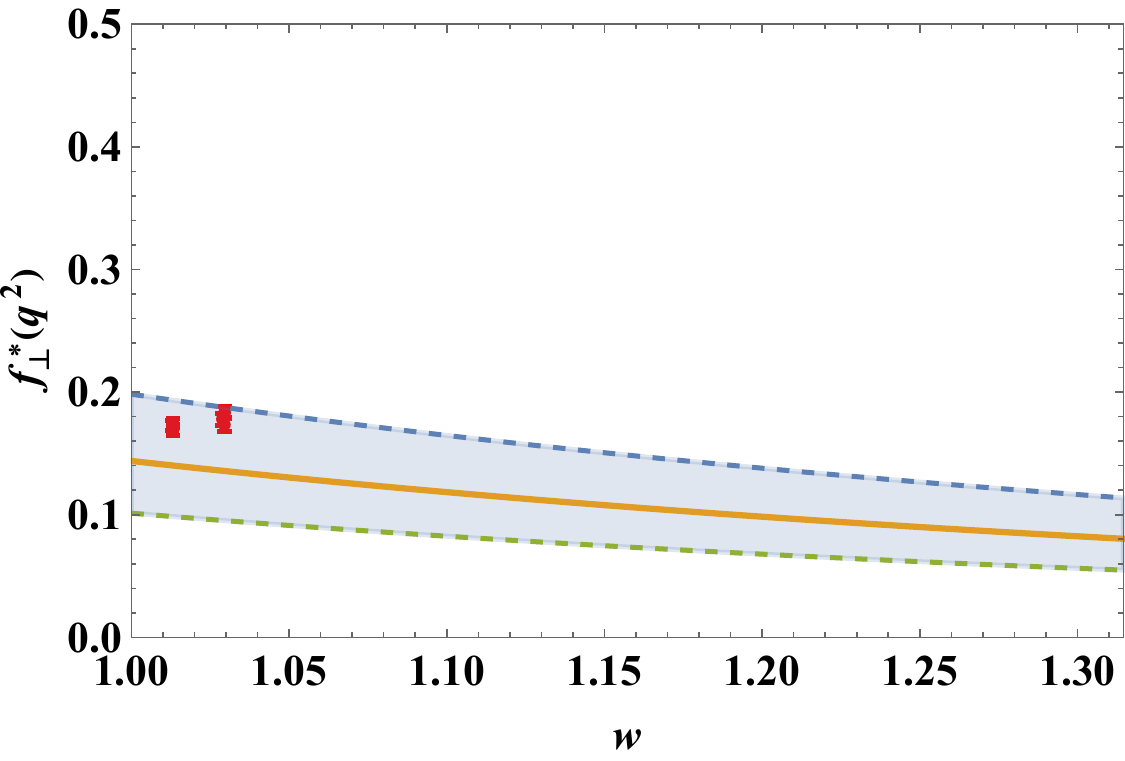} \\
   	\includegraphics[width=0.3\textwidth]{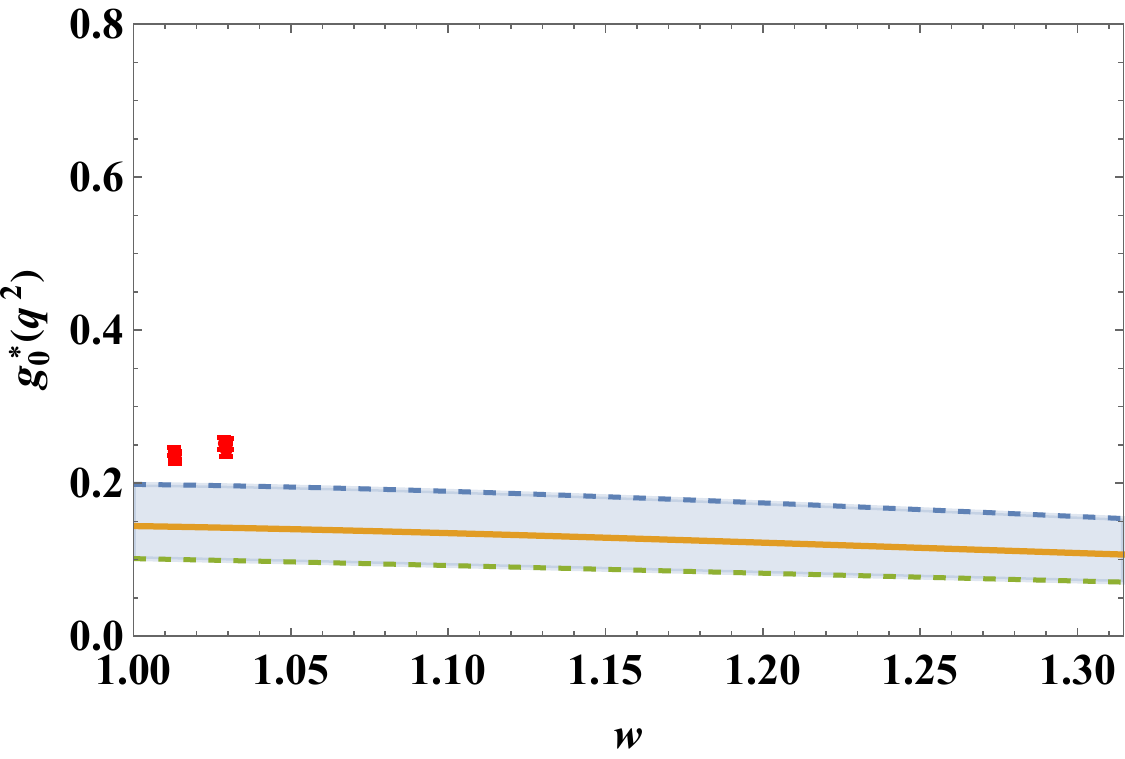} 
   	\quad  
   	\includegraphics[width=0.3\textwidth]{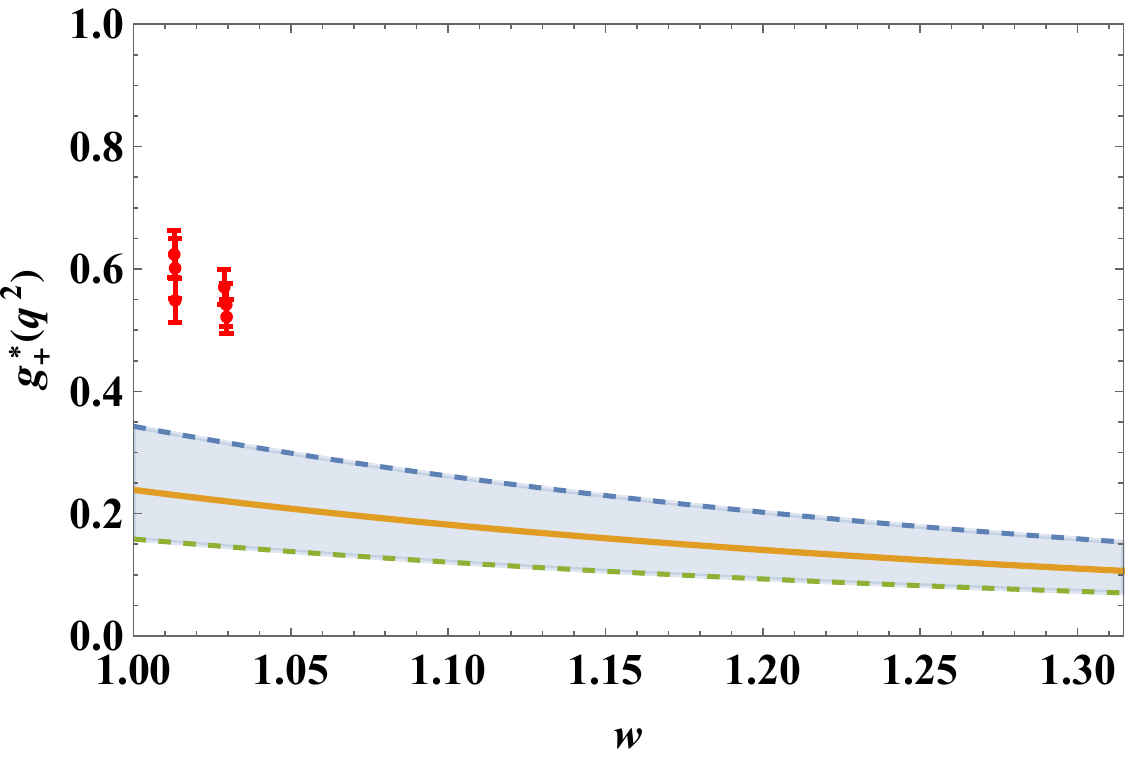} 
   	\quad 
   	\includegraphics[width=0.3\textwidth]{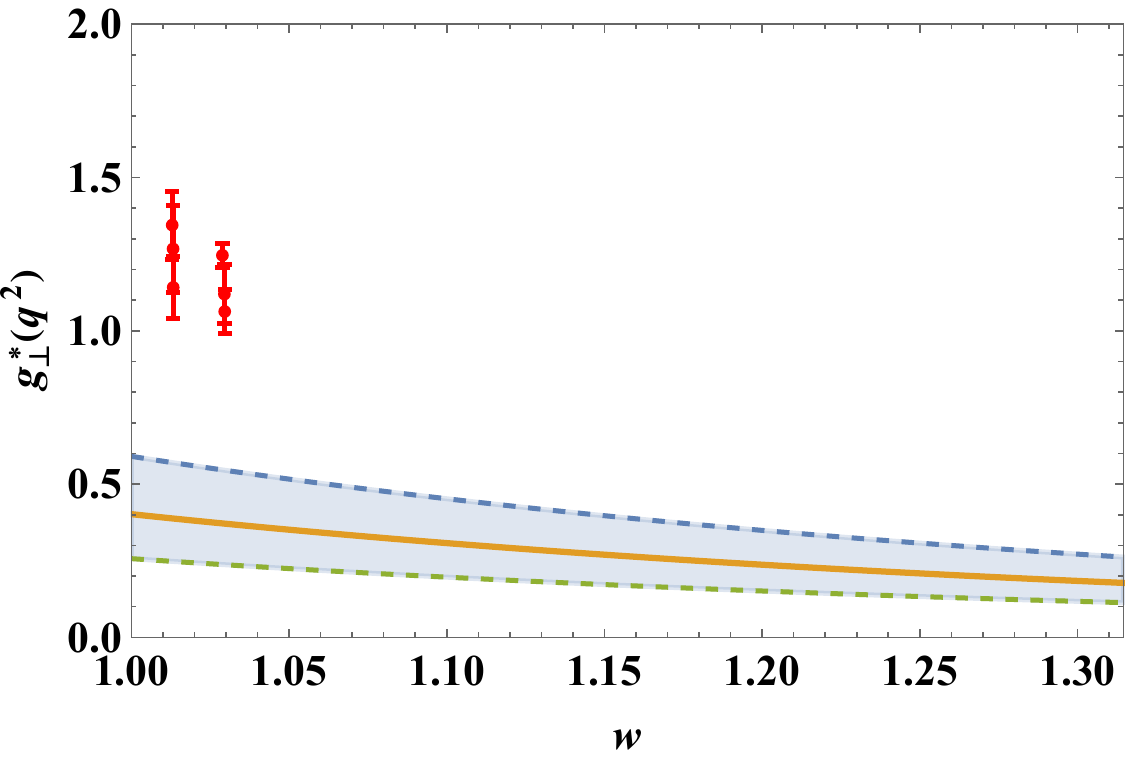}
   	\caption{Similar to Fig. \ref{fig1.2}, but for the leading order perturbative QCD sum rule LCDA models described in reference \cite{Ball:2008fw} (Type IV).}\label{fig1.5}  
   \end{figure*}

\begin{figure*}
   	\centering  
   	\includegraphics[width=0.3\textwidth]{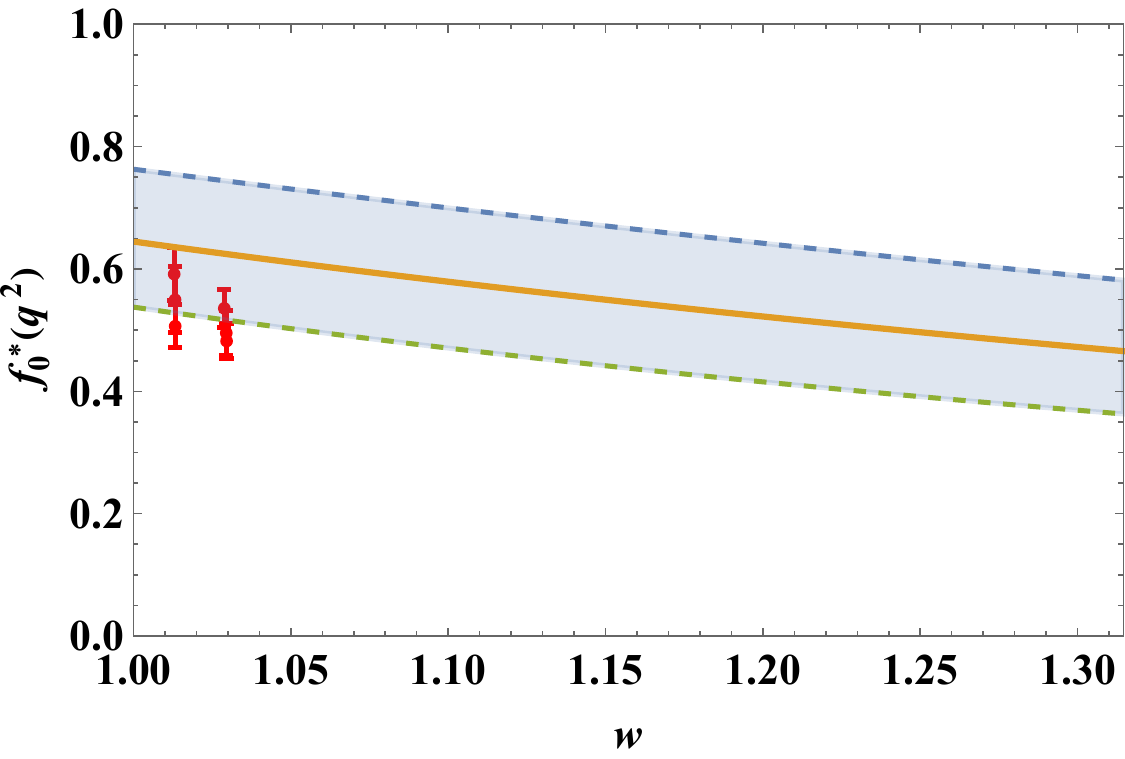} 
   	\quad  
   	\includegraphics[width=0.3\textwidth]{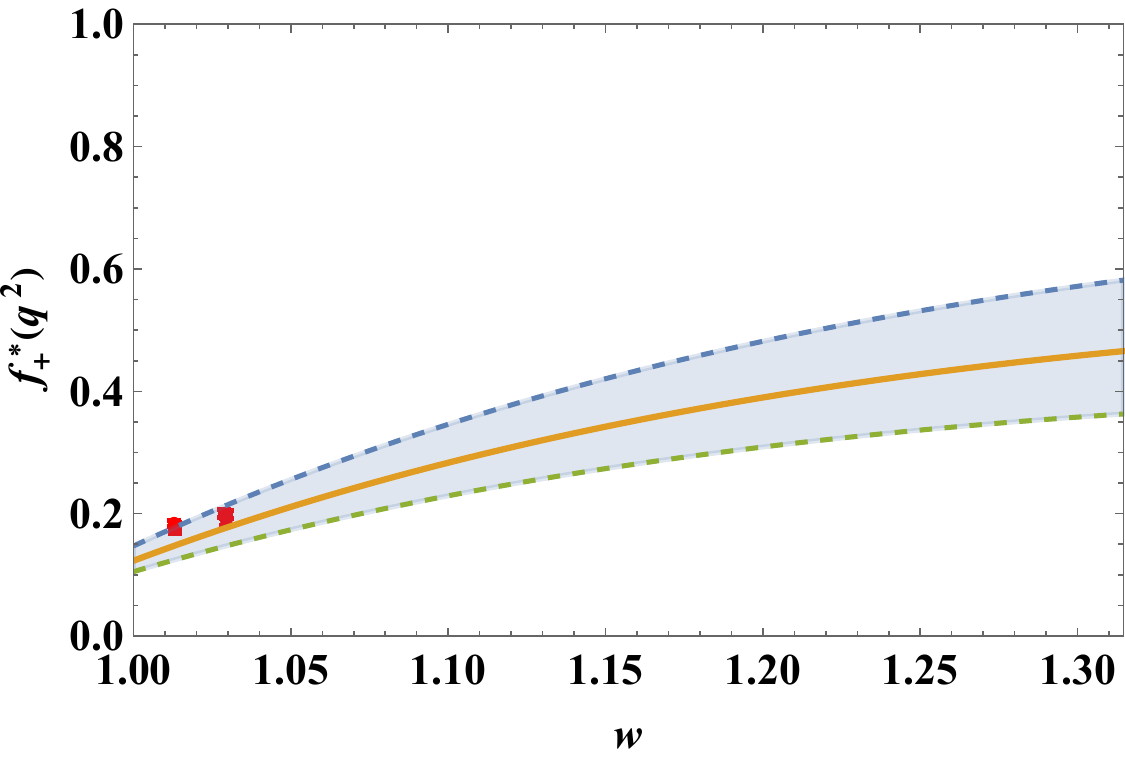} 
   	\quad 
   	\includegraphics[width=0.3\textwidth]{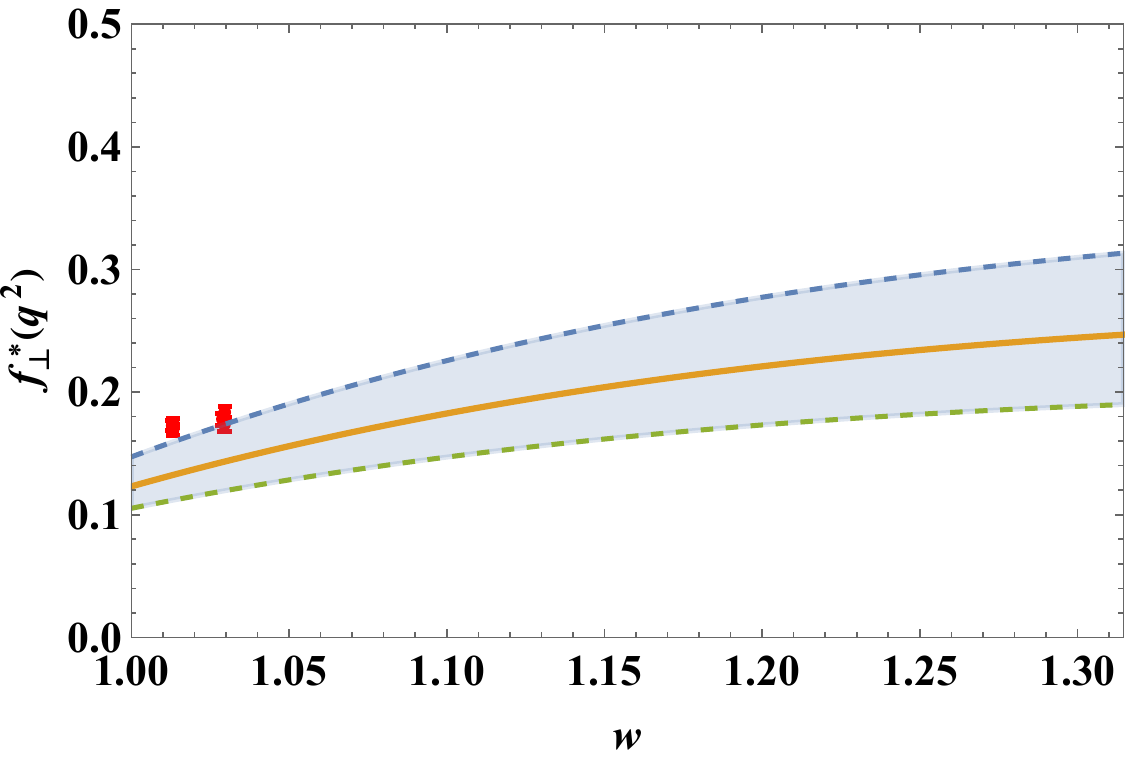} \\
   	\includegraphics[width=0.3\textwidth]{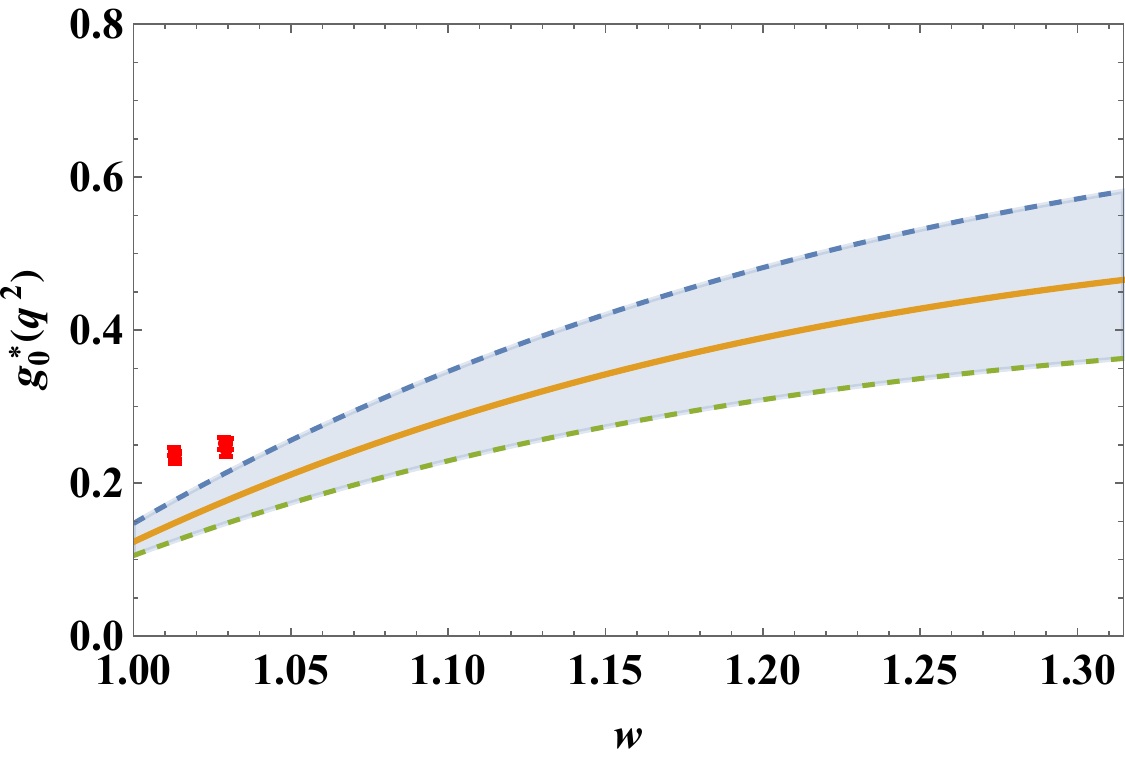} 
   	\quad  
   	\includegraphics[width=0.3\textwidth]{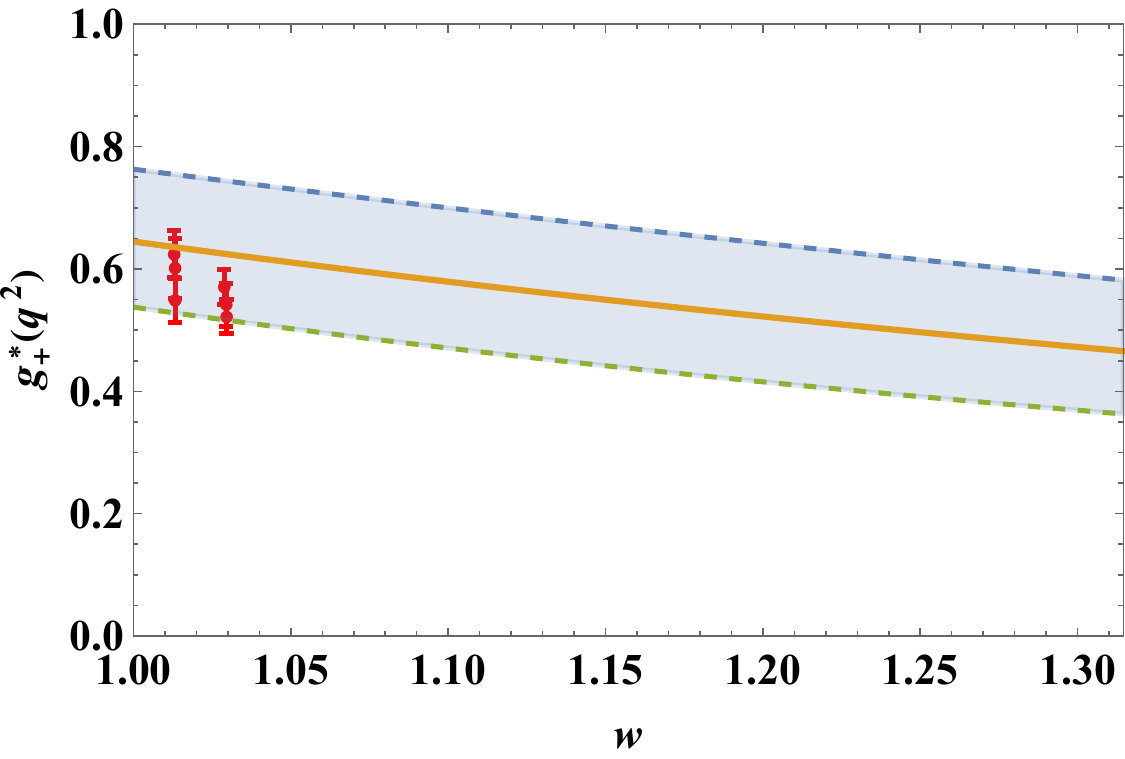} 
   	\quad 
   	\includegraphics[width=0.3\textwidth]{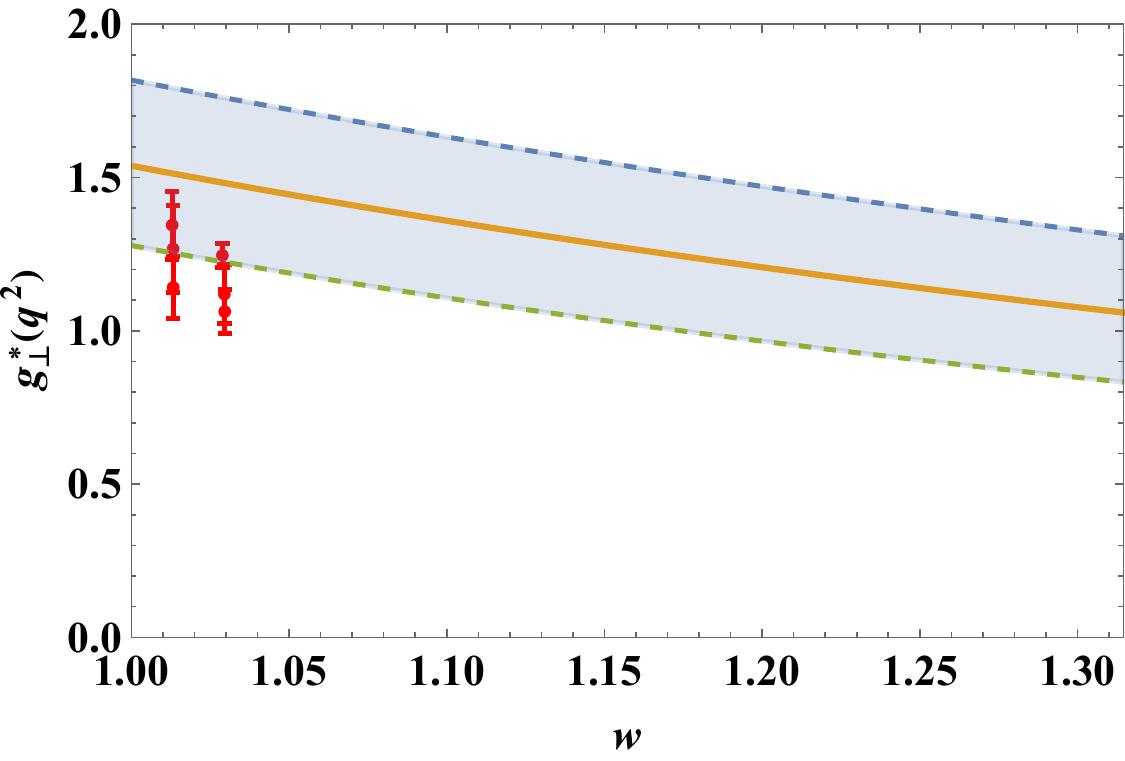}
   	\caption{Similar to Fig. \ref{fig1.2}, but for the Gegenbauer moments LCDA models described in reference \cite{Ball:2008fw} (Type V).}\label{fig1.6}  
   \end{figure*}

    The plots of form factors on the entire physical region within different LCDAs models of the $\Lambda_b$ baryon are plotted in Fig. \ref{fig1.2} to Fig. \ref{fig1.6}. In these plots, only the form factors deduced by $j^A$ type $\Lambda_c$ interpolating current are depicts. Additionally, comparisions with LQCD data from Ref. \cite{Meinel:2021rbm} are displayed in these figures as discrete points. 
   
   \subsection{Semileptonic decay of $\Lambda_b^0 \to \Lambda_c^{*+}\ell^-\bar{\nu}_\ell$} \label{sec:IV}
	
   In order to obtain the decay widths and branching ratios of $\Lambda_b^0 \to \Lambda_c^* \ell^- \bar{\nu}_\ell$, we will utilize the helicity amplitudes representation of baryon transition matrix element. The formula for differential decay width of semileptonic decay, with the inclusion of lepton mass, can be written as referenced in \cite{Faessler:2009xn,Gutsche:2017wag}:  
   \begin{align}  
	\frac{d\Gamma}{dq^2} = \frac{G_F^2 |V_{\text{CKM}}|^2}{192\pi^3} \frac{M_{\Lambda_b}^2 - M_{\Lambda_c^*}^2}{2M_{\Lambda_b}^3} \frac{(q^2 - m_\ell^2)^2}{q^2} \mathcal{H}_{\frac{1}{2} \to -\frac{1}{2}}. 
   \end{align}  
   Where the total decay width is obtained by integrating the differential decay width over the entire physical range:  
   \begin{align}  
	\Gamma = \int_{m_\ell^2}^{M_-^2} \frac{d\Gamma}{dq^2}.  
   \end{align}  
   Here, $\mathcal{H}_{\frac{1}{2} \to -\frac{1}{2}}$ is defined as:  
   \begin{align}  
	\mathcal{H}_{\frac{1}{2} \to -\frac{1}{2}} =& |H_{\frac{1}{2},0}|^2 + |H_{-\frac{1}{2},0}|^2 + |H_{\frac{1}{2},1}|^2 + |H_{-\frac{1}{2},-1}|^2 \notag \\  
	& + \frac{m_\ell^2}{2q^2}\left(3|H_{\frac{1}{2},t}|^2 + 3|H_{-\frac{1}{2},t}|^2 + |H_{\frac{1}{2},1}|^2\right.  \notag \\ 
	& \left.+ |H_{-\frac{1}{2},-1}|^2 + |H_{\frac{1}{2},0}|^2 + |H_{-\frac{1}{2},0}|^2\right).  \label{helicity am}
   \end{align}  
   The helicity amplitudes $H_{\frac{1}{2},0}$, $H_{\frac{1}{2},1}$, and $H_{\frac{1}{2},t}$ connected to the form factors for the $\frac{1}{2}^+ \to \frac{1}{2}^-$ processes are expressed as:  
   \begin{align}  
	H_{\frac{1}{2},0}^V &= \sqrt{\frac{Q_+}{q^2}} \left[ f_1^*(q^2)M_- + f_2^*(q^2)\frac{q^2}{M_{\Lambda_b}} \right], \notag \\
	H_{\frac{1}{2},0}^A &= \sqrt{\frac{Q_-}{q^2}} \left[ M_+g_1^*(q^2) - \frac{q^2}{M_{\Lambda_b}}g_2^*(q^2) \right], \notag \\  
	H_{\frac{1}{2},1}^V &= \sqrt{2Q_+} \left[ -f_1^*(q^2) - \frac{M_-}{M_{\Lambda_b}}f_2^*(q^2) \right], \notag \\
	H_{\frac{1}{2},1}^A &= \sqrt{2Q_-} \left[ -g_1^*(q^2) + \frac{M_+}{M_{\Lambda_b}}g_2^*(q^2) \right], \notag \\  
	H_{\frac{1}{2},t}^V &= \sqrt{\frac{Q_-}{q^2}} \left[ M_+f_1^*(q^2) - \frac{q^2}{M_{\Lambda_b}}f_3^*(q^2) \right], \notag \\
	H_{\frac{1}{2},t}^A &= \sqrt{\frac{Q_+}{q^2}} \left[ M_-g_1^*(q^2) + \frac{q^2}{M_{\Lambda_b}}g_3^*(q^2) \right].  \label{helicity amp2}
   \end{align}
   The negative helicity amplitudes can be derived from the positive ones in the following transformation:  
   \begin{equation}  
	H_{-\lambda,-\lambda_W}^V = -H_{\lambda,\lambda_W}^V, \qquad 
	H_{-\lambda,-\lambda_W}^A = H_{\lambda,\lambda_W}^A.  
   \end{equation}
    Here, $\lambda$ and $\lambda_W$ represent the polarizations of the final $\Lambda_c^*$ baryon and W-Boson, respectively. Employing the $V-A$ current, the total helicity amplitudes are formulated as:  
    \begin{align}  
	H_{\lambda,\lambda_W} = H_{\lambda,\lambda_W}^V - H_{\lambda,\lambda_W}^A.  
    \end{align}  
   In the preceding equations Eq. (\ref{helicity am}) and Eq. (\ref{helicity amp2}), $Q_\pm$ is defined as $(M_{\Lambda_b} \pm M_{\Lambda_c^*})^2 - q^2$, $M_\pm$ stands for $M_{\Lambda_b} \pm M_{\Lambda_c^*}$, and $m_\ell$ denotes the mass of leptons.
 
	\begin{figure*}
	\begin{center}
		\includegraphics[width=0.31\textwidth]{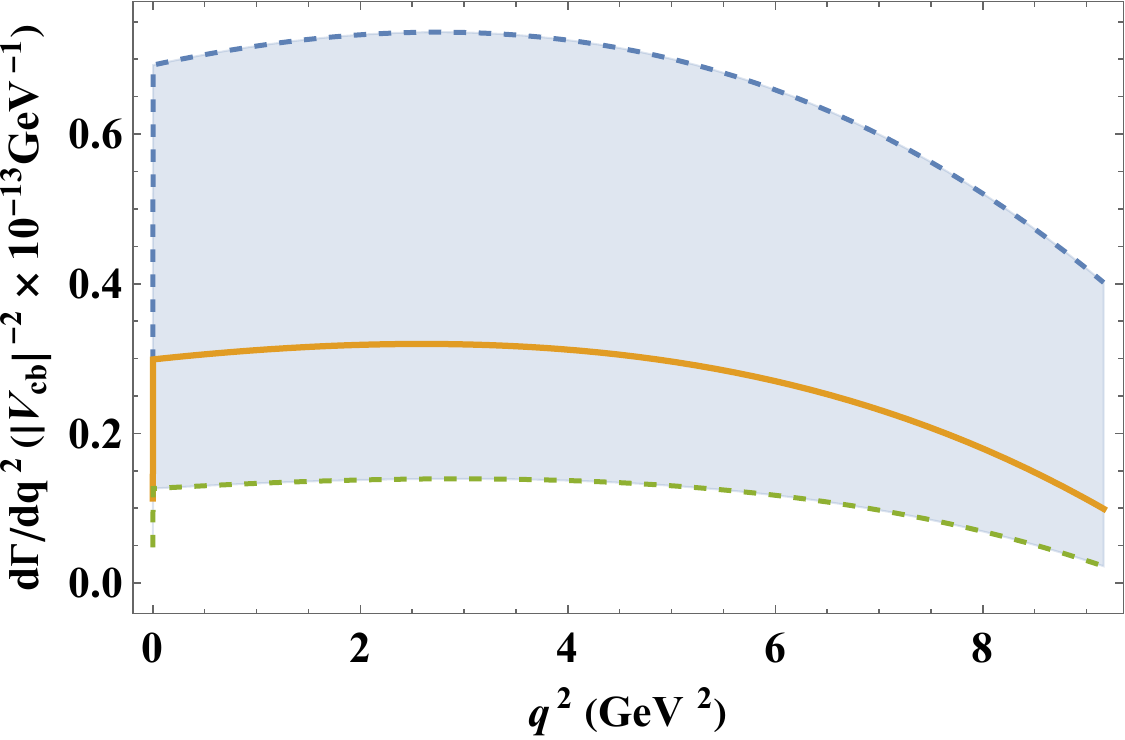}		\quad		\includegraphics[width=0.31\textwidth]{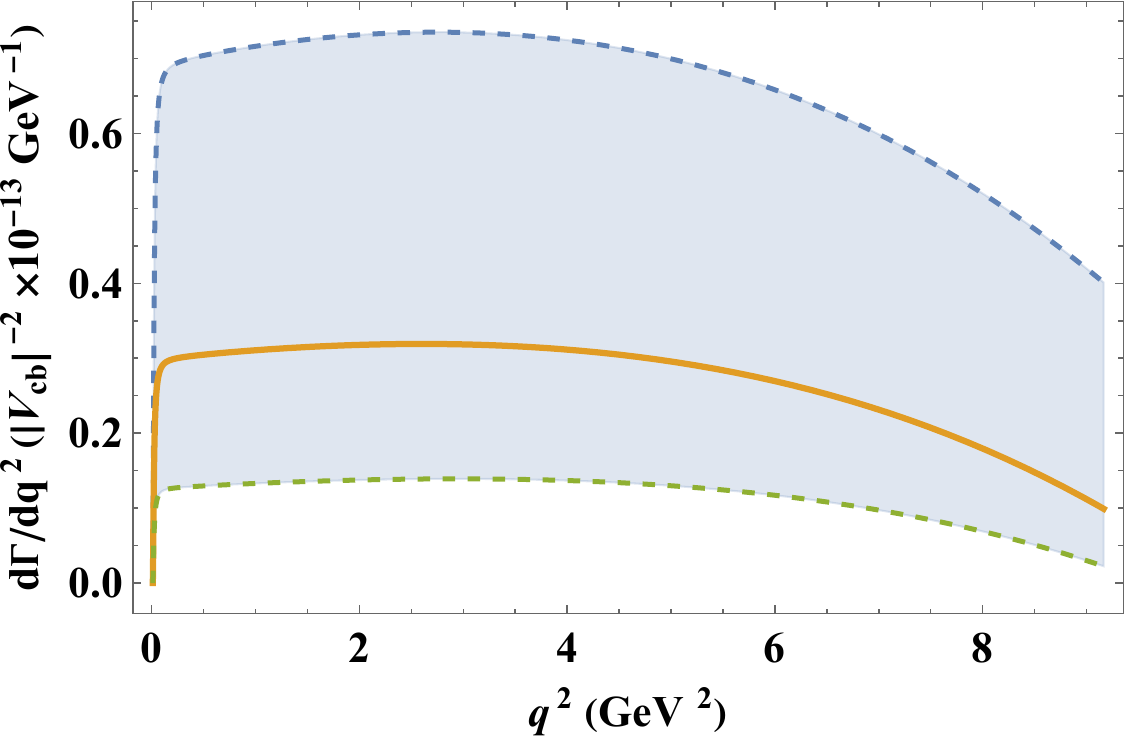}      \quad    	\includegraphics[width=0.31\textwidth]{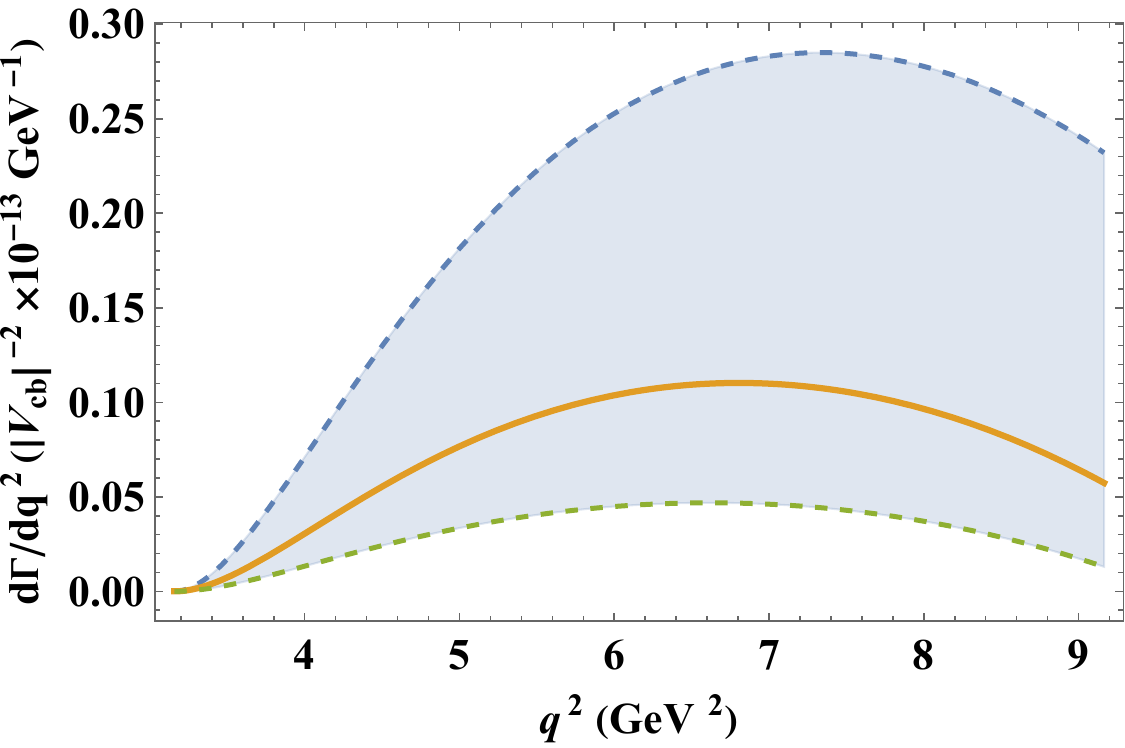} \\
		\includegraphics[width=0.31\textwidth]{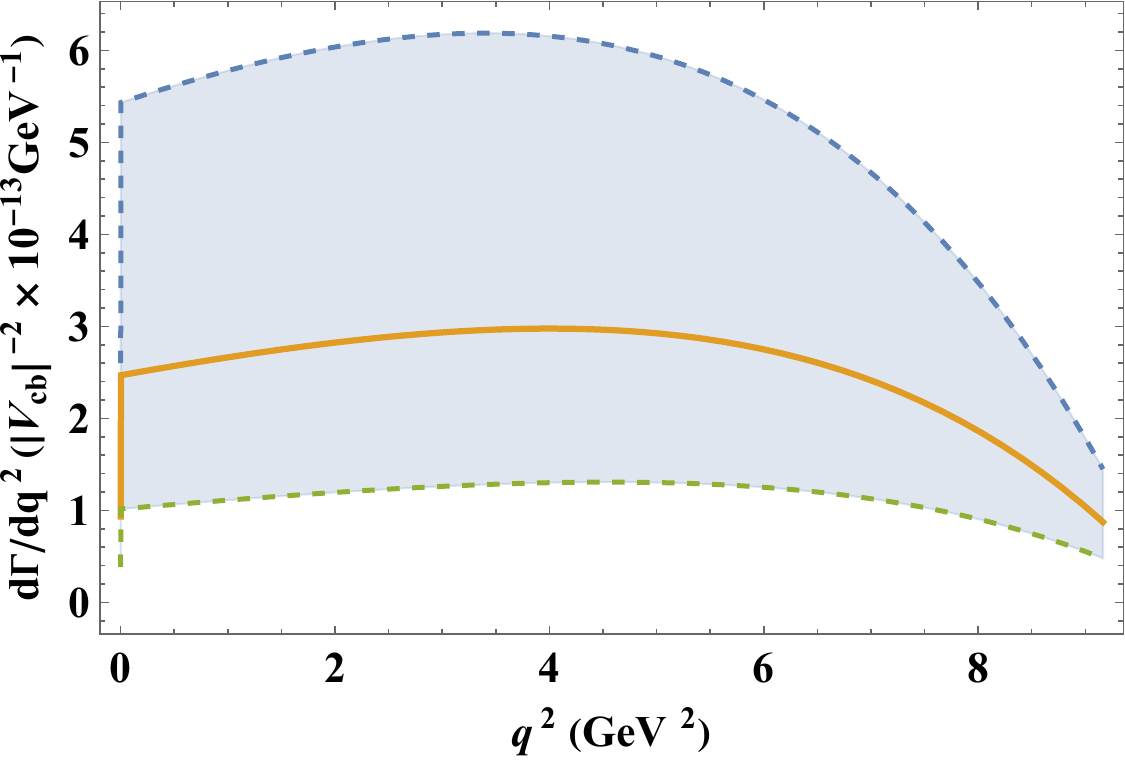}        \quad      	\includegraphics[width=0.31\textwidth]{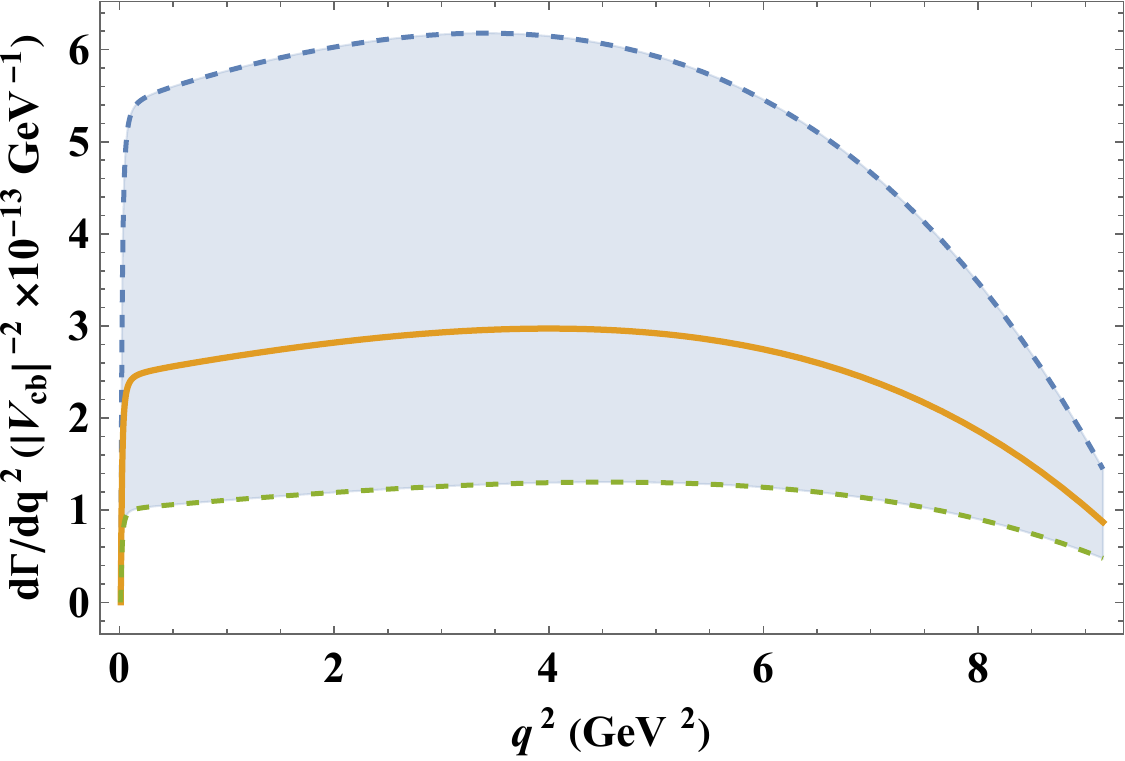}  	   \quad        \includegraphics[width=0.31\textwidth]{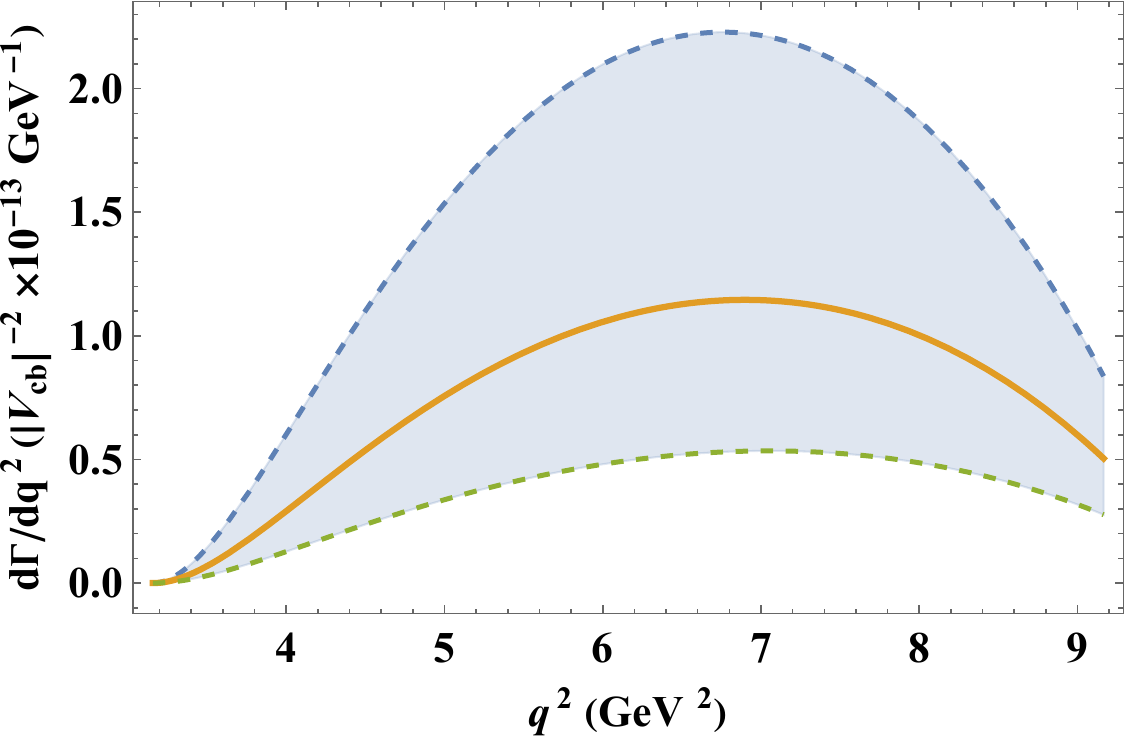} 
	\end{center}
	\caption{Differential decay widths of $\Lambda_b^0 \to \Lambda_c^* \ell^- \bar{\nu}_\ell$ ($\ell=e$ (left column), $\mu$ (middle column), $\tau$(right column)) with $\Lambda_c$ interpolating current $j_{\Lambda_c}^P$ (top row) and $j_{\Lambda_c}^A$ (bottom row), within the LCDAs of $\Lambda_b$ described in reference \cite{Ali:2012pn}.}\label{fig:ddwp1}
\end{figure*}	

	\begin{figure*}
	\begin{center}
		\includegraphics[width=0.31\textwidth]{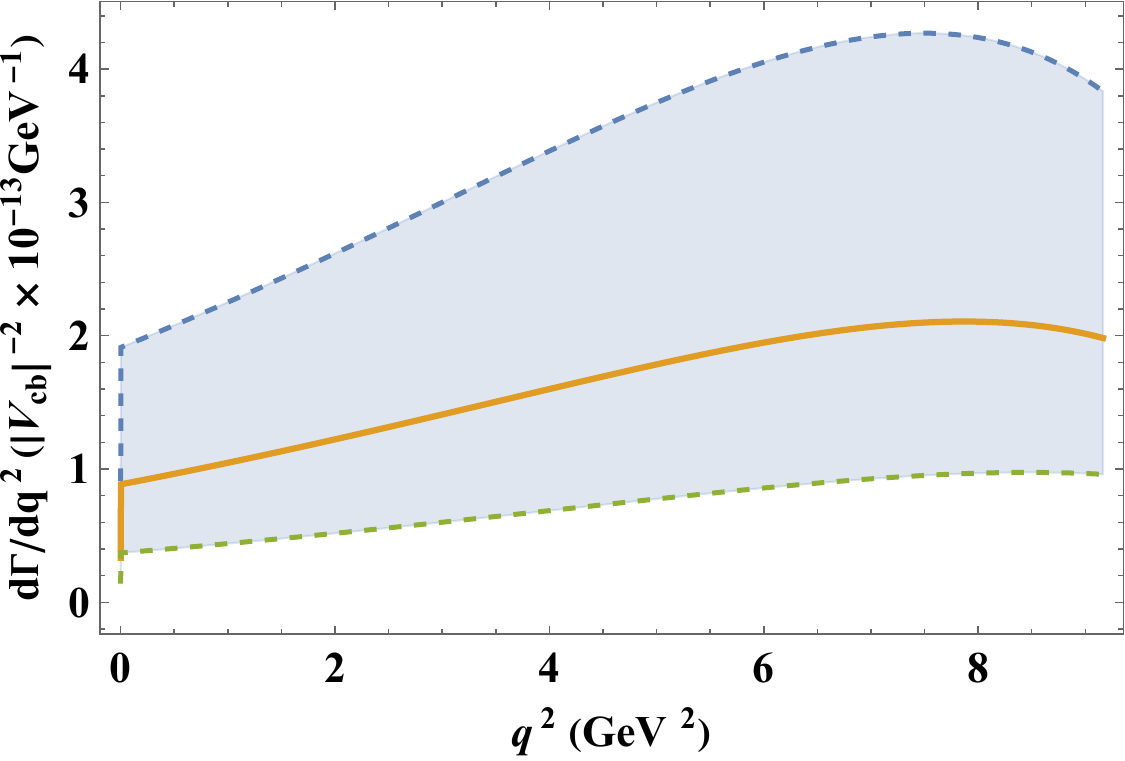}		\quad		\includegraphics[width=0.31\textwidth]{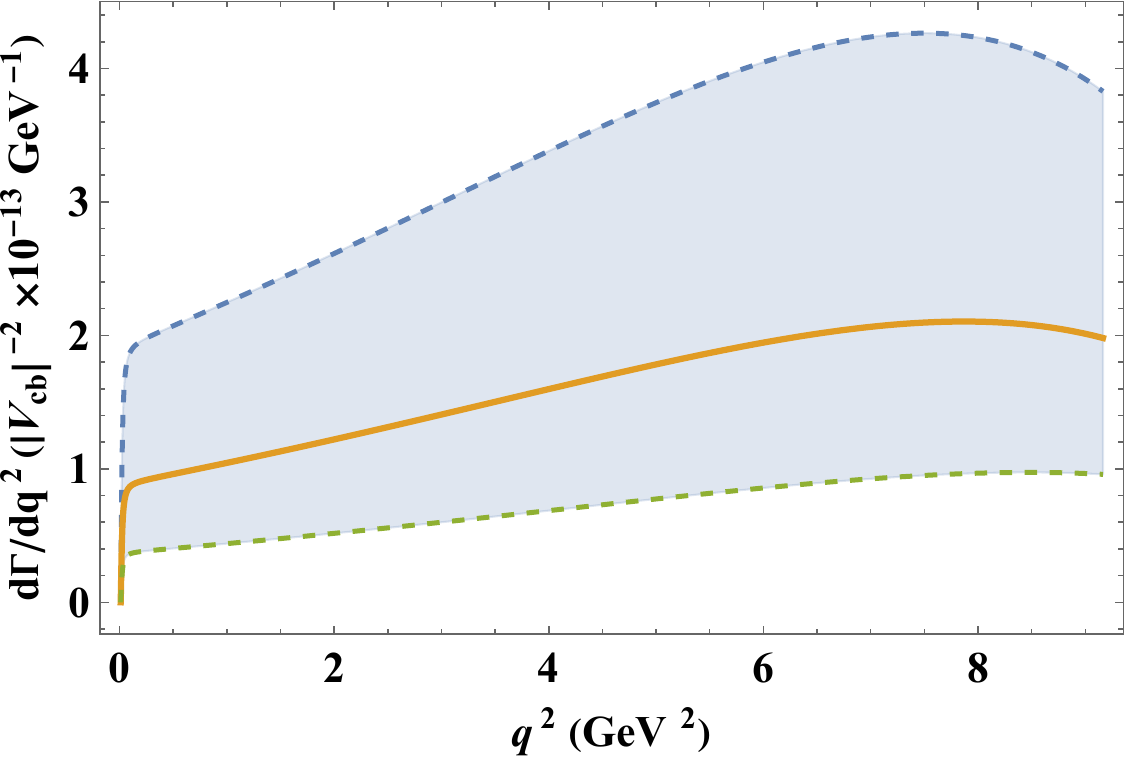}      \quad    	\includegraphics[width=0.31\textwidth]{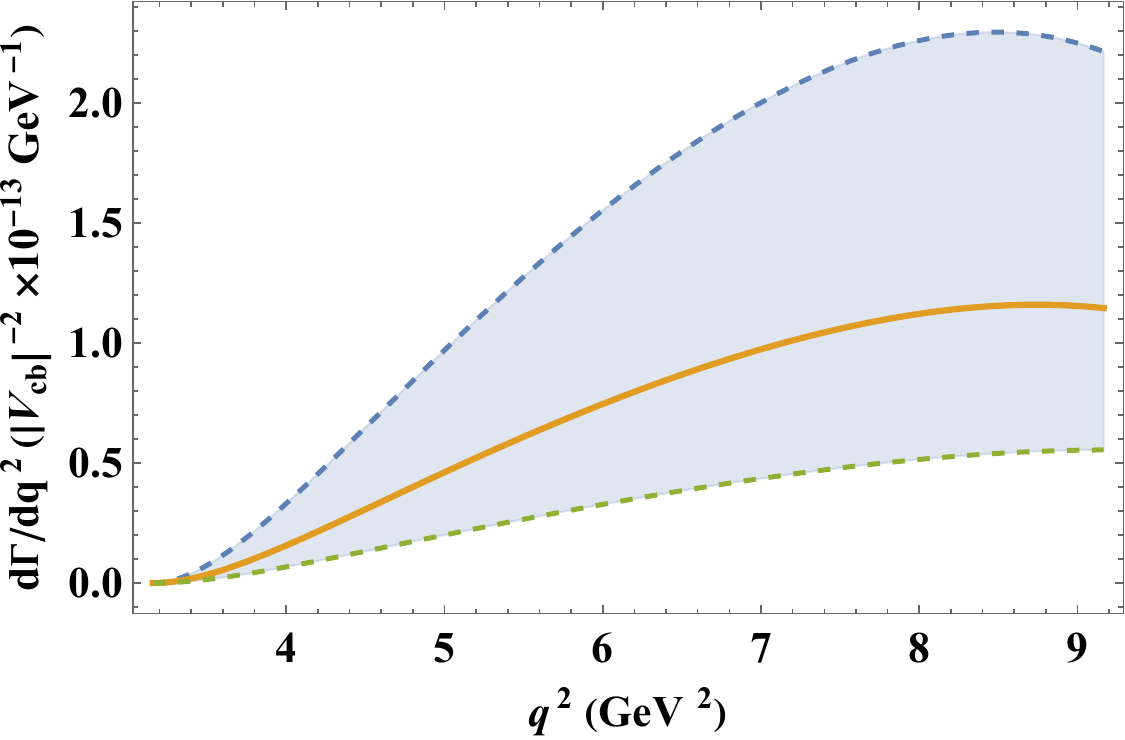} \\
		\includegraphics[width=0.31\textwidth]{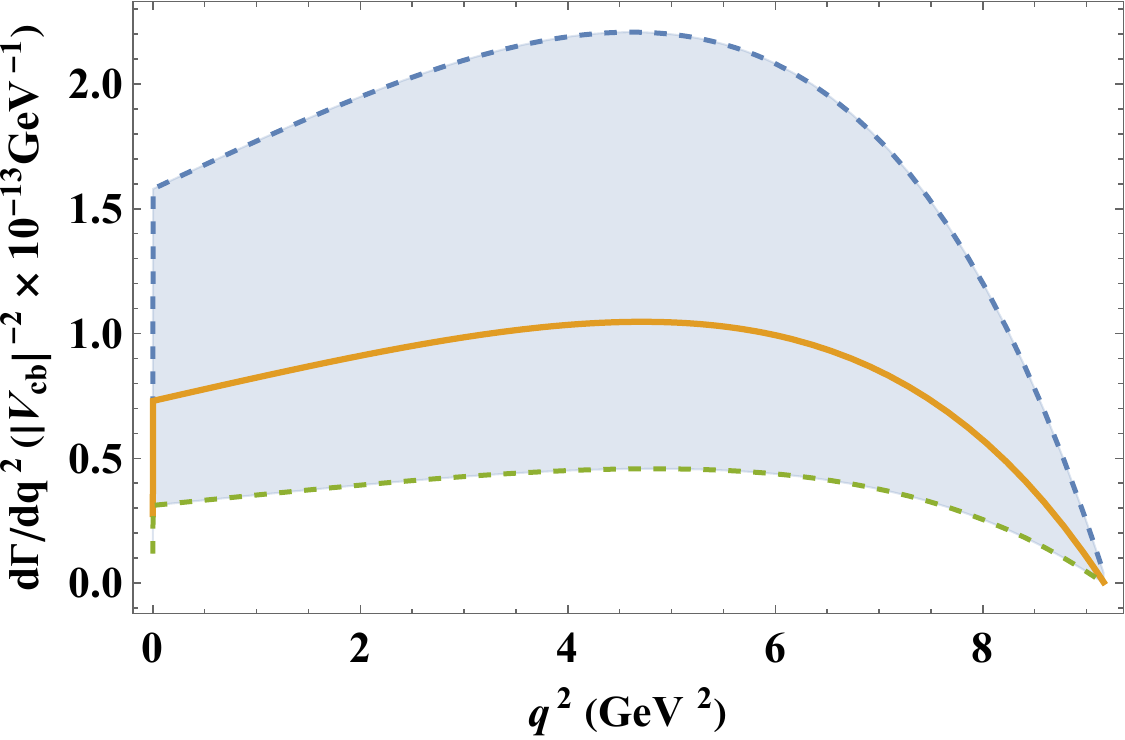}        \quad      	\includegraphics[width=0.31\textwidth]{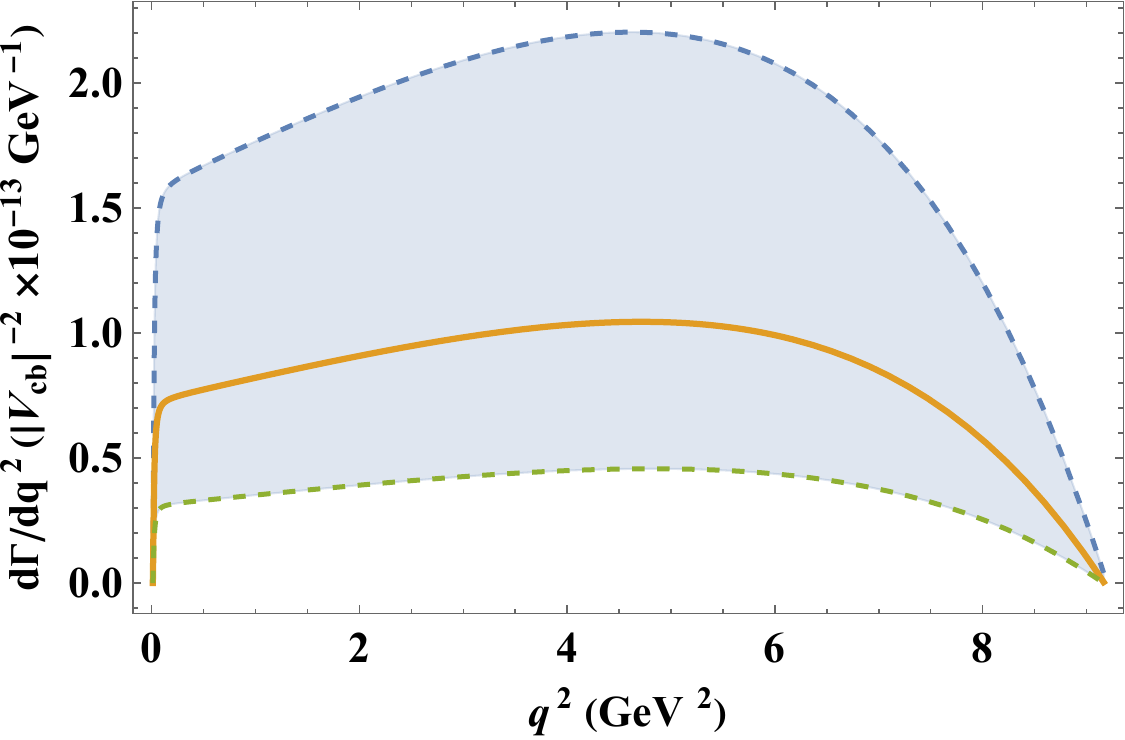}  	   \quad        \includegraphics[width=0.31\textwidth]{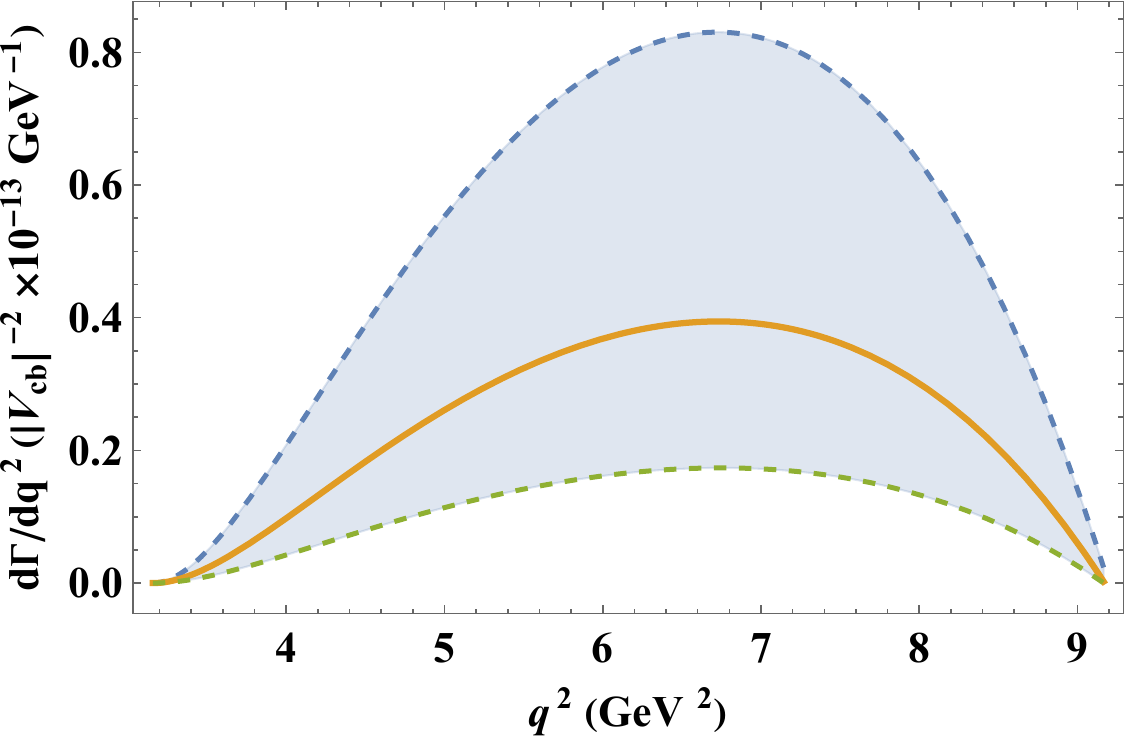} 
	\end{center}
	\caption{Similar to figure \ref{fig:ddwp1}, but for the free parton LCDAs models described in reference \cite{Bell:2013tfa}.}\label{fig:ddwp2}
\end{figure*}	   
    
     Based on these calculational procedures and input parameters, along with the Fermi constant $G_F = 1.166 \times 10^{-5} \text{ GeV}^{-2}$, the graphical representations of the differential decay widths for the semileptonic decay channels of $\Lambda_b^0 \to \Lambda_c^{*+}\ell^- \bar{\nu}_\ell$ can be obtained. As examples, we plot the differential decay widths graphics in Fig. \ref{fig:ddwp1} within the Type I LCDAs models of $\Lambda_b$ and the interpolating current types $j_{\Lambda_c}^P$ and $j_{\Lambda_c}^A$, in Fig. \ref{fig:ddwp2} within the Type III LCDAs of $\Lambda_b$. 
			
    The absolute branching fractions of the semileptonic decays $\Lambda_b^0 \to \Lambda_c^* \ell^- \bar{\nu}_\ell$ depend on the mean lifetime of the $\Lambda_b^0$ baryon and the value of the CKM matrix element $|V_{cb}|$. For this work, according to the PDG average values, the lifetime of $\Lambda_b^0$ baryon is adopted as $\tau_{\Lambda_b^0} = (1.471 \pm 0.009) \times 10^{-12} \text{ s}$. However, $|V_{cb}|$ is not a precise value up to now. Therefore, for a rough estimation, we resort to the values obtained from the fits of inclusive and exclusive semileptonic decays of the $B$ meson in the PDG, specifically $|V_{cb}| = (40.8 \pm 1.4) \times 10^{-3}$.  
	 
	  \begin{table*}  
	 	\centering  
	 	\caption{Decay widths and branching fractions of $\Lambda_b^0 \to \Lambda_c(2595)^+ \ell^-\bar{\nu}_\ell$ for ours and compared with others.} \label{table5}  
	 	\begin{tabular}{cccccc} \hline
	 	   \multirow{2}*{$\Lambda_b$ LCDAs model} &	\multirow{2}*{Decay channel}                        &     \multicolumn{2}{c}{Decay width $\Gamma$ ($\times|V_{cb}|^2 \times 10^{-12}~\rm{GeV}$)}       &          \multicolumn{2}{c}{Branching fraction($\times 10^{-3})$}              \\ 
	 	&	&       $j_{\Lambda_c}^P$                     &  $j_{\Lambda_c}^A$                         &    $j_{\Lambda_c}^P$             &  $j_{\Lambda_c}^A$                          \\  \midrule
	 	\multirow{4}*{Type I}&	$\Lambda_b^0 \to \Lambda_c(2595)^+e^-\bar{\nu}_e$             &    $0.247^{+0.356}_{-0.142}$                &  $2.336^{+2.441}_{-1.300}$                    &    $0.918^{+1.324}_{-0.528}$     &  $8.692_{-4.836}^{+9.082}$                     \\
	 	                          &	$\Lambda_b^0 \to \Lambda_c(2595)^+\mu^-\bar{\nu}_\mu$         &    $0.245^{+0.354}_{-0.141}$                &  $2.325^{+2.428}_{-1.293}$                    &    $0.913^{+1.317}_{-0.526}$     &  $8.649_{-4.811}^{+9.033}$                     \\
	 	&	$\Lambda_b^0 \to \Lambda_c(2595)^+\tau^-\bar{\nu}_\tau$       &    $0.046^{+0.077}_{-0.027}$                &  $0.468^{+0.439}_{-0.249}$                    &    $0.172^{+0.285}_{-0.102}$     &  $1.741_{-0.926}^{+1.632}$           \\ \midrule
	 	\multirow{4}*{Type II}&	$\Lambda_b^0 \to \Lambda_c(2595)^+e^-\bar{\nu}_e$             &    $0.021^{+0.023}_{-0.011}$                &  $0.784^{+0.671}_{-0.391}$                    &    $0.079^{+0.084}_{-0.043}$     &  $2.917^{+2.496}_{-1.453}$                     \\
	 	&	$\Lambda_b^0 \to \Lambda_c(2595)^+\mu^-\bar{\nu}_\mu$         &    $0.021^{+0.023}_{-0.011}$                &  $0.781^{+0.668}_{-0.389}$                    &    $0.078^{+0.084}_{-0.042}$     &  $2.905_{-1.448}^{+2.486}$                     \\
	 	&	$\Lambda_b^0 \to \Lambda_c(2595)^+\tau^-\bar{\nu}_\tau$       &    $0.004^{+0.004}_{-0.002}$                &  $0.201^{+0.167}_{-0.098}$                    &    $0.015^{+0.016}_{-0.008}$     &  $0.747_{-0.365}^{+0.622}$           \\ \midrule
	 		\multirow{4}*{Type III}&	$\Lambda_b^0 \to \Lambda_c(2595)^+e^-\bar{\nu}_e$             &    $0.770^{+0.862}_{-0.434}$                &  $1.499^{+1.617}_{-0.838}$                    &    $2.864^{+3.206}_{-1.615}$     &  $5.576^{+6.015}_{-3.117}$                     \\
	 	&	$\Lambda_b^0 \to \Lambda_c(2595)^+\mu^-\bar{\nu}_\mu$         &    $0.766^{+0.858}_{-0.432}$                &  $1.494^{+1.611}_{-0.835}$                    &    $2.850^{+3.190}_{-1.608}$     &  $5.558_{-3.106}^{+5.995}$                     \\
	 	&	$\Lambda_b^0 \to \Lambda_c(2595)^+\tau^-\bar{\nu}_\tau$       &    $0.148^{+0.165}_{-0.083}$                &  $0.423^{+0.439}_{-0.231}$                    &    $0.550^{+0.615}_{-0.309}$     &  $1.572_{-0.861}^{+1.633}$           \\ \midrule
	 		\multirow{4}*{Type IV}&	$\Lambda_b^0 \to \Lambda_c(2595)^+e^-\bar{\nu}_e$             &    $0.025^{+0.028}_{-0.014}$                &  $0.754^{+0.759}_{-0.407}$                    &    $0.092^{+0.104}_{-0.052}$     &  $2.804_{-1.513}^{+2.824}$                     \\
	 	&	$\Lambda_b^0 \to \Lambda_c(2595)^+\mu^-\bar{\nu}_\mu$         &    $0.024^{+0.028}_{-0.014}$                &  $0.751^{+0.757}_{-0.405}$                    &    $0.091^{+0.103}_{-0.052}$     &  $2.794_{-1.508}^{+2.814}$                     \\
	 	&	$\Lambda_b^0 \to \Lambda_c(2595)^+\tau^-\bar{\nu}_\tau$       &    $0.005^{+0.005}_{-0.003}$                &  $0.214^{+0.208}_{-0.113}$                    &    $0.018^{+0.020}_{-0.010}$     &  $0.795^{+0.775}_{-0.421}$           \\ \midrule
	 		\multirow{4}*{Type V}&	$\Lambda_b^0 \to \Lambda_c(2595)^+e^-\bar{\nu}_e$             &    $0.111^{+0.102}_{-0.054}$                &  $6.418^{+3.253}_{-2.324}$                    &    $0.414^{+0.380}_{-0.200}$     &  $23.875_{-8.647}^{+12.101}$                     \\
	 	&	$\Lambda_b^0 \to \Lambda_c(2595)^+\mu^-\bar{\nu}_\mu$         &    $0.111^{+0.102}_{-0.053}$                &  $6.380^{+3.232}_{-2.310}$                    &    $0.412^{+0.378}_{-0.199}$     &  $23.735_{-8.593}^{+12.024}$                     \\
	 	&	$\Lambda_b^0 \to \Lambda_c(2595)^+\tau^-\bar{\nu}_\tau$       &    $0.020^{+0.020}_{-0.010}$            &$1.049^{+0.490}_{-0.354}$    &  $0.075^{+0.073}_{-0.037}$                    &    $3.903^{+1.824}_{-1.316}$     \\  \midrule     
	 	   PDG \cite{ParticleDataGroup:2022pth,ParticleDataGroup:2024cfk}                  &  $\Lambda_b^0 \to \Lambda_c(2595)^+\ell^-\bar{\nu}_\ell$    & \multicolumn{2}{c}{$-$} & \multicolumn{2}{c}{$7.9^{+4.0}_{-3.5}$} \\ 
	 	   \midrule
	 	    \multirow{3}*{LFQM \cite{Li:2021qod}}  &  $\Lambda_b^0 \to \Lambda_c(2595)^+e^-\bar{\nu}_e$    & \multicolumn{2}{c}{$-$} & \multicolumn{2}{c}{$17.3\pm5.9$} \\
	 	    & $\Lambda_b^0 \to \Lambda_c(2595)^+\mu^-\bar{\nu}_\mu$    & \multicolumn{2}{c}{$-$} & \multicolumn{2}{c}{$17.2\pm5.8$} \\
	 	    & $\Lambda_b^0 \to \Lambda_c(2595)^+\tau^-\bar{\nu}_\tau$    & \multicolumn{2}{c}{$-$} & \multicolumn{2}{c}{$2.4\pm1.1$} \\
	 	    \midrule
	 	     \multirow{3}*{CCQM \cite{Gutsche:2018nks}}  &  $\Lambda_b^0 \to \Lambda_c(2595)^+e^-\bar{\nu}_e$    & \multicolumn{2}{c}{$-$} &\multicolumn{2}{c}{$8.6\pm1.7$} \\
	 	    & $\Lambda_b^0 \to \Lambda_c(2595)^+\mu^-\bar{\nu}_\mu$    & \multicolumn{2}{c}{$-$} & \multicolumn{2}{c}{$8.5\pm1.7$} \\
	 	    & $\Lambda_b^0 \to \Lambda_c(2595)^+\tau^-\bar{\nu}_\tau$    & \multicolumn{2}{c}{$-$} & \multicolumn{2}{c}{$1.1\pm0.2$} \\
	 	   \hline
	 	\end{tabular}
	 \end{table*}
	 
	 Taking these input parameters into account, we derived the decay widths of the process $\Lambda_b^0 \to \Lambda_c^*\ell^-\bar\nu_\ell$ and calculated the absolute branching fractions for both the $j_{\Lambda_c}^P$ and $j_{\Lambda_c}^A$ types interpolating currents. These results are tabulated in Table \ref{table5}.

    With these absolute branching fractions of semileptonic decays $\Lambda_b^0 \to \Lambda_c^* \ell^-\bar{\nu}_\ell$, we can examine the value of lepton flavor universality $\mathcal{R}(\Lambda_c^*)$. $\mathcal{R}(\Lambda_c^*)$ is the ratio of absolute branching fractions of $\Lambda_b^0\to\Lambda_c^*\tau^-\bar{\nu}_\tau$ and $\Lambda_b^0\to\Lambda_c^*\mu^-\bar{\nu}_\mu$, given by $\mathcal{R}(\Lambda_c^*)=\mathcal{B}r(\Lambda_b^0\to\Lambda_c^*\tau^-\bar{\nu}_\tau)/\mathcal{B}r(\Lambda_b^0\to\Lambda_c^*\mu^-\bar{\nu}_\mu)$. In this work, for the $j_{\Lambda_c}^P$ and $j_{\Lambda_c}^A$ type baryon interpolating current, we give their $\mathcal{R}(\Lambda_c^*)$ values for each LCDA model in Table \ref{table6}. We also list a comparison with other research values in the same table. Notably, the four values of $\mathcal{R}(\Lambda_c^*)$ in Table \ref{table6} (from Ref. \cite{Pervin:2005ve}) correspond to four distinct models discussed in that reference.
    
    \begin{table*}
    	\centering
    	\caption{Lepton flavor universality ratio $\mathcal{R}(\Lambda_c^*)$ and compared with other works.} \label{table6}
    	\begin{tabular}{ccccccc} \hline
    	                               This work            &                    &  Type I  &  Type II  &  Type III &  Type IV  &  Type V      \\ 
    	                                           \midrule
    	\multirow{2}*{$\mathcal{R}(\Lambda_c^*)$}  & $j_{\Lambda_c}^A$  &   $0.201^{+0.011}_{-0.011}$       &   $0.257^{+0.005}_{-0.003}$        &   $0.253^{+0.007}_{-0.005}$        &   $0.285^{+0.006}_{-0.005}$        &     $0.164^{+0.006}_{-0.004}$       \\ 
    	                                           & $j_{\Lambda_c}^P$  &  $0.188^{+0.016}_{-0.008}$        &    $0.189^{+0.001}_{-0.000}$       &    $0.193^{+0.001}_{-0.000}$       &     $0.193^{+0.001}_{-0.001}$      &    $0.182^{+0.006}_{-0.002}$         \\       \hline
          References          &      \multicolumn{2}{c}{\cite{Li:2021qod}}   &   \multicolumn{2}{c}{\cite{Pervin:2005ve}}  &  \multicolumn{2}{c}{\cite{Gutsche:2018nks}}              \\  \midrule
    	    $\mathcal{R}(\Lambda_c^*)$             &           \multicolumn{2}{c}{$0.14\pm0.01$}  & \multicolumn{2}{c}{0.21, 0.22, 0.26, 0.31} &   \multicolumn{2}{c}{$0.13\pm0.03$}     \\ 
    	                                               \hline
    	\end{tabular}
    \end{table*} 
 
    \section{Conclusion and disscussion} \label{sec:V}

     In the framework of light-cone QCD sum rules, we derived the theoretical form of $\Lambda_b^0 \to \Lambda_c^*$ transition form factors within five different LCDAs models for the $\Lambda_b$ baryon and two types interpolating currents for the $\Lambda_c$ baryon (pseudoscalar $j^P_{\Lambda_c}$ and axial-vector $j^A_{\Lambda_c}$). For all $j_{\Lambda_c}^P$ and $j_{\Lambda_c}^A$ type interpolating currents, these form factors satisfy the relations $f_1^* = g_1^*$ and $f_2^* = f_3^* = g_2^* = g_3^*$ . However, different interpolating current types of the $\Lambda_c$ baryon introduce opposing signs of form factors due to a phase difference, but these signs will cancel out in the calculation of decay widths due to the squaring of the form factors. The form factors obtained in this work are listed in Table \ref{table3}. For a comparison with Lattice data, we also insert the Lattice data into our form factors plots in Figs. \ref{fig1.2}, \ref{fig1.3}, \ref{fig1.4}, \ref{fig1.5} and \ref{fig1.6}.
	
	 With the helicity amplitudes of baryon decay, we also provided the decay widths for semileptonic decays $\Lambda_b^0 \to \Lambda_c^* \ell^- \bar{\nu}_\ell$ for each model of $\Lambda_b$ LCDAs and two types of $\Lambda_c$ interpolating currents. Comparing with experimental values, other theoretical predictions, and the branching ratios obtained from five LCDAs models of $\Lambda_b$ and two types interpolating current for the $\Lambda_c$ baryon (as expressed in Table \ref{table5}), we discovered that the $j^A_{\Lambda_c}$ interpolating current gives good consistent branching ratios, except for the Type V LCDAs of $\Lambda_b$. However, the Type V LCDAs of $\Lambda_b$ give large values for $j^A_{\Lambda_c}$, but these are consistent with the LFQM results in \cite{Li:2021qod}.  All the branching ratios for $j^P_{\Lambda_c}$ are smaller than those obtained from experiment, other theoretical models and the results obtained using $j^A_{\Lambda_c}$. But for the free parton models of $\Lambda_b$ LCDAs (Type III), both $j^P_{\Lambda_c}$ and $j^A_{\Lambda_c}$ give good predictions for our sum rules. The parameter $A$ in the LCDAs of bottom baryons with Type I only functions effectively within a narrow range in the semileptonic decay channels $\Lambda_b^0 \to \Lambda_c^* \ell^-\bar{\nu}_\ell$. For small LCDAs parameter $A$ in \cite{Ali:2012pn}, we observed that the $\Lambda_b$ baryon current $\bar{J}_2 = \epsilon_{abc}[\bar{d}(x)\slashed{v}\gamma_5\mathcal{C}^T\bar{u}^b(x)]\bar{h}_v^c(x)$ and the $\Lambda_c$ interpolating current $j_{\Lambda_c}^A$ make significant contributions to the light-cone sum rules for $\Lambda_b^0\to\Lambda_c^*\ell^-\bar{\nu}_\ell$. On the other hand, form factors for the $\Lambda_b^0$ to $\Lambda_c^*$ transition and branching fractions of semileptonic decays $\Lambda_b^0 \to \Lambda_c^* \ell^- \bar{\nu}_\ell$ are sensitive to the mass of the charm quark.  
		
    The ratio of branching ratios of semileptonic decay $\mathcal{R}(\Lambda_c^*)$ is irrelevant to the CKM matrix element, and it is an important parameter to test the lepton flavor universality. The value derived in this work falls within the range of other studies for all the $\Lambda_b$ LCDAs models and the two types of $\Lambda_c$ interpolating currents, offering a viable region for future exploration.
    
    We can see that in the context and other works \cite{Duan:2022uzm, Li:2021qod, Miao:2022bga}, although applying the LCDAs of $\Lambda_b$ in the $\frac{1}{2}^+ \to \frac{1}{2}^+$ processes yields satisfactory results, the results in the $\frac{1}{2}^+ \to \frac{1}{2}^-$ processes depend on the models we choose. This phenomenology needs more in-depth investigation into the LCDAs of heavy baryons and the interpolating currents of hadrons in future works.
	
	\section*{Acknowledgments}
	This work was supported by the Postdoctoral Fellowship Program of CPSF under Contract No. GZC20230738, National Natural Science Foundation of China under Grants No. 12275067 and Natural Science Foundation of Henan Province under Grant No. 225200810030.

	\begin{appendices}
	\section{Hadronic representation of correlation function}
	\setcounter{equation}{0}
	\renewcommand\theequation{A.\arabic{equation}}
	 \begin{align}
		T_\mu(p,q)=&\frac{f_{\Lambda_c}}{M_{\Lambda_c}^2-p^2}\Bigg\{2M_{\Lambda_b}f_1 v_\mu-\Big[(M_{\Lambda_b} -M_{\Lambda_c})f_1   \notag \\
		           &-\frac{M_{\Lambda_b}^2-M_{\Lambda_c}^2}{M_{\Lambda_b}}f_2\Big]\gamma_\mu-\Big[2f_1-\frac{M_{\Lambda_b}+M_{\Lambda_c}}{M_{\Lambda_b}}(f_2 \notag \\
		           &+f_3)\Big]q_\mu-2f_2 v_\mu \slashed{q}+\Big(f_1-\frac{M_{\Lambda_b}+M_{\Lambda_c}}{M_{\Lambda_b}}f_2\Big)\gamma_\mu\slashed{q} \notag \\
		           & +\frac{1}{M_{\Lambda_b}}(f_2-f_3)q_\mu\slashed{q}-2M_{\Lambda_b}g_1 v_\mu\gamma_5-\Big[(M_{\Lambda_b} \notag \\
		           &+M_{\Lambda_c})g_1+\frac{M_{\Lambda_b}^2-M_{\Lambda_c}^2}{M_{\Lambda_b}}g_2\Big]\gamma_\mu \gamma_5+\Big[2 g_1 \notag \\
		           &+\frac{M_{\Lambda_b}-M_{\Lambda_c}}{M_{\Lambda_b}}(g_2 +g_3)\Big]q_\mu\gamma_5+2g_2v_\mu\slashed{q}\gamma_5 \notag \\
		           &-\Big(g_1+\frac{M_{\Lambda_b}-M_{\Lambda_c}}{M_{\Lambda_b}}g_2\Big)\gamma_\mu\slashed{q}\gamma_5-\frac{1}{M_{\Lambda_b}}(g_2\notag \\
		           &-g_3)q_\mu\slashed{q}\gamma_5\Bigg\} u_{\Lambda_b}(v)  \notag \\
		           &+\frac{f_{\Lambda_c^*}}{M_{\Lambda_c^*}^2-p^{\prime 2}}\Bigg\{2M_{\Lambda_b}f_1^*v_\mu-\Big[(M_{\Lambda_b}+M_{\Lambda_c^*})f_1^* \notag \\
		           &+\frac{M_{\Lambda_b}^2-M_{\Lambda_c}^2}{M_{\Lambda_b}}f_2^*\Big]\gamma_\mu-\Big[2f_1^*+\frac{M_{\Lambda_b}-M_{\Lambda_c^*}}{M_{\Lambda_b}}(f_2^* \notag \\ 
		           &+f_3^*)\Big]q_\mu+2f_2^*v_\mu\slashed{q}+\Big(f_1^*+\frac{M_{\Lambda_b}-M_{\Lambda_c^*}}{M_{\Lambda_b}}f_2^*\Big)\gamma_\mu\slashed{q} \notag \\
		           &-\frac{1}{M_{\Lambda_b}}(f_2^*-f_3^*)q_\mu\slashed{q}-2M_{\Lambda_b}g_1^*v_\mu\gamma_5-\Big[(M_{\Lambda_b} \notag \\
		           &-M_{\Lambda_c^*})g_1^*-\frac{M_{\Lambda_b}^2-M_{\Lambda_c^*}^2}{M_{\Lambda_b}}g_2^*\Big]\gamma_\mu\gamma_5+\Big[2g_1^* \notag \\
		           &-\frac{M_{\Lambda_b}+M_{\Lambda_c^*}}{M_{\Lambda_b}}(g_2^*+g_3^*)\Big]q_\mu\gamma_5-2g_2^*v_\mu\slashed{q}\gamma_5 \notag \\
		           &-\Big(g_1^*-\frac{M_{\Lambda_b}+M_{\Lambda_c^*}}{M_{\Lambda_b}}g_2^*\Big)\gamma_\mu\slashed{q}\gamma_5 \notag \\
		           &+\frac{1}{M_{\Lambda_b}}(g_2^*-g_3^*)q_\mu\slashed{q}\gamma_5\Bigg\}u_{\Lambda_b}(v). \label{a1}
	\end{align}
	We have omitted the higher excited states and the continuum states in Eq. (\ref{a1}); they will be eliminated by the quark-hadron duality and Borel transformations. 
	
    \section{Detail expressions of $\rho_{_{\Gamma_i}}$ in the coefficience of correlation function on quark-gluon level}
    \setcounter{equation}{0}
    \renewcommand\theequation{B.\arabic{equation}}		
		For the interpolating current of $\Lambda_c$ baryon with the scalar type $j_{\Lambda_c}^P$, we have
		\begin{align}
			\rho_{v_\mu}&=-2\sigma M_{\Lambda_b}^3(1-\sigma)\tilde{\psi}_3^s(\omega,u), \notag \\
			\rho_{\gamma_\mu}&=\sigma M_{\Lambda_b}^2[M_{\Lambda_b}(1-\sigma)-m_c]\tilde{\psi}_3^s(\omega, u),  \notag \\
			\rho_{q_\mu}&=2\sigma M_{\Lambda_b}^2\tilde{\psi}_3^s(\omega, u), \notag \\
			\rho_{v_\mu\slashed{q}}&=0, \notag \\
			\rho_{\gamma_\mu\slashed{q}}&=-\sigma M_{\Lambda_b}^2 \tilde{\psi}_3^s(\omega, u), \notag \\
			\rho_{q_\mu\slashed{q}}&=0,  \notag \\
			\rho_{v_\mu\gamma_5}&=2\sigma(1-\sigma)M_{\Lambda_b}^3\tilde{\psi}_3^s(\omega, u), \notag \\
			\rho_{\gamma_\mu\gamma_5}&=\sigma M_{\Lambda_b}^2(M_{\Lambda_b}-\sigma M_{\Lambda_b}+m_c)\tilde{\psi}_3^s(\omega, u), \notag \\
			\rho_{q_\mu\gamma_5}&=-2\sigma M_{\Lambda_b}^2 \tilde{\psi}_3^s(\omega, u), \notag \\
			\rho_{v_\mu\slashed{q}\gamma_5}&=0, \notag \\
			\rho_{\gamma_\mu\slashed{q}\gamma_5}&=\sigma M_{\Lambda_b}^2\tilde{\psi}_3^s(\omega, u), \notag \\
			\rho_{q_\mu\slashed{q}\gamma_5}&=0.
		\end{align}	
		
		For $j_{\Lambda_c}^A$, we have
		\begin{align}
			\rho_{v_\mu}^1=&-2\sigma M_{\Lambda_b}^2 m_c \tilde{\psi}_2(\omega,u),  \notag \\
			\rho_{v_\mu}^2=&-2(1-\sigma)M_{\Lambda_b}^2 m_c [\hat{\psi}_2(\omega, u)-\hat{\psi}_4(\omega, u)], \notag  \\
			\rho_{\gamma_\mu}^1=&\sigma M_{\Lambda_b} [\sigma M_{\Lambda_b}^2-(M_{\Lambda_c}^2-q^2)+M_{\Lambda_b} m_c]\tilde{\psi}_2(\omega, u) \notag \\ & +2 M_{\Lambda_b}[\hat{\psi}_2(\omega, u)-\hat{\psi}_4(\omega, u)], \notag  \\
			\rho_{\gamma_\mu}^2=&M_{\Lambda_b}[\sigma(1-\sigma)M_{\Lambda_b}^2-(1-\sigma)M_{\Lambda_c}^2-\sigma q^2  \notag \\& +(1-\sigma)M_{\Lambda_b}m_c][\hat{\psi}_2(\omega, u)-\hat{\psi}_4(\omega, u)], \notag  \\
			\rho_{q_\mu}^1=&2\sigma M_{\Lambda_b}^2\tilde{\psi}_2(\omega, u), \notag  \\
			\rho_{q_\mu}^2=&2M_{\Lambda_b}m_c[\hat{\psi}_2(\omega, u)-\hat{\psi}_4(\omega, u)],  \notag  \\
			\rho_{v_\mu\slashed{q}}^1=&-2\sigma M_{\Lambda_b}^2\tilde{\psi}_2(\omega, u),  \notag  \\
			\rho_{v_\mu\slashed{q}}^2=&0, \notag  \\
			\rho_{\gamma_\mu\slashed{q}}^1=&-\sigma M_{\Lambda_b}^2\tilde{\psi}_2(\omega, u), \notag  \\
			\rho_{\gamma_\mu\slashed{q}}^2=&-M_{\Lambda_b}m_c[\hat{\psi}_2(\omega, u)-\hat{\psi}_4(\omega, u)], \notag  \\
			\rho_{q_\mu\slashed{q}}^1=&0, \notag  \\
			\rho_{q_\mu\slashed{q}}^2=&0, \notag  \\
			\rho_{v_\mu\gamma_5}^1=&-\rho_{v_\mu}^1, \notag  \\
			\rho_{v_\mu\gamma_5}^2=&-\rho_{v_\mu}^2,  \notag   \\
			\rho_{\gamma_\mu\gamma_5}^1=&-\sigma M_{\Lambda_b}(\sigma M_{\Lambda_b}^2-M_{\Lambda_c}^2+q^2-M_{\Lambda_b}m_c)\tilde{\psi}_2(\omega, u) \notag \\& -2M_{\Lambda_b}[\hat{\psi}_2(\omega, u)-\hat{\psi}_4(\omega, u)], \notag  \\
			\rho_{\gamma_\mu\gamma_5}^2=&M_{\Lambda_b}[(1-\sigma)M_{\Lambda_c}^2-\sigma(1-\sigma)M_{\Lambda_b}^2+\sigma q^2 \notag \\& +(1-\sigma)M_{\Lambda_b}m_c][\hat{\psi}_2(\omega, u)-\hat{\psi}_4(\omega, u)], \notag  \\
			\rho_{q_\mu\gamma_5}^1=&2\sigma M_{\Lambda_b}^2\tilde{\psi}_2(\omega, u),\notag  \\
			\rho_{q_\mu\gamma_5}^2=&-2M_{\Lambda_b}m_c[\hat{\psi}_2(\omega, u)-\hat{\psi}_4(\omega, u)], \notag  \\
			\rho_{v_\mu\slashed{q}\gamma_5}^1=&-\rho_{v_\mu\slashed{q}}^1, \notag  \\
			\rho_{v_\mu\slashed{q}\gamma_5}^2=&-\rho_{v_\mu\slashed{q}}^2, \notag  \\
			\rho_{\gamma_\mu\slashed{q}\gamma_5}^1=&-\sigma M_{\Lambda_b}^2\tilde{\psi}_2(\omega, u), \notag  \\
			\rho_{\gamma_\mu\slashed{q}\gamma_5}^2=&M_{\Lambda_b}m_c[\hat{\psi}_2(\omega, u)-\hat{\psi}_4(\omega, u)], \notag  \\
			\rho_{q_\mu\slashed{q}\gamma_5}^1=&0, \notag  \\
			\rho_{q_\mu\slashed{q}\gamma_5}^2=&0.
		\end{align}
		Where
		\begin{gather}
			\hat{\psi}_i(\omega, u)=\int_{0}^{\omega}d\tau \tau \tilde{\psi}_i(\tau, u).
		\end{gather}
		
		\section{Light-cone distribution amplitudes models of $\Lambda_b$ baryon}
		\setcounter{equation}{0}
		\renewcommand\theequation{C.\arabic{equation}}
		
		The exponatial model, Type II \cite{Bell:2013tfa}:
		\begin{align}
			\psi_2(\omega, u)=&\frac{\omega^2 u (1-u)}{\omega_0^4} e^{-\omega/\omega_0}, \notag \\
			\psi_3^s(\omega, u)=&\frac{\omega}{2\omega_0^3} e^{-\omega/\omega_0},  \notag \\
			\psi_3^\sigma(\omega, u)=&\frac{\omega(2u-1)}{2\omega_0^3}e^{-\omega/\omega_0}, \notag \\
			\psi_4(\omega, u)=&\frac{1}{\omega_0^2} e^{-\omega/\omega_0},
		\end{align}
		 where $\omega_0 \simeq 0.4~\rm{GeV}$.
		 
		 The free parton model, Type III \cite{Bell:2013tfa}
		 \begin{align}
		 	\psi_2(\omega, u)=&\frac{15\omega^2 u (1-u) (2\bar{\Lambda}-\omega)}{4\bar{\Lambda}^5} \theta(2 \bar{\Lambda}-\omega), \notag \\
		 	\psi_3^s(\omega, u)=&\frac{15\omega(2\bar{\Lambda}-\omega)^2}{16 \bar{\Lambda}^5} \theta(2\bar{\Lambda}-\omega),  \notag \\
		 	\psi_3^\sigma(\omega, u)=&\frac{15\omega(2 \bar{\Lambda}-\omega)^2 (2u-1)}{8 \bar{\Lambda}^5}\theta(2\bar{\Lambda}-\omega), \notag \\
		 	\psi_4(\omega, u)=&\frac{5(2\bar{\Lambda}-\omega)^3}{8\bar{\Lambda}^5}\theta(2\bar{\Lambda}-\omega),
		 \end{align}
		 where $\theta(2\bar{\Lambda}-\omega)$ is step-function and $\bar{\Lambda}=M_{\Lambda_b}-m_b \simeq 0.8~\rm{GeV}$.
		 
		  The leading order pertubative contribution model to the sum rules, Type IV \cite{Ball:2008fw}:
		 \begin{align}
		 	\tilde{\psi}_2(\omega,u)=&\frac{15}{2}\mathcal{N}^{-1}\omega^2(1-u)u\int_{\omega/2}^{s_0^{\Lambda_b}}ds e^{-s/\tau}(s-\omega/2), \notag \\
		 	\tilde{\psi}_3^s(\omega,u)=&\frac{15}{4}\mathcal{N}^{-1}\omega\int_{\omega/2}^{s_0^{\Lambda_b}}ds e^{-s/\tau}(s-\omega/2)^2, \notag \\
		 	\tilde{\psi}_3^\sigma(\omega,u)=&\frac{15}{4}\mathcal{N}^{-1}\omega(2u-1)\int_{\omega/2}^{s_0^{\Lambda_b}}ds e^{-s/\tau}(s-\omega/2)^2, \notag \\
		 	\tilde{\psi}_4(\omega,u)=&5\mathcal{N}^{-1}\int_{\omega/2}^{s_0^{\Lambda_b}}dse^{-s/\tau}(s-\omega/2)^3,
		 \end{align}
		 with $s_0^{\Lambda_b} \simeq 1.2~\rm{GeV}$, $\tau\simeq 0.6~\rm{GeV}$, and $$\mathcal{N}=\int_{0}^{s_0^{\Lambda_b}}ds s^5 e^{-s/\tau}.$$
		 
		 The Gegenbauer moment model within QCD sum rules, Type V \cite{Ball:2008fw}:
		  \begin{align}
		 	\tilde{\psi}_2(\omega,u)=&\omega^2u(1-u)[\frac{1}{\epsilon_0^4}e^{-\omega/\epsilon_0} \notag \\ & +a_2C_2^{3/2}(2u-1)\frac{1}{\epsilon_1^4}e^{-\omega/\epsilon_1}], \notag \\
		 	\tilde{\psi}_3^s(\omega,u)=&\frac{\omega}{2\epsilon_3^3}e^{-\omega/\epsilon_3}, \notag \\
		 	\tilde{\psi}_3^\sigma(\omega,u)=&\frac{\omega}{2\epsilon_3^3}(2u-1)e^{-\omega/\epsilon_3}, \notag \\
		 	\tilde{\psi}_4(\omega,u)=&5\mathcal{N}^{-1}\int_{\omega/2}^{s_0^{\Lambda_b}}dse^{-s/\tau}(s-\omega/2)^3,
		 \end{align}
		 with $\epsilon_0 \simeq 200~\rm{MeV}$, $\epsilon_1 \simeq 650~\rm{MeV}$, $\epsilon_3 \simeq 230~\rm{MeV}$ and $a_2 \simeq 0.333$.
		 
  \end{appendices}
  
   \bibliographystyle{unsrt}
   
   \bibliography{references}
  
\end{document}